\documentclass[12pt]{article}
\usepackage{etex}
\usepackage[utf8]{inputenc}
\usepackage{microtype}

\def\lsim{\raise0.3ex\hbox{$<$\kern-0.75em\raise-1.1ex\hbox{$\sim$}}}
\def\gsim{\raise0.3ex\hbox{$>$\kern-0.75em\raise-1.1ex\hbox{$\sim$}}}
\def\noi{\noindent}

\def\bea{\begin{eqnarray}}
\def\eea{\end{eqnarray}}
\def\beq{\begin{equation}}
\def\eeq{\end{equation}}

\def\beeq{\begin{eqnarray}}
\def\eeeq{\end{eqnarray}}
\def\R{ {\rm R\kern -.31cm I \kern .15cm}}
\def\C{ {\rm C \kern -.15cm \vrule width.5pt \kern .12cm}}
\def\Z{ {\rm Z \kern -.27cm \angle \kern.02cm}}
\def\N{ {\rm N \kern -.26cm \vrule width.4pt \kern .10cm}}
\def\1{{\rm 1\mskip-4.5mu l} }

\usepackage[pdftex]{graphicx}
\usepackage{amsmath,caption}

\newlength{\graphewidth}
\newlength{\grapheheight}
\begin{document}

\begin{center} 

\par 
\vskip 5 truemm

{\large \bf Heavy baryon wave functions,\\[1mm]
Bakamjian-Thomas approach to form factors,\\[2mm]
and observables in $\boldsymbol{\Lambda_b \to \Lambda_c\left({1 \over 2}^\pm \right) \ell \overline{\nu}}$ transitions}

\par \vskip 5 truemm

 {D. Be\v{c}irevi\'c $^a$, A. Le Yaouanc $^a$,  V. Mor\'enas $^b$ and L. Oliver $^a$}

\vskip 2 truemm

$^a$ {\it Laboratoire de Physique Th\'eorique} \footnote{Unit\'e Mixte de Recherche UMR 8627 - CNRS}
\vskip 1 truemm
{\it Universit\'e de Paris XI, B\^atiment 210, 91405 Orsay Cedex, France}

\vskip 2 truemm

$^b$ {\it Laboratoire de Physique de Clermont-Ferrand \footnote{Unit\'e Mixte de Recherche UMR 6533 - CNRS}\par
Campus des C\' ezeaux\par
4 Avenue Blaise Pascal, TSA 60026, CS 60026, 
63178 Aubi\`ere Cedex, France}

\end{center}

\begin{abstract}

Motivated by the calculation of observables in the decays $\Lambda_b \to \Lambda_c\left({1 \over 2}^\pm \right) \ell \overline{\nu}$, as possible tests of Lepton Flavor Universality (LFU), we present a calculation of the necessary form factors in the quark model. Our scheme combines a spectroscopic model, providing the internal wave functions, and the Bakamjian-Thomas (BT) relativistic formalism to deduce the wave functions in motion, and then the current matrix elements, that amount in the heavy quark limit to the Isgur-Wise (IW) function. This limit is covariant and satisfies a large set of sum rules.
This framework has been successfully applied to mesons. On the other hand, for baryons, we meet difficulties using standard spectroscopic models. This leads us to propose a provisory spectroscopic phenomenological model : a Q-pointlike-diquark model, non-relativistic, with harmonic oscillator forces, flexible enough to give both a reasonable low-lying spectrum and the expected slope of the IW function through the BT formalism. To begin, we extract this slope from Lattice QCD data and find it to be around $\rho_\Lambda^2 \sim 2$, which we use as a guideline. Then we find and try to explain why we are not able to reproduce the right $\rho_\Lambda^2$ when using certain typical standard linear + Coulomb potential models, both with three quarks $Qqq$ or in a Q-pointlike-diquark picture, since we get too large or too small $\rho_\Lambda^2$. 
These difficulties do not question the BT formalism itself, but seem to derive from the high sensitivity of $\rho_\Lambda^2$ to the structure of the light quark subsystem in a relativistic scheme, in contrast with a non-relativistic treatment. Finally we present the interim model, and after fixing its parameters to yield the correct spectrum and $\rho_\Lambda^2 \sim 2$, we apply it to the calculation of observables. By studying Bjorken sum rule we show that the inelastic IW function has to be large, and therefore the transitions $\Lambda_b \to \Lambda_c \left({1 \over 2}^-, {3 \over 2}^- \right) \ell \overline{\nu}$ could be studied at LHCb. Interestingly, some observables in the $\tau$ case present zeroes for specific values of $q^2$ that could be tests of the Standard Model. For example, the forward-backward asymmetry for both $\Lambda_b \to \Lambda_c \left({1 \over 2}^\pm \right) \tau \nu$ presents a zero for $q^2 \simeq m_\tau \sqrt {m_b^2-m_c^2}$.

\end{abstract}

\par \noi LPT-Orsay May 2020 \par 
\vskip 3 truemm

\noi Damir.Becirevic@th.u-psud.fr, Alain.Le-Yaouanc@th.u-psud.fr, \par 
\noi morenas@in2p3.fr, Luis.Oliver@th.u-psud.fr

\section{Introduction}

Possible physics beyond the Standard Model (SM), suggesting Lepton Flavor Universality Violation (LFUV), has been pointed out by data of different experiments on $\overline{B} \to D^{(*)} \ell \nu$ \cite{BaBarLFUV, BelleLFUV, LHCbLFUV}, summarized in \cite{HFAG-LFUV}. These experimental results have attracted much attention in terms of analyses within the Standard Model (SM) and also using models for New Physics (NP)  \cite{SAKAKI-LEPTOQUARKS-LFUV} - \cite{LI-YANG-ZHANG}. In particular, following the lattice calculations of form factors in the SM \cite{DETMOLD ET AL.},  ref. \cite{MEINEL-TAU-NU} examines  $\Lambda_b \to \Lambda_c \tau \overline{\nu}_\tau$ with NP operators.\par
With the intention of providing predictions for observables in $\Lambda_b \to \Lambda_c^{(*)} \ell \overline{\nu}$, we have considered the quark model, which can describe simultaneously the ground state and the excitations, not calculated in the present works on Lattice QCD. Moreover, we use the BT relativistic framework, that yields a Lorentz invariant Isgur-Wise (IW) function in terms of internal hadron wave functions. The latter are deduced from a quark model spectroscopic Hamiltonian describing the states at rest, and fitting the observed spectrum. The resulting IW function gives the leading order of the form factors in a heavy quark expansion.\par 
We have used this guideline in the meson case $\overline{B} \to D^{(*)} \ell \overline{\nu}$, for ground state \cite{COVARIANT-QM} and orbitally excited $D$ mesons \cite{MORENAS-1}. In the meson case we did use, as spectroscopic Hamitonian, the one of Godfrey and Isgur (GI), that describes a wealth of meson data for the $q \overline{q}$ and $Q \overline{q}$ systems \cite{GODFREY-ISGUR}. In this way, it was obtained a reasonable and theoretically founded description of IW functions, both elastic and inelastic \cite{MORENAS-2}.\par 
\subsection{Relevance of the BT scheme for hadron form factors}

One must underline in detail the relevance of the BT scheme for the calculation of form factors of heavy hadron transitions by heavy currents. BT is an approach to hadron motion which can be combined with any internal (rest frame) wave function. In quark model calculations, like \cite{NIEVES-2, PERVIN ET AL.}, the spectroscopic model providing these wave functions could be either non-relativistic, as in \cite{NIEVES-2}, or possibly relativistic, as in Pervin {\itshape et al.} \cite{PERVIN ET AL.}, that consider both cases. But whatever the type of spectroscopic equation, both groups apply the usual non-relativistic treatment for the hadron motion. In the calculation of \cite{NIEVES-2}, although a very careful calculation of the spectrum and the wave functions is done, a very small IW slope is found $\rho_\Lambda^2 \simeq 0.6 - 1.$, instead of $\rho_\Lambda^2 \simeq 2$ for the $\Lambda_b$, as indicated by our fit below to lattice QCD \cite{DETMOLD ET AL.} and the LHCb data \cite{IW-LHCb}.\par
As has been shown previously in detail in the meson case \cite{COVARIANT-QM, MORENAS-2, MORENAS-3}, the BT calculation gives a large enhancement for the IW slope with respect to the non-relativistic calculation with the same internal wave functions. This is due to the Lorentz transformation of the spatial arguments (i.e. quark momenta) of the wave function for hadrons in motion. This effect gives, with respect to the non-relativistic slope ($\rho^2_{NR} = {1 \over 2}\ m^2 R^2$ for a Gaussian), an additional contribution that is (i) independent of the wave function shape and parameters, and (ii) very large since it is roughly $\delta\rho^2 \simeq 0.75$ for a model with a scalar light quark, and $\delta\rho^2 \simeq 1$ for a meson (see for instance our discussion in \cite{MORENAS-3}). The Bonn group \cite{BONN} seems to find such an enhancement in a Bethe-Salpeter approach by applying also the full Lorentz transformation.\par
We can write an illustrating explicit simple formula for the slope in the BT scheme if we consider a Gaussian wave function $\exp\left(- R^2 p^2/2 \right)$ for a $Q-\overline{q}$ bound state, neglecting the light quark spin and the Jacobian factor. 
The product of the initial and final wave functions with Lorentz transformation along $Oz$ gives
\begin{multline*}
\exp\left[-R^2\left (w~(p^z)^2+\frac{w+1}{2} (p^T)^2+\frac{w-1}{2}~m^2\right)\right]\\
= \exp\left[-R^2~ w~(p^z)^2\right]~\exp\left[-R^2 ~\frac{w+1}{2} (\vec{p^T})^2\right]~\exp\left(-R^2~\frac{w-1}{2}~m^2\right)
\end{multline*}
Integrating over $p$, and expanding around $w=1$, one finds the slope with neglect of the Jacobian factor. The neglect of the latter factor allows to get a completely analytic result :
\begin{eqnarray}
\rho^2=\frac{1}{2}\  m^2 R^2+1
\end{eqnarray}
This $+1$ is the enhancement with respect to the non relativistic result, which is the first term. The Jacobian can be expanded in terms of the internal velocity, and the lowest term gives $-0.25$, whence the final enhancement $0.75+{\cal O}(v^2/c^2)$ for a $j^P=0^+$
light cloud (diquark model of baryons, see Section 4). On the other hand, for a meson, another contribution $+\frac{1}{4}$ must be added, corresponding to the heavy quark current. Finally, the slope for a meson is around $1$, as observed.\par
For baryons, the more complex structure of the three-quark wave functions and of the BT expression for the IW function makes this enhancement effect more difficult to evaluate, and strongly dependent on this structure. However, the general expectation is that the enhancement of $\rho^2$ should be larger than for mesons.\par
In the simple case of wave functions factorised in $\mid \vec{r}_2\mid, \mid \vec{r}_3\mid$ one would have an enhancement for two light quarks twice the one for one light quark, $\delta\rho^2 \simeq 2 \times 0.75 = 1.5$, pointing naturally towards $\rho_\Lambda^2 \simeq 2$ or more. But as we show below in Section 4, it could be much larger (and too large) for a wave function of the type of ref. \cite{PERVIN ET AL.}, which causes new problems. As a general fact, one observes a very strong dependence of $\rho^2_{\Lambda}$ on the structure of the wave function, for instance for a gaussian in the relative $\vec{\lambda},\vec{\rho}$ coordinates, it depends strongly on the ratio $R_{\rho}/R_{\lambda}$, and may acquire much too large values. See the analysis of Subsection 3.2.\par
It should be noted that the Lorentz transformation also implies Wigner rotations of spins, but their effect is found to be small for the ground state IW function.\par
Another important feature of the BT approach is that it implements automatically the HQET sum rules like Bjorken's or the curvature sum rules, which help to constrain rather efficiently the contributions of higher states. 

\subsection{Failures in the attempt to calculate the baryon IW function from standard spectroscopic models}

In trying to apply this scheme to heavy baryons, we have found a number of problems. There are several quark model approaches which could provide the required internal wave functions.\par 
Among the most standard ones (i.e. with linear+Coulomb potential), we quote first the work parallel to GI for mesons, the {\it relativistic} Hamiltonian of Capstick and Isgur for the $Qqq$ system \cite{CAPSTICK-ISGUR}. Unfortunately, this is a rather complicated model, which reproduces a very large spectrum of states, but for which it is not easy to obtain the corresponding wave functions.
Second, the work of Albertus {\itshape et al.} \cite{NIEVES-2}, using a non relativistic kinetic energy, with a very complete study of the states, and which writes explicitly the wave functions, but these are not easy to use in our calculation.
Third, there is the quark model study of Pervin, Roberts and Capstick for $\Lambda_Q$ baryons \cite{PERVIN ET AL.}, more manageable than the former two models, and to which we refer now.\par
In the present paper we have computed the IW function $\Lambda_b \to \Lambda_c$ in terms of a generic internal  $Qqq$ wave function. Then, we have used one of the internal wave functions given by ref. \cite{PERVIN ET AL.} in an harmonic oscillator basis in order to compute numerically the IW function and the corresponding slope. As pointed out in detail below, using the parameters of Pervin {\itshape et al.} \cite{PERVIN ET AL.}, we have found a slope $\rho_{\Lambda}^2 \simeq 4$. \par
This is much larger than the estimate by LHCb, $\rho_\Lambda^2 \simeq 1.8$ \cite{IW-LHCb}, and the value that follows from Lattice QCD calculations. Indeed, we describe below a fit to the Lattice data of Detmold {\itshape et al.} \cite{DETMOLD ET AL.}, that gives $\rho_{\Lambda}^2 \simeq 2$. \par 
Note that LHCb does not perform properly a determination or measurement of the true $\rho_\Lambda^2$ that we need, since it would require an extraction of the $1/m_Q$ corrections for each form factor, which do not separate. It is, as qualified by the authors, a ``measurement of the shape of the differential decay rate''.\par
We identify the mathematical origin of the large value of the slope obtained from the spectrum and the BT scheme, and we comment on the related work by Cardarelli and Simula in the Light Front formulation of the BT approach \cite{CARDARELLI-SIMULA}.\par
Then, we turn to the simpler scheme of a quark-diquark model, a bound state of a heavy quark and a color triplet pointlike diquark. This model has been widely used in the literature to compute the heavy baryon spectrum and heavy baryon form factors appearing in different processes \cite{Q-DIQUARK MODELS}. Concerning the spectrum, there is the interesting paper by Bing Chen {\itshape et al.} \cite{BING-CHEN}, a non-relativistic model with QCD-inspired potential, that, as we will show, presents also problems for the description of the IW slope, that turns out to be too small.\par 

\subsection{A simple provisory model for calculation of observables}

On the other hand, within the quark-diquark scheme, but renouncing to QCD-inspired potentials, a simple non relativistic harmonic oscillator model can be adjusted to give reasonable level spacings, and one can get also the IW slope in the BT scheme $\rho_{\Lambda}^2 \simeq 2$. In this paper we will adopt, for the moment, this simple model for the internal wave functions in view of the computation of $\Lambda_b \to \Lambda_c^{(*)} \ell \overline{\nu}$ observables.\par
When this paper was in progress, the Mainz group has issued a paper \cite{MAINZ LAMBDAb LAMBDAcSTAR} on some observables that could be useful to test LFUV in $\Lambda_b \to \Lambda_c \left({1 \over 2}^\pm , {3 \over 2}^- \right) \ell \overline{\nu}$ transitions, one of the objects of the present paper. However, as our approach is different, we still present our results, and compare with their work and other related literature.\par

\subsection{Plan of the paper}

The paper is organized as follows. In Section 2 we present a fit to the Lattice QCD data, that yields a slope of the IW function $\rho_{\Lambda}^2 \simeq 2$, and we quote the value of $\rho_{\Lambda}^2$ given by LHCb. In Section 3 we expose the numerical problem that we find on computing the IW slope for $Qqq$ baryons with the wave functions of ref. \cite{PERVIN ET AL.}, and we trace back the mathematical origin of this difficulty using a generic gaussian wave function in the spirit of Cardarelli and Simula \cite{CARDARELLI-SIMULA}. As an alternative model, we turn to the quark-diquark model in Section 4, we compute the IW functions for the elastic case and for the $L = 0 \to L = 1$ transitions in the BT scheme from the wave functions of the Bing Chen {\itshape et al.} Hamiltonian \cite{BING-CHEN}, and we find a much too low value compared to the lattice result. In Section 5, in front of these difficulties, to compute the observables \cite{MAINZ OBSERVABLES} that could be sensitive to LFUV for ${1 \over 2}^+ \to {1 \over 2}^\pm$ transitions, we renounce to models with QCD-inspired potentials, and use quark-diquark wave functions deduced from a non-relativistic harmonic oscillator quark model, adjusted to give the desired $\rho_\Lambda^2 \simeq 2$. In Appendix A we define the baryon form factors using different needed conventions, in B we give some details of the involved calculation of the $Qqq$ elastic IW function in the BT scheme, in C we compute the elastic and inelastic quark-diquark IW functions in the BT scheme, in D we make explicit the quark-diquark wave functions within the Bing Chen {\itshape et al.} scheme, and finally in E we give the expressions for the helicity amplitudes and the observables as formulated by the Mainz group, that we have used for our applications in Section 5. 

\section{LHCb measurement of the $\boldsymbol{d\Gamma / dq^2}$ shape, Lattice form factors and slope of the IW function}

\subsection{LHCb measurement of the $\boldsymbol{d\Gamma / dq^2}$ shape}

The differential rate of the decay $\Lambda_b \to \Lambda_c + \ell^- + \overline{\nu}_\ell$ writes, for $m_\ell = 0$ and heavy quark limit form factors,
\beq
\label{LHCb-1e}
{d\Gamma \over dw} = {G^2_F \over 12 \pi^3} \mid V_{cb} \mid^2 m^2_{\Lambda_b}  m^3_{\Lambda_c} \sqrt{w^2-1} \left[ 3w(1-2rw+r^2)+2r(w^2-1) \right] \mid \xi(w) \mid^2
\eeq

\noi where $r = m_{\Lambda_c}/m_{\Lambda_b}$.\par

Of course, here $\xi(w)$ is not the real IW function, but a rough approximation to it, with unspecified $1/m_Q$ errors. In next subsection, we try to extract the IW function from the form factors calculated in Lattice QCD by taking into account the $1/m_Q$ corrections. This is the way one must actually define the IW function $\xi_\Lambda (w)$ and its slope, in accordance with the $1/m_Q$ expansion, and it is this $\rho_\Lambda^2$ which  we use in the discussions which follows.

With the ``dipole'' ansatz
\beq
\label{LHCb-2e}
\xi(w) = \left({2 \over {w+1}} \right)^{2 \rho_{dip}^2}
\eeq

\noi LHCb finds the value of his $\rho_{dip}^2$ parameter \cite{IW-LHCb}
\beq
\label{LHCb-3e}
\rho_{dip}^2 = -\xi'(1)  = 1.82 \pm 0.03
\eeq

\noi and the curvature
\beq
\label{LHCb-4bise}
\sigma_{dip}^2 = \xi_\Lambda''(1)  = 4.22 \pm 0.12
\eeq

\noi Of course, the very small errors in eqns. (\ref{LHCb-3e},\ref{LHCb-4bise}) are not to be taken as the actual errors on the real IW slope and curvature.

The main reason for adopting the shape (\ref{LHCb-2e}) is that a number of theorems have been obtained that constrain the successive zero recoil derivatives of the baryon IW function \cite{CURVATURE-BARYONS,DERIVATIVES-IW-RAYNAL}, in particular the bound on the curvature
\beq
\label{LHCb-4e}
\sigma_\Lambda^2 \geq {3 \over 5} \left [\rho_\Lambda^2 + (\rho_\Lambda^2)^2 \right ]
\eeq

It has been established  \cite{DERIVATIVES-IW-RAYNAL} that the ``dipole'' form (\ref{LHCb-2e}), that depends on a single parameter, satisfies these constraints provided that $\rho_\Lambda^2 \geq {1 \over 4}$.\par

\subsection{Fits to lattice data on form factors}

Early studies of the $\Lambda_b \to \Lambda_c \ell \nu$ lattice form factors in the quenched approximation were provided in refs. \cite{UKQCD-SLOPE} and \cite{MILC-SLOPE}. In the former study, a value was given for the slope of the IW function, $\rho_\Lambda^2 \simeq 2.4$ with a $15 \%$ error, showing no dependence on the heavy quark masses and a value was also obtained for the HQET parameter $\overline{\Lambda} \simeq 0.75$ with a $20 \%$ error.\par

A great wealth of new precise data in Lattice QCD has been obtained recently by W. Detmold, C. Lehner and S. Meinel \cite{DETMOLD ET AL.}, that have given results for all the form factors entering in the process $\Lambda_b \to \Lambda_c \ell \overline{\nu}_\ell$ within the Standard Model. \par

Our aim is now to try to extract information on the slope of the IW function $\xi_\Lambda (w)$ and other parameters, $\overline{\Lambda}$ and the heavy quark masses $m_Q\ (Q = b, c)$, from these lattice calculations, that are summarized in Fig. 12 of ref. \cite{DETMOLD ET AL.}.\par
We adopt a simple HQET model, keeping the form factors {\it up to first order in $1/m_Q$ included}, as given in the formulas of Appendix A. Unlike ref. \cite{LIGETI} we do not take into account explicitly the QCD perturbative corrections to HQET, and therefore it must be understood that our slope $\rho_\Lambda^2$  accounts by itself roughly for such effects.\par 
Inspection of the formulas of Appendix A shows that, at this first order, besides the dependence of the form factors on the IW function $\xi_\Lambda(w)$, on the heavy quark masses $m_Q\ (Q = b, c)$ and on the HQET parameter $\overline{\Lambda}$, there is another subleading function $A(w)$ that, due to Luke's theorem \cite{LUKE-THEOREM} must vanish at $w = 1$. For this function, since the domain in $w$ is not large, we adopt the parametrization
\beq
\label{2-3e}
A(w) = A'(1) (w-1)
\eeq

\noi Moreover, we will adopt the explicit ``dipole'' form (\ref{LHCb-2e}) for the leading IW function.

\subsubsection{The IW function slope from lattice form factors}

In the approximation that we adopt, HQET up to first order in $1/m_Q$ included, there are two quantities that isolate the IW function, where all dependence on $\overline{\Lambda}$ and the parameter $A'(1)$ defined by (\ref{2-3e}) cancels. These quantities are differences of ratios that, up to $O(1/m_Q^2)$ corrections, are identical to the IW function $\xi_\Lambda (w)$,
\beq
\label{2-4ae}
R_1(w) = {w+1 \over 2} {f_\perp (w)-g_\perp (w) \over f_\perp (1)-g_\perp (1)} 
\eeq
\beq
\label{2-4be}
R_2(w) = {f_\perp (w)-f_+ (w) \over f_\perp (1)-f_+ (1)} 
\eeq

\noi Inspection of the formulas of Appendix A shows indeed that these ratios do not depend on $\overline{\Lambda}$ and on the parameter $A'(1)$, that cancel in these quantities,
\beq
\label{2-5ae}
R_1(w) = \xi_\Lambda (w) + O(1/m_Q^2)\, \qquad \qquad R_2(w) = \xi_\Lambda (w) + O(1/m_Q^2)
\eeq

To have information on the IW function, we will use the $z$-expansion parametrization in \cite{DETMOLD ET AL.}, that we will fit with our HQET model of form factors, that includes up to $O(1/m_Q)$ corrections, made explicit in Appendix A. The lattice data is parametrized by the $z$-expansion \cite{BOURRELY} for each form factor 
\beq
\label{2-5be}
f(q^2) = {1 \over 1 - {q^2 \over (m_{pole}^f)^2}} \left[a_0^f + a_1^f z^f(q^2) + ...\right]\ , \ \ \  z^f(q^2) = {\sqrt{t^f_+ - q^2} - \sqrt{t^f_+ - t_0} \over \sqrt{t^f_+ - q^2} -+ \sqrt{t^f_+ - t_0}}
\eeq

\noi where $t_0 = (m_{\Lambda_b} - m_{\Lambda_c})^2$, $m_{pole}^f$ and $t^f_+$ are given in Table VII, $a_0^f, a_1^f$ up to $O(z)$ in Table VIII, and $a_0^f, a_1^f, a_2^f$ up to $O(z^2)$ in Table X of \cite{DETMOLD ET AL.}.\par 
We do not pretend to make a fit on the two ratios (\ref{2-4ae},\ref{2-4be}) with their errors. We just take the $z$-expansion central values {\it at face value}, to see if for the two ratios we find reasonable consistent values for the IW function slope $\rho_\Lambda^2$, using both expansions up to $O(z)$ and up to $O(z^2)$. For the IW function we adopt the ``dipole'' parametrization (\ref{LHCb-2e}), that satisfies the rigurous results that constrain the successive zero recoil derivatives of the baryon IW function \cite{CURVATURE-BARYONS,DERIVATIVES-IW-RAYNAL}.\par 
To perform the fits we select a number of points of the $z$-expansions for the lattice values for these ratios, up to first order and up to second order in $z$, and we use the {\it Mathematica} package {\it FindFit}. For the IW function $\xi(w)$ we consider the domain $1 \leq w \leq 1.2$ where there are data points measured on the lattice. For the individual form factors we will consider below the $z$-expansions and the fits for the whole phase space.\par 
\begin{figure}[hbt]
\begin{center}
\includegraphics[width=\graphewidth]{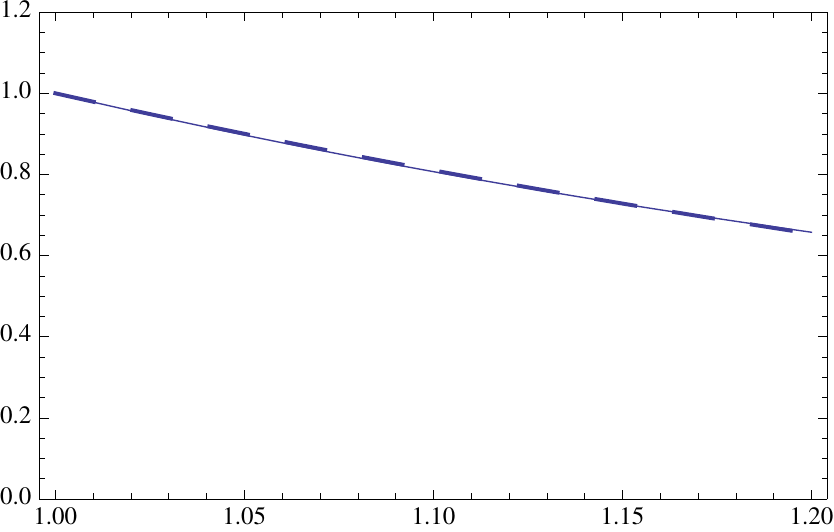}
$\quad$
\includegraphics[width=\graphewidth]{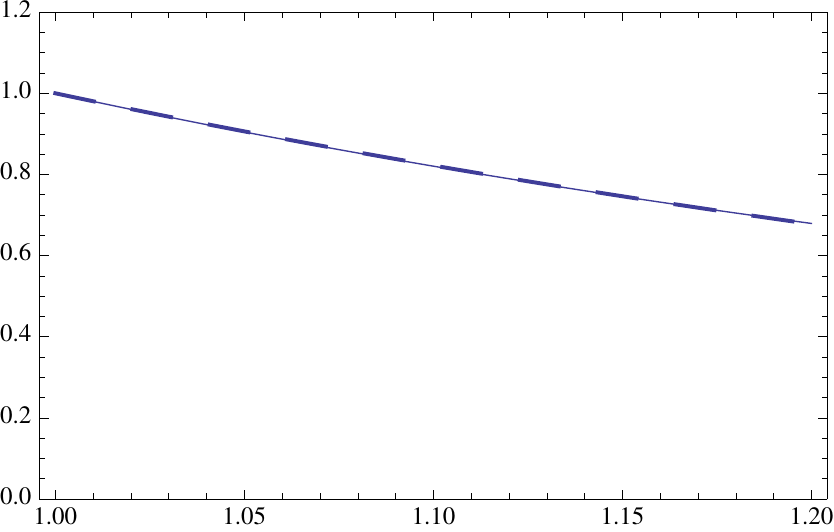}
\end{center}
\caption{Fits to the ratio $R_1(w) \simeq \xi(w)$ (\ref{2-4ae},\ref{2-5ae}) with the Isgur-Wise function (\ref{LHCb-2e}) (continuous curve) using the parametrization of the lattice data up to first order in the $z$ expansion (dashed curve, left), that yields the slope $\rho^2_\Lambda \simeq 2.20$, and up to second order (dashed curve, right), that gives $\rho^2_\Lambda \simeq 2.03$.}
\end{figure}
%
%
%
%
%
\begin{figure}[htb]
\begin{center}
\includegraphics[width=\graphewidth]{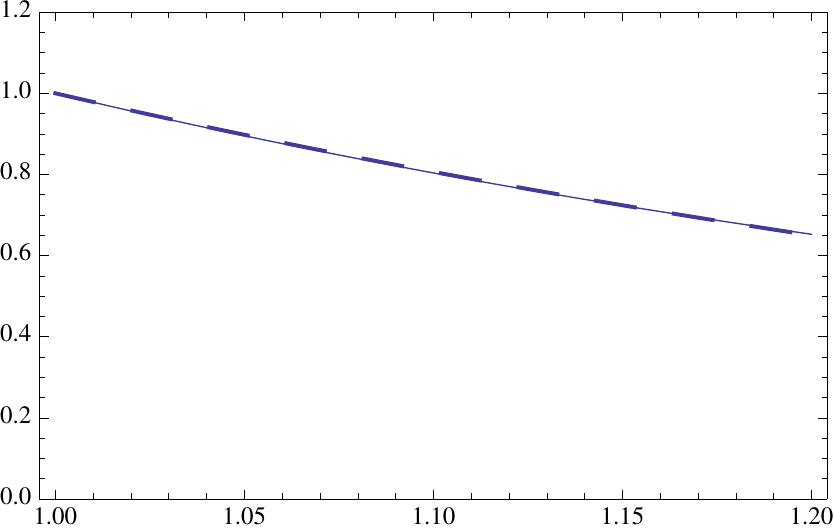}
$\quad$
\includegraphics[width=\graphewidth]{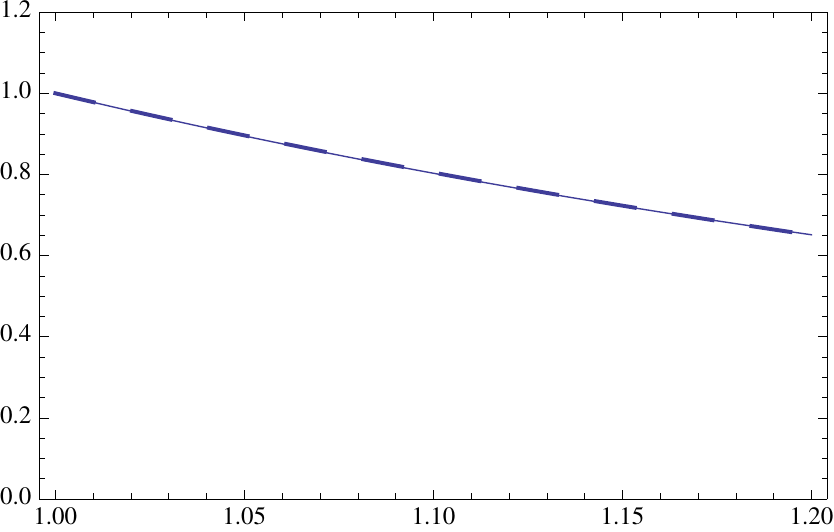}
\end{center}
\caption{Fits to the ratio $R_2(w) \simeq \xi(w)$ (\ref{2-4be},\ref{2-5ae}) with the Isgur-Wise function (\ref{LHCb-2e}) (continuous curve) using the parametrization of the lattice data up to first order in the $z$ expansion (dashed curve, left), that yields the slope $\rho^2_\Lambda \simeq 2.24$, and up to second order (dashed curve, right), that gives $\rho^2_\Lambda \simeq 2.25$.}
\end{figure}
%
%
%
%
%
The results are given in Figs. 1 and 2. From the ratio $R_1(w)$ we get from the fit in the region $1 \leq w \leq 1.2$, $\rho^2_\Lambda \simeq 2.20$ for the first order expansion in $z$, to be compared with the true slope of the IW function (\ref{2-4ae}) $-\xi'(1) = 2.11$, and $\rho^2_\Lambda \simeq 2.03$ for second order in $z$, to be compared with the true slope $-\xi'(1) = 1.99$. On the other hand, from the ratio $R_2(w)$ we get results that are close in both cases, $\rho^2_\Lambda \simeq 2.25$, compared to the true slope $-\xi'(1) = 2.16$ at first order in $z$, and $-\xi'(1) = 2.21$ at second order.\par
We can safely conclude that the slope is consistent with the following ranges obtained from the fit. For the first order $z$ expansion we get the domain
\beq
\label{2-6ae}
\rho^2_\Lambda \simeq 2.20 - 2.24
\eeq

\noi while for the $z^2$ order we obtain the range 
\beq
\label{2-6be}
\rho^2_\Lambda \simeq 2.03 - 2.25
\eeq

Although our fits are somewhat naive, from (\ref{2-6ae},\ref{2-6be}) we can safely conclude that the data on $\Lambda_b \to \Lambda_c$ form factors  \cite{DETMOLD ET AL.} can be described in HQET up to $O(1/m_Q)$ included, with the slope of the ``dipole'' form for the IW function (\ref{LHCb-2e}), 
\beq
\label{2-26tere}
\rho^2_\Lambda \simeq 2.15 \pm 0.10
\eeq

\subsubsection{Fits to the different form factors}

We do not want to make an overall fit on the whole set of form factors with their errors (errors that we do not master), but just to study individually each form factor of Fig.~12 of \cite{DETMOLD ET AL.} by making a fit to the central values of these domains, given by the $z$-expansion up to order $z$.\par
We take these central values {\it at face value}, to see if for each form factor we find reasonable results for the set of parameters, 
\beq
\label{2-7e}
\rho_\Lambda^2 \ , \ m_c\ ,\  m_b\ ,  \ \overline{\Lambda}\ , \ A'(1)
\eeq

\noi and how these parameters compare between the different form factors, i.e. how dispersed they are.\par 
Since we have an independent estimate of the slope of the IW function (\ref{2-6ae},\ref{2-6be}), we now fix 
\beq
\label{2-8bise}
\rho_\Lambda^2 \simeq 2.15
\eeq

\noi and we use {\it FindFit} to perform {\it constrained fits} for $m_c$ and $m_b$ and $\overline{\Lambda}$, assuming the following domains, from different analyses within HQET \cite{LEIBOVICH-STEWART,OLIVER-JUGEAU}, in GeV units :
\beq
\label{2-8e}
1.15 \leq m_c \leq 1.35 \ , \qquad 4.10 \leq m_b \leq 4.40 \ ,\qquad 0.60 \leq \overline{\Lambda} \leq 0.90
\eeq

\noi and the slope of the $1/m_Q$ form factor $A'(1)$ as a free parameter.

We take now the data for each form factor, the central values of Fig.~12 of \cite{DETMOLD ET AL.}, that are fitted by the first order $z$-expansion. Choosing a number of values of these $z$-expansion curves and the HQET model up to $O(1/m_Q)$ of Appendix A, the resulting fit obtained with {\it FindFit} gives the plots of Fig. 3 and the parameters of Table 1. 
\begin{figure}[!tbp]
\begin{center}
\begin{minipage}{\graphewidth}
\includegraphics[width=\graphewidth]{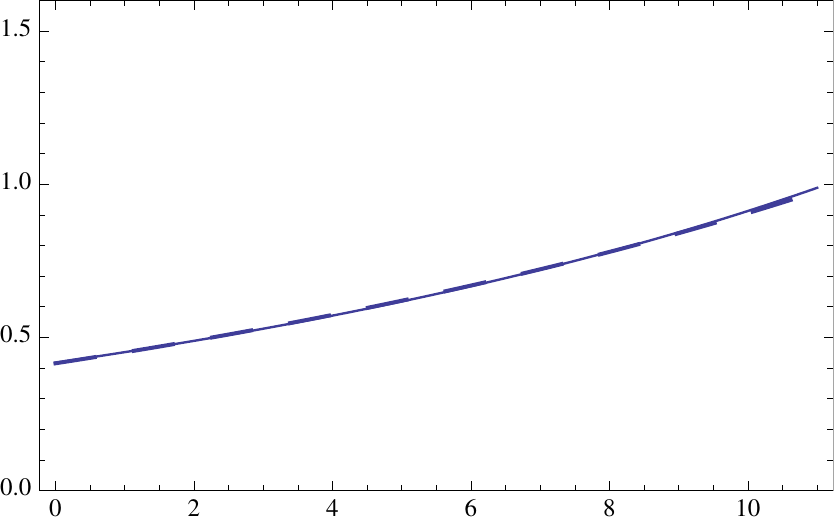}\\\centerline{$f_0(q^2)$}
\end{minipage}
$\quad$
\begin{minipage}{\graphewidth}
\includegraphics[width=\graphewidth]{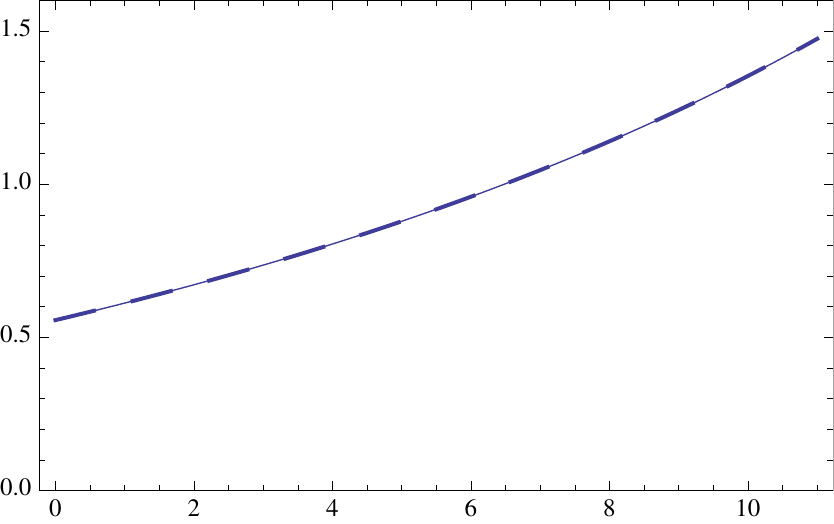}\\\centerline{$f_\perp(q^2)$}
\end{minipage}
\end{center}
\begin{center}
\begin{minipage}{\graphewidth}
\includegraphics[width=\graphewidth]{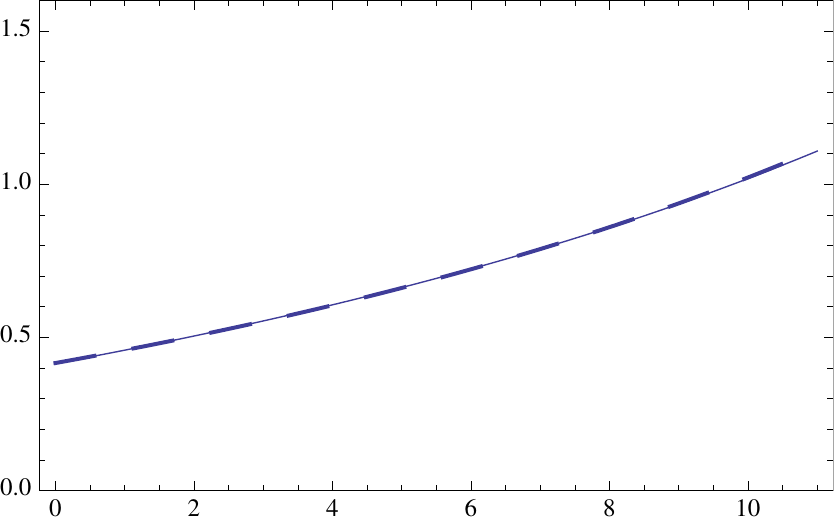}\\\centerline{$f_+(q^2)$}
\end{minipage}
$\quad$
\begin{minipage}{\graphewidth}
\includegraphics[width=\graphewidth]{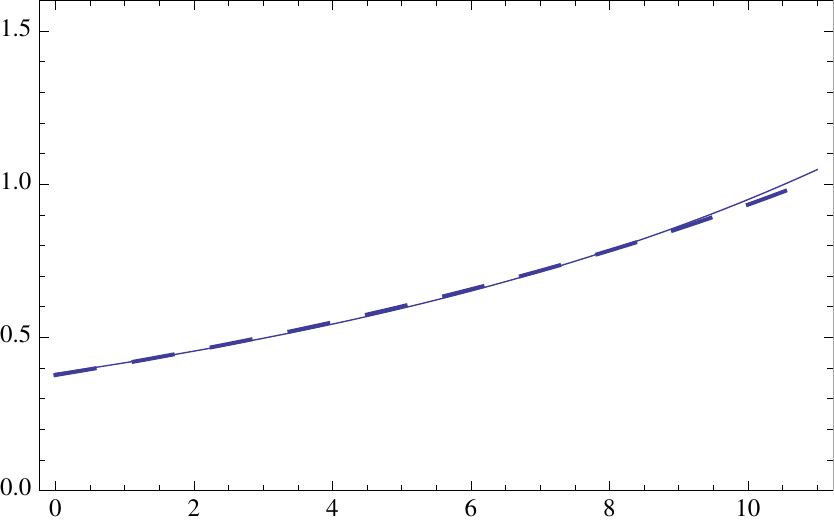}\\\centerline{$g_0(q^2)$}
\end{minipage}
\end{center}
\begin{center}
\begin{minipage}{\graphewidth}
\includegraphics[width=\graphewidth]{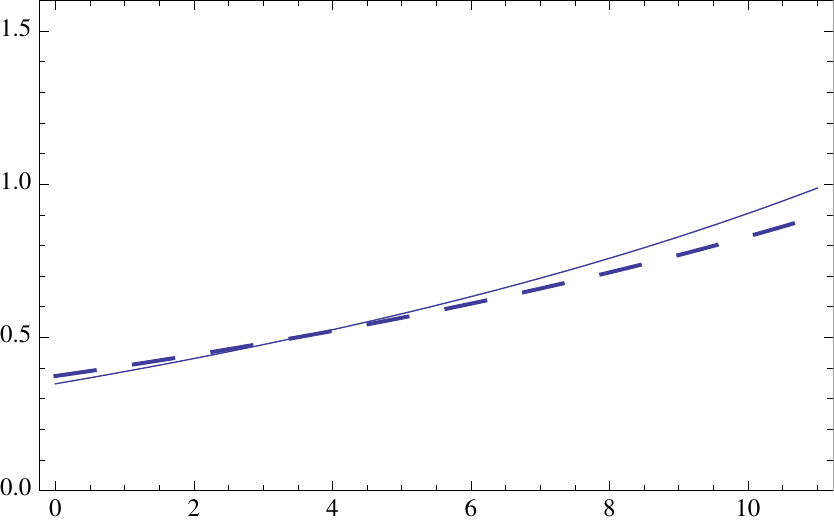}\\\centerline{$g_\perp(q^2)$}
\end{minipage}
$\quad$
\begin{minipage}{\graphewidth}
\includegraphics[width=\graphewidth]{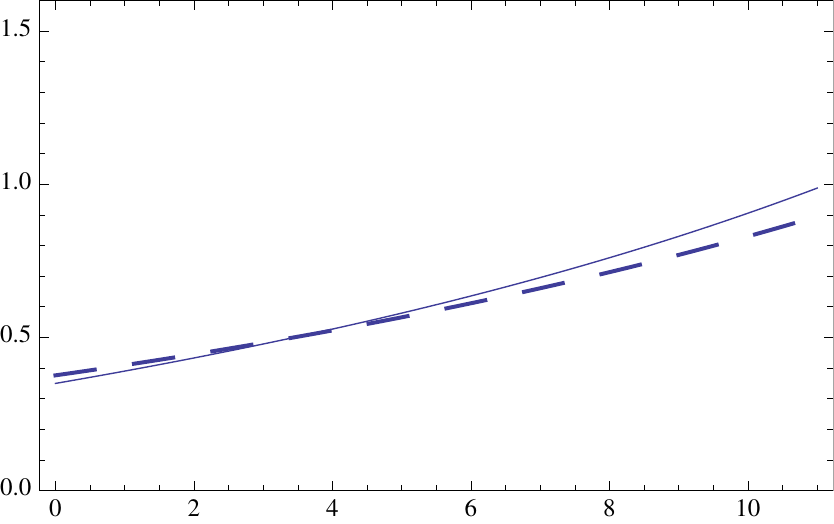}\\\centerline{$g_+(q^2)$}
\end{minipage}
\end{center}
\caption{Center values of lattice form factors in first order of the $z$-expansion \cite{DETMOLD ET AL.} (dashed lines) compared to the fit using the HQET model up to $O(1/m_Q)$ (continuous curves) obtained from {\it FindFit}.}
\end{figure}
%
%
%

%
%
%
%
%
%
%
%
We summarize the values for the parameters obtained from the fits for the different form factors in Table 1.
%
%
\begin{table}[htb]
\begin{center}
\begin{tabular}{|c|c|c|c|c|}
\hline
{Form factor} & $m_c \ {\rm (GeV)}$ & $m_b \ {\rm (GeV)}$ & $\overline{\Lambda} \ {\rm (GeV)}$ & $A'(1) \ {\rm (GeV)}$ \\
\hline
$f_0(\Lambda_b \to \Lambda_c)$ & $1.33$ & $4.34$ & $0.66$ & $-0.21$ \\
\hline
$f_\perp(\Lambda_b \to \Lambda_c)$ & $1.15$  & $4.13$ & $0.90$ & $-0.35$ \\
\hline
$f_+(\Lambda_b \to \Lambda_c)$ & $1.15$  & $4.40$ & $0.90$ & $-0.25$ \\
\hline
$g_0(\Lambda_b \to \Lambda_c)$ & $1.35$  & $4.10$ & $0.60$ & $-0.41$ \\
\hline
$g_\perp(\Lambda_b \to \Lambda_c)$ & $1.35$  & $4.38$ & $0.60$ & $-0.49$ \\
\hline
$g_+(\Lambda_b \to \Lambda_c)$ & $1.35$  & $4.10$ & $0.60$ & $-0.50$ \\
\hline
\end{tabular}
\end{center}
\caption{Fits with the constraints (\ref{2-8e}) and arbitrary value for $A'(1)$.}
\end{table}
%
%
%

Let us comment on Fig. 3 and Table 1.
First, the values obtained for the parameters $m_c$, $m_b$ and $\overline{\Lambda}$ are of course within the imposed limits (\ref{2-8e}).
The fits are quite good for all form factors, except for $g_+(\Lambda_b \to \Lambda_c)$ and $g_\perp(\Lambda_b \to \Lambda_c)$ at large $q^2$ or near zero recoil $w = 1$. In the lattice data one sees that $g_\perp(q^2_{max}), g_+(q^2_{max}) < 1$, while the calculation of the model gives $g_\perp(q^2_{max}) = g_+(q^2_{max}) = 1$. The discrepancy is due to the same $1/m_Q^2$ correction at $w = 1$, since one has
\beq
\label{2-26e}
g_+(q^2_{max}) = g_\perp(q^2_{max})
\eeq

\noi Indeed, it is well-known that at zero recoil $w = 1$ there is a {\it negative} $1/m_Q^2$ correction for $g_\perp(q^2_{max})$, and this explains the discrepancy between the lattice data and the model. The $1/m_Q^2$ correction satisfies a sum rule that gives its sign, see for example the discussion of the meson form factor $F_{D^*}$ at zero recoil in the review paper \cite{URALTSEV}. \par

As a numerical example, from the range (\ref{2-26tere}), we adopt for $\rho^2_\Lambda \simeq 2.15$ we find in Table 1 the following ranges for the quark masses and subleading parameters, 
$$\qquad m_c \simeq 1.25 \pm 0.10\ {\rm GeV}\ , \qquad m_b \simeq 4.25 \pm 0.15\ {\rm GeV}$$
\beq
\label{2-26bise}
\overline{\Lambda} \simeq 0.75 \pm 0.15\ {\rm GeV} \ ,  \qquad A'(1) \simeq -0.35 \pm 0.15\ {\rm GeV}
\eeq

It is worth to emphasize that  the parameter $A'(1)$ turns out to be negative and sizeable. This is a new result from the present analysis of form factors. Interestingly, the sign and magnitude is in qualitative agreement with the expression obtained in the non-relativistic quark model, $A'(1) = - {\overline{\Lambda} \over 2} \rho^2_\Lambda$ (formula (126) of ref. \cite{OLIVER-JUGEAU}).\par

\subsubsection{Correlation between $\boldsymbol{\rho^2_\Lambda}$ and the slope parameter $\boldsymbol{A'(1)}$}

There is a correlation between $\rho^2_\Lambda$ and the slope parameter $A'(1)$. Indeed, taking the heavy quark limit in the expression of the form factors except for $A'(1)$, one finds, for all 6 form factors, for small $w-1$,
\beq
\label{2-9e}
F(w) \simeq 1 + \left[\left({1 \over 2m_b} + {1 \over 2m_c}\right) A'(1) - \rho_\Lambda^2\right](w-1) + O \left({\overline{\Lambda} \over m_Q} \right)
\eeq 

\noi that yields an effective slope 
\beq
F'(1) \simeq \left[ \left({1 \over 2m_b} + {1 \over 2m_c}\right) A'(1) - \rho_\Lambda^2 \right] + O \left({\overline{\Lambda} \over m_Q} \right)
\eeq 

Of course, one must take into account that the terms $O \left({\overline{\Lambda} \over m_Q} \right)$ contribute to the coefficient of $\rho^2_\Lambda$, so that the relation is somewhat different according to the form factors (Appendix A). For instance
\beq
f_\perp'(1) \simeq \left({1 \over 2m_b} + {1 \over 2m_c}\right) A'(1) - \left[ 1 + \overline{\Lambda} \left({1 \over 2m_b} + {1 \over 2m_c}\right)\right] \rho_\Lambda^2
\eeq 

\noi and different expressions depending on ${\overline{\Lambda} \over 2 m_Q}  \rho_\Lambda^2$ for the derivatives of the other form factors.\par

These differences allow us to determine separately $\rho_\Lambda^2$ and $A'(1)$, but the tendency of the correlation remains the same. This correlation indicates that with some increase of $\rho_\Lambda^2$, as it is possible from the above discussion, $A'(1)$, which is found negative for $\rho_\Lambda^2 \simeq 2.15$, should decrease in magnitude, or even change its sign.\par
Note that in the paper \cite{LIGETI}, the parameter $A'(1)$ has been absorbed into the slope, introducing a combination $\rho_\Lambda^2 - \left({1 \over 2m_b} + {1 \over 2m_c}\right) A'(1)$, dependent on the quark masses, and common to all form factors. Their $\zeta'(1)$ therefore differs from our $-\rho_\Lambda^2$. But also, as we have shown above, $A'(1)$ can be estimated separately from the lattice data, although with a rather large error.\par
The values of Table 1 correspond to the choice $\rho_\Lambda^2 = 2.15$, central value of the domain (\ref{2-26tere}). Had we adopted a higher value for the slope, then $A'(1)$ would be negative but with a smaller absolute magnitude, and for $\rho_\Lambda^2 \simeq 2.5-2.6$ there is a change of sign for $A'(1)$, although this depends on the particular form factor.\par
On the other hand, in ref. \cite{LIGETI} the curvature appears to be rather small in comparison with our fits. Indeed, with a $\rho_\Lambda^2$ around 2.15 with our dipole fit, which satisfies the well established lower bound (\ref{LHCb-4e}) on the curvature, we find that $\sigma_\Lambda^2 \simeq 5.7$, i.e. a term $+ 2.8 (w-1)^2$ in the expansion. However, this is partly compensated by our next negative term $-3.0 (w-1)^3$, which is still not negligible.\par

\section{Elastic IW function for the $\boldsymbol{Qqq}$ system in the Bakamjian-Thomas quark model}

For the ground state we have the total wave function
\beq
\label{A-1e}
\psi^{s \mu} = {1 \over \sqrt{3}} \sum_{P(231)} \varphi^{\Lambda_Q}_{231} \varphi^s_{231} \chi_{231}^{'\mu}
\eeq

\noi The flavor wave function writes $\varphi^{\Lambda_Q}_{231} = {1 \over \sqrt{2}}\ (d_2u_3-u_2d_3)Q_1$, the spin wave function $\chi_{231}^{'\mu}$ is antisymmetric in the 2,3 quarks and the full antisymmetry of the baryon wave function follows from the antisymmetry of the color singlet wave function. 

For the simple case of the non-relativistic harmonic oscillator, the ground state internal wave function (see for example Appendix A of ref. \cite{JARFI-LAZRAK}) writes :
\beq
\label{1-44e}
\varphi(\{ {\bf p}_i \}) = (2\pi)^3 \left({3\sqrt{3} R^3_\rho R^3_\lambda \over \pi^3} \right)^{1/2} \exp\left(- {{\bf p}^2_\rho R_\rho^2 + {\bf p}^2_\lambda R_\lambda^2 \over 2} \right)
\eeq

\noi where the relative momentum variables ${\bf p}_\rho$ and ${\bf p}_\lambda$ for $m_2 = m_3 = m$ are defined by
\beq
\label{1-45e}
{\bf p}_\rho = {1 \over \sqrt{2}}\ ({\bf p}_2 - {\bf p}_3)\ , \qquad {\bf p}_\lambda = \sqrt{3 \over 2}\ {m_1 ({\bf p}_2 + {\bf p}_3) - 2m {\bf p}_1 \over m_1+2m}
\eeq

\noi The wave function (\ref{1-44e}) is normalized according to 
\beq
\label{1-36e}
\int \prod_{i=1}^{n} {d {\bf p}_i \over (2 \pi)^3}\ \delta\left( \sum_{i=1}^{n} {\bf p}_i \right) \mid \varphi(\{ {\bf p}_i \} \mid^2\ = 1
\eeq 

\noi or equivalently, 
\beq
\label{1-36bise}
{1 \over 3 \sqrt{3}} \int {d {\bf p}_\rho \over (2 \pi)^3}  {d {\bf p}_\lambda \over (2 \pi)^3}\ \mid \varphi(\{ {\bf p}_\rho,  {\bf p}_\lambda\} \mid^2\ = 1
\eeq 

Some words of caution concerning the wave function (\ref{1-44e}). First, this expression is valid in the limiting case of equal masses for the two light quarks \cite{JARFI-LAZRAK} but, in general, crossed terms of the form ${\bf p}_\rho.{\bf p}_\lambda$ could appear. Here we restrict ourselves to the simplest case of the non-relativistic harmonic oscillator with two light quarks of equal mass.  

Assuming that the harmonic oscillator spring constant is flavor-independent, the reduced radii $R_\rho$ and $R_\lambda$ are given, in terms of the {\it equal mass} baryon radius $R$, by the expressions :
\beq
\label{1-46e}
R_\rho^4 = R^4\ , \qquad \qquad R_\lambda^4 = {m_1+2m \over 3 m_1} R^4
\eeq

\noi In the center-of-mass ${\bf p}_1 + {\bf p}_2 + {\bf p}_3 = 0$, relations (\ref{1-45e}) give
\beq
\label{1-45bise}
{\bf p}_\rho = {1 \over \sqrt{2}}\ ({\bf p}_2 - {\bf p}_3)\ , \qquad {\bf p}_\lambda = \sqrt{3 \over 2}\ ({\bf p}_2 + {\bf p}_3)
\eeq

\noi In the heavy quark limit $m << m_1$, the reduced radii (\ref{1-46e}) become
\beq
\label{1-47e}
R_\rho^4 = R^4\ , \qquad \qquad R_\lambda^4 = {R^4 \over 3}
\eeq

Finally we can obtain the explicit form of the baryon IW function with harmonic oscillator wave functions by replacing the expression (\ref{1-44e}) for the initial and final states in the general formula (\ref{1-35e}) of Appendix B.\par 
We find
\beq
\label{1-48e}
\xi_\Lambda(v.v') =  (2\pi)^6\ {3 \sqrt{3} R_\rho^3 R_\lambda^3 \over \pi^3} \int {d {\bf p}_2 \over (2 \pi)^3}\ {1 \over p_2^0}\ {d {\bf p}_3 \over (2 \pi)^3}\ {1 \over p_3^0}\ \sqrt{(p_2.v)(p_3.v)(p_2.v')(p_3.v')}
\eeq
$$\exp\left\{-\left( {3R_\lambda^2 + R_\rho^2 \over 4} \left[ (p_2.v')^2 + (p_3.v')^2 - 2 m^2 
\right] + {3R_\lambda^2 - R_\rho^2 \over 4}\ 2 \left[ (p_2.v') (p_3.v') - (p_2.p_3) \right] \right) \right\}$$
$$\exp\left\{-\left( {3R_\lambda^2 + R_\rho^2 \over 4} \left[ (p_2.v)^2 + (p_3.v)^2 - 2 m^2 
\right] + {3R_\lambda^2 - R_\rho^2 \over 4}\ 2 \left[ (p_2.v) (p_3.v) - (p_2.p_3) \right] \right) \right\}$$
$$\times\ {\left[m^2(1+v.v')+m(v+v').(p_2+p_3)+(p_2.v)(p_3.v')+(p_3.v)(p_2.v')+(p_2.p_3)(1-v.v') \right] \over 2 \sqrt{(p_2.v+m)(p_3.v+m)(p_2.v'+m)(p_3.v'+m)}}$$

\vskip 3 mm
\noi where the factor $\sqrt{(p_2.v)(p_3.v)(p_2.v')(p_3.v')}$ in the first line is due to the Jacobian of the change of variables in the BT scheme, and the complicated factor in the last line comes from the Wigner rotations \cite{COVARIANT-QM}, computed in the baryon case in Appendix B.

\subsection{An attempt to a concrete calculation of the IW function}

In Pervin {\itshape et al.} \cite{PERVIN ET AL.}, the spectrum of heavy baryonic states has been studied with a linear + Coulomb Hamiltonian, with diagonalisation in harmonic oscillator (HO) or pseudocoulombic (PC) bases and with kinetic energy either non-relativistic or relativistic. In practice, the ground state wave function seems to be well represented by one gaussian or one PC wave function. Choosing the Hamiltonian with relativistic energy, and the HO basis, we read from their tables in the HOSR entry, the light quark mass and the ground state internal wave function necessary for the BT calculation, the latter being well approximated by one gaussian, eqn. (\ref{1-44e}). They are given as follows :
\beq
\label{1-48bise}
m = 0.38\ \rm{GeV}\ , \qquad \qquad R_\rho^2 = R_\lambda^2 = 2.16\ \rm{GeV^{-2}} 
\eeq

We have computed the slope of the IW function (\ref{1-48e}) and have found
\beq
\label{1-48tere}
\rho_\Lambda^2 = 4.01
\eeq 

\noi This value is much larger than the ranges (\ref{2-6ae},\ref{2-6be}) determined in Section 2.\par
Notice that the the last factor in (\ref{1-48e}), due to the Wigner rotations, gives a very small numerical contribution to the slope.\par
It must be emphasized that this value is different and larger from the one given by the authors,
\beq
\rho_\Lambda^2 = 1.33
\eeq 

\noi the reason being that they use a non-relativistic treatment to calculate the form factors, where in principle $\rho_\Lambda^2 = 3 m^2 R_\lambda^2$. This shows the tendency of the relativistic BT treatment to enhance the slope, which is what one would like. But of course the enhancement is too large, and it is worse with the PC basis.\par
On trying to understand this disapointing result one notices that, as found by the authors of \cite{PERVIN ET AL.}, there could be artefacts due to the smallness of the HO or PC expansion bases. On the other hand, in BT there is for baryons, in contrast with mesons, a particular sensitivity of the value of $\rho_\Lambda^2$ to the detailed structure of the wave function, as we argue below. This emphasizes the need for larger bases.

\subsection{General discussion}

In the meantime, in view of this conclusion concerning the above gaussian wave function, we proceed as follows. We pursue the investigation with the gaussian shape (\ref{1-44e}) now considered as a model, with the objective of investigating the dependence of $\rho_\Lambda^2$ on the shape of the generic internal wave functions in the BT scheme, and in particular to understand the high value obtained above, $\rho_\Lambda^2 \simeq 4$. In fact, a somewhat similar discussion has been done numerically by Cardarelli and Simula \cite{CARDARELLI-SIMULA} in the null plane formalism, which is known to be equivalent to the BT formalism in the heavy quark limit. However, one must avoid to give a physical interpretation to the gaussian wave function, as we will see below.
Here we will rather develop a mathematical analysis to understand the variations of the slope $\rho_\Lambda^2$.\par
The formula (\ref{1-48e}) and its expansion at small velocity to extract $\rho_\Lambda^2$, keeping for simplicity the terms coming from the gaussians, and disregarding the contributions from the Jacobian and from the Wigner rotations, gives :
\beq
\label{1-49e}
\rho_\Lambda^2 = 3m^2R_\lambda^2 + {3R_\lambda^2+R_\rho^2 \over 3} \left< \vec{p_2}^2+\vec{p_3}^2 \right>+  {3R_\lambda^2-R_\rho^2 \over 2} \left< {1 \over 3}\ \vec{p_2}.\vec{p_3} +p_2^0 p_3^0 - m^2 \right>
\eeq

\noi with 2, 3 labelling the two light quarks, and $\left< ... \right>$ denoting averages on the wave functions. The first term is the non-relativistic contribution, the two others are relativistic corrections. The third term corresponds to crossed terms that, of course, are absent in mesons.\par
This formula shows that $\rho_\Lambda^2 $ depends on two parameters, instead of one in the non-relativistic limit (the term $\rho_\Lambda^2 = 3m^2R_\lambda^2$), and one can get very high values because for $R_\rho^2 < 3 R_\lambda^2$ the last two terms in the expression are positive, and when $R_\rho^2$ becomes small, the $\left< ... \right>$ averages become large. Indeed, 
\beq
\label{1-49-1e}
\left< \vec{p_2}^2 \right> = \left< \vec{p_3}^2 \right> = {3 \over 4} \left( {1 \over R_\rho^2} + {1 \over 3} {1 \over R_\lambda^2} \right)
\eeq
\beq
\label{1-49-2e}
\left< \vec{p_2}.\vec{p_3}\right> = {3 \over 4} \left(- {1 \over R_\rho^2} + {1 \over 3} {1 \over R_\lambda^2} \right)
\eeq

\noi i.e. the momenta are equal, large and antiparallel in average (and of course $\left< p_2^0 p_3^0 \right>$ becomes also large). Then, the magnitude of $\rho_\Lambda^2$ is controlled by the ratio $R_\lambda^2/R_\rho^2$, and $\rho_\Lambda^2$ diverges when $R_\rho^2 \to 0$ at fixed $R_\lambda$, in agreement with the numerical findings of Cardarelli and Simula in the null plane formalism \cite{CARDARELLI-SIMULA}.\par 
Though, the interpretation of this limit as corresponding to the point-like diquark model given in this reference is at odds with our analysis of the quark-diquark model, which gives a small $\rho_\Lambda^2$, analogous to mesons (Section 4 of the present paper). This is understandable, since in this limit $R_\rho^2 \to 0$ the gaussian is not the physical solution calculated from a QCD inspired Hamiltonian.\par
In fact, the last two terms in eqn. (\ref{1-49e}) diverge for $R_\rho/R_\lambda \to 0\ \rm{or}\  \infty$, but in one case they have the same sign, and whence $\rho_\Lambda^2$ diverges, while in the other case the divergences cancel when $R_\lambda$ is help fixed, and $\rho_\Lambda^2$ tends to a finite positive value.\par 
It can be seen that these large values are related to the crossed term in the arguments of the two gaussians with the coefficient $- 2 (\vec{p_2}.\vec{p_3}) \left( 3 R_\lambda^2-R_\rho^2 \right)$ give a large positive contribution balancing the factorisable one $-( \vec{p_2}^2+\vec{p_3}^2) \left( 3R_\lambda^2 + R_\rho^2 \right)$ when $R_\rho^2$ approaches 0.\par
On the other hand, imposing 
\beq
R_\rho^2 = 3R_\lambda^2
\eeq

\noi which corresponds to cancelling the ``crossed'' terms, the expression (\ref{1-49e}) simplifies very much and one finds :
\beq
\label{1-49-4e}
\rho_\Lambda^2 = 3m^2 R_\lambda^2 + 2
\eeq

\noi corresponding to the factorization of the wave function in $p_2, p_3$. The value (\ref{1-49-4e}) is not at odds with the slope determined from the lattice data in Section 2, $\rho_\Lambda^2 \simeq 2$.\par
To repeat, the BT result is quite unlike the non-relativistic treatment, which gives always $\rho_\Lambda^2 = 3m^2 R_\lambda^2$, independently of $R_\rho^2$ : it depends now strongly on $R_\rho/R_\lambda$.\par
One sees that in the relativistic treatment {\it $\rho_\Lambda^2$ can get arbitrary large values, while none of the two radii is supposed to be large}.\par
Of course, let us recall that there is no claim to a dynamical calculation in all this discussion, but only an analysis of the relation between a generic gaussian internal wave function and $\rho_\Lambda^2$, specific to the relativistic BT formalism.\par
However, it is interesting to note that the condition (\ref{1-49-4e}), which corresponds to a reasonable value of $\rho_\Lambda^2$, corresponds also to a situation where the distance between the two light quarks would be larger than the distance between each light quark and the heavy quark. This seems consistent with the intuition that the Compton wave length of each light quark is large, and this is in fact the situation which seems to be found in dynamical calculations, like the one of Hernandez {\itshape et al.} in the non-relativistic quark model \cite{NIEVES},
as well as in lattice studies \cite{GREEN}\cite{DE-FORCRAND}. Indeed, it is very important to recall that also in lattice QCD calculations one finds a $qq$ system with a large separation. Let us emphasize that in conclusion of these calculations, the term {\it diquark} must be taken with care since it is often meant on the contrary as a pointlike diquark, especially when speaking of diquark {\it models}. And, of course, these calculations question the very idea of a point-diquark model, at least when claiming to QCD inspired models, as we discuss in the next section.\par
Let us recall now another important conclusion coming from the above discussion. In the BT scheme, the value of the IW function slope for the $Qqq$ system depends strongly on the spatial configuration of the light diquark. This illustrates strikingly the contrast between the BT scheme and the non-relativistic treatment of the center-of-mass motion of hadrons, for which there is no dependence of the slope on $R_\rho$, but only on $R_\lambda$. Therefore, in this relativistic scheme there is a need to have a very good calculation of the wave function.\par
Interestingly, in ref. \cite{NIEVES-2}) the wave function has been calculated very carefully, although in a spectroscopic model with non-relativistic kinetic energy, which may be less worrying for a baryon. As to the authors themselves, they propose a rather low value $\rho_\Lambda^2 \simeq 0.6 - 1.$, too low of course. But this value derives from the non-relativistic treatment of the center-of-mass motion of the baryons.\par 
It would be worth applying the BT formalism to the wave function of \cite{NIEVES-2}) to see whether it yields a correct slope. Indeed, the relativistic BT treatment could enlarge the value appreciably, as explained above and in Subsection 1.1.

\section{The Q-pointlike-diquark models}

As a possible alternative, the models with a point-like diquark instead of two light quarks would be attractive because of their simplicity. 
One must note that the diquark may be also considered as extended, like in the works of Ebert {\itshape et al.} \cite{Q-DIQUARK MODELS}, but this is a different idea, outside of the present discussion (see also ref. \cite{GUO-MUTA}). The quark-diquark model has been widely used to compute properties of the baryon spectrum, and also relevant form factors in heavy baryon transitions \cite{Q-DIQUARK MODELS}.\par
Nevertheless, considering the several findings that have been recalled in the previous section, showing definitely that the $qq$ light quark subsystem has a large size, comparable with the one of the whole baryon, it is paradoxical to appeal to a point-like diquark model. And indeed, our conclusion below in subsection 4.2 is that such a model is not valid in the context of the QCD-inspired potentials, since it leads to a too low value $\rho_\Lambda^2 \simeq 1$ as for mesons, which is quite logical since they are both two-body bound states with one heavy quark, and the potential is quite similar to the one for mesons.\par
On the other hand, this negative argument does not apply if we renounce to a QCD-inspired potential and introduce a non standard harmonic oscillator potential, whose strength can be freely adjusted. And indeed, we develop such a model as a provisory solution in the next section.\par
In subsection 4.1 we first develop the general BT framework for models with scalar point-like diquark model, which will serve for both sections and then apply it to the model of Bing Chen {\itshape et al.}, with a standard QCD-inspired potential, in subsection 4.2.\par

\subsection{Isgur-Wise functions in the BT scheme} 

Let us indeed present the general calculation of the IW functions for a scalar $0^+$ and $\overline{3}$ under color, point-like particle, in the field of a heavy quark. As we will see now, there are no Wigner rotations in this case, and the BT results for IW functions simplify enormously.

\subsubsection{Elastic IW function}

One finds the simple expression (\ref{D-3e}) of Appendix C,
\beq
\label{1-50-11e}
\xi_\Lambda(v.v') = \int {d {\bf p}_2 \over (2 \pi)^3}\ {1 \over p_2^0}\ \varphi(\overrightarrow{{\bf B}^{-1}_{v'} p_2})^* \varphi(\overrightarrow{{\bf B}^{-1}_{v} p_2})\ \sqrt{(p_2.v)(p_2.v')}
\eeq

\noi with
\beq
\label{1-50-12e}
\varphi(\overrightarrow{{\bf B}^{-1}_{v'} p_2})^* \varphi(\overrightarrow{{\bf B}^{-1}_{v} p_2}) = \varphi((p_2.v')^2-m_D^2)^* \varphi((p_2.v)^2-m_D^2)
\eeq

\noi where $m_D$ denotes is the scalar diquark mass, of the order of twice the light quark mass, $m_D \simeq 2m$.

\subsubsection{IW function for $\boldsymbol{L = 1}$ excited states}

In this case one finds expression (\ref{D-8e}) of Appendix C,
\beq\label{1-50-13e}
	\sigma_\Lambda(v.v') = {\sqrt{3} \over w^2-1} \int {d {\bf p}_2 \over (2 \pi)^3}\ {1 \over p_2^0}\ \varphi_1(\overrightarrow{{\bf B}^{-1}_{v'} p_2})^* \varphi(\overrightarrow{{\bf B}^{-1}_{v} p_2})
\ {p_2.(v-wv') \over \sqrt{(p_2.v')^2-m_D^2}}\ \sqrt{(p_2.v)(p_2.v')}
\eeq

\noi where one can see that the $1/(w-1)$ singularity in the overall factor cancels with the numerator $p_2.(v-wv')$, that vanishes when $w \to 1$.

\subsubsection{Bjorken sum rule}

From (\ref{1-50-11e}), the slope of the elastic IW function is
\beq
\label{1-50-14e}
\rho_\Lambda^2 = -\xi'_\Lambda(1) = - {1 \over 24 \pi^2} \int_0^\infty dp\ {1 \over \sqrt{m_D^2+p^2}}
\eeq
$$\times\ p \varphi(p)^* \left\{(6m_D^2p+5p^3)\varphi(p)+4(m_D^2+p^2)\left[2(m_D^2+2p^2)\varphi'(p)+p(m_D^2+p^2)\varphi''(p) \right] \right\}$$

\noi and using (\ref{1-50-13e}) and the completeness relation for radial wave functions,
\beq
\label{1-50-15e}
\sum_n \varphi_1^{(n)}(p^2)\ \varphi_1^{(n)*}(k^2) = 2 \pi^2\ {\delta(p-k) \over p^2} 
\eeq 

\noi we compute the sum $\sum\limits_n \mid \sigma^{(n)}(1) \mid^2$, that gives 
\begin{multline}
\label{1-50-16e}
\sum_n \mid \sigma^{(n)}(1) \mid^2 =  {1 \over 24 \pi^2} \int_0^\infty dp\ {1 \over \sqrt{m_D^2+p^2}}\\ \times
p^2 \left[p\varphi(p)^*+2(m_D^2+p^2) \varphi^{'*}(p) \right] \left[p\varphi(p)+2(m_D^2+p^2) \varphi'(p) \right]
\end{multline}

Integrating by parts this expression, one finds precisely the r.h.s.~of (\ref{1-50-14e}), {\itshape i.e.} one finds the Bjorken sum rule
\beq
\label{1-50-17e}
\rho_\Lambda^2 = \sum_n \mid \sigma^{(n)}(1) \mid^2
\eeq

\noi Moreover, from the positivity of (\ref{1-50-16e}), one recovers the lower bound already established by heavy quark symmetry in ref. \cite{YOUSSEFMIR}
\beq
\label{1-50-18e}
\rho_\Lambda^2 \geq 0
\eeq

\subsubsection{An improved bound on the slope}

However, in this $Q$-diquark model, one can demonstrate a better lower bound due to the absence of the Wigner rotations, just by using the careful analysis of the different contributions to the meson IW slope given in ref. \cite{MORENAS-3}.\par
One finds that the expression for the slope writes
\beq
\label{1-50-19e}
\rho_\Lambda^2 = {1 \over 2 \pi^2} \int_0^\infty dp\ p^2\ \varphi(p)^* \left({p^0z+zp^0 \over 2} \right)^2 \varphi(p)
\eeq

\noi where
\beq
\label{1-50-20e}
p^0 = \sqrt{p^2+m_D^2} \qquad \qquad \qquad z = i {d \over dp^z}
\eeq

\noi and one has demonstrated that the lower bound of expression (\ref{1-50-19e}) is given by
\beq
\label{1-50-21e}
\rho_\Lambda^2 \geq {1 \over 3} \ ,
\eeq

\noi i.e., an improved bound relatively to the general bound (\ref{1-50-18e}).\par
Moreover, this bound was obtained on general grounds for a heavy baryon with light cloud $j = 0$ \cite{DERIVATIVES-IW-RAYNAL} for the shape (\ref{LHCb-2e}) of the IW function.

\subsection{Elastic and inelastic IW functions from wave functions in a QCD-inspired potential model} 

The heavy baryon spectrum has been studied by Bing Chen {\itshape et al.} within the Q-diquark description with a QCD-inspired Hamiltonian \cite{BING-CHEN},
\beq
\label{1-50-30bise}
\left( {{\bf p}^2 \over 2 \mu} - {4 \alpha_s \over 3 r} + b r + C + {\rm spin\ dependent\ terms} \right)\psi = E \psi
\eeq

\noi where $\mu$ is the reduced mass, 
\beq
\label{1-50-30-2e}
\mu = {m_Dm_Q \over m_D + m_Q}
\eeq

\noi and ${\bf p}$ is the relative momentum of the heavy quark $Q$ and light point-like diquark of mass $m_D$.\par 

One notices that the potential in (\ref{1-50-30bise}) is very similar to the one for a meson. This is easily understood : the interquark potential inside baryons is known to be half the one inside mesons, but on the other hand there are two quarks on a diquark. This leads to a similarity in wave functions and finally for $\rho_\Lambda^2 $, except that the mass here is heavier than for a light quark.\par

The wave functions corresponding to the spin-independent part of the Hamiltonian (\ref{1-50-30bise}) are given in Appendix D for the heavy quark limit, and with the free $\beta$ parameter characterizing the variational basis chosen to $\beta = 0.4$.

\begin{figure}[htb]
\centerline{\includegraphics[height=\grapheheight]{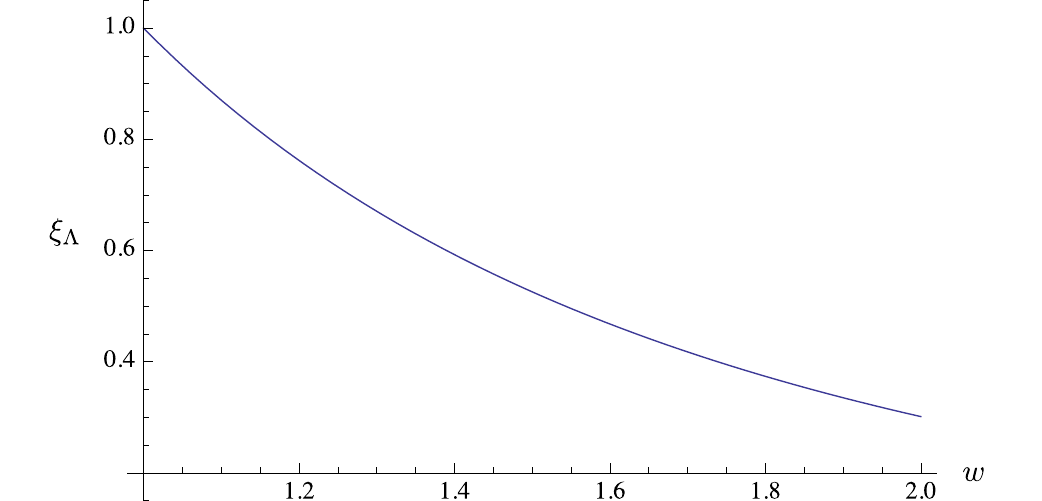}}
\caption{Elastic Isgur-Wise function $\xi_{\Lambda}(w)$ obtained in the $Q$-diquark model with the Bing Chen {\itshape et~al.} wave function (\ref{1-50-32e}).}
\end{figure}
Inserting the heavy quark limit wave function $\varphi^{(n)}_0 ({\bf p})$ (\ref{1-50-32e}) into the expression (\ref{1-50-11e}) and with the reduced mass parameter that describes the spectrum within the Bing Chen {\itshape et al.} Hamiltonian
\beq
\label{1-50-33bise} 
\mu = 0.45\ \rm{GeV}
\eeq
\noi one finds the elastic IW function $\xi_{\Lambda}(w)$ of Fig.~4.
%
%
%
%
An excellent fit is the ``dipole'' function (\ref{LHCb-2e})
\beq
\label{1-50-35e} 
\xi_\Lambda(1) = \left({2 \over {w+1}}\right)^{2\rho_\Lambda^2}
\eeq

\noi with the slope and curvature

\beq
\label{1-50-36bise} 
\rho_\Lambda^2 = 1.27\ , \qquad \qquad \sigma_\Lambda^2 = 2.25
\eeq

\noi The slope is lower than the value $\rho_\Lambda^2 \simeq 2$ obtained from the data of Lattice QCD described in Section 2. This low value is easily understood because this model amounts to a meson system, except for details of spin and for the light diquark mass which should be larger than for one quark. One must also take into account the difference of definition for baryons against mesons, that have a $+ 1/4$ for the slope.\par
Consequently, the much too low value of $\rho_\Lambda^2$ (\ref{1-50-36bise}) compels us to abandon the model, at least for form factors, as this is seen to be an unavoidable consequence of the scheme.\par 
\begin{figure}[htb]
\centerline{\includegraphics[height=\grapheheight]{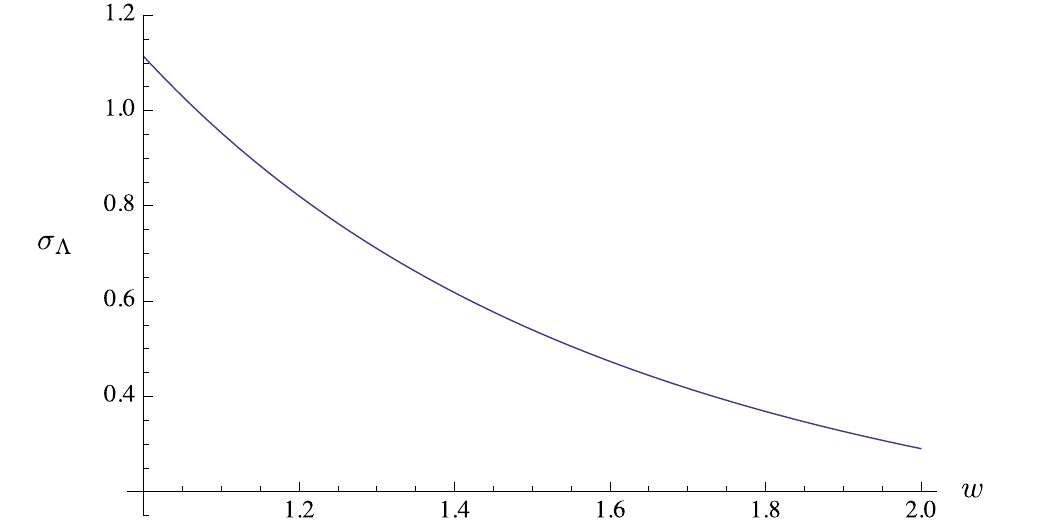}}
\caption{Inelastic Isgur-Wise function $\sigma_\Lambda (w)$ obtained in the Q-diquark model with the Bing Chen {\itshape et al.} wave functions (\ref{1-50-32e},\ref{1-50-33e}).}
\end{figure}
Nevertheless, we add for completeness the predictions of the model for the $L = 1$ excitation. In the Bing Chen {\itshape et al.} model, from $\varphi^{(n)}_0 ({\bf p})$ (\ref{1-50-32e}) and $\varphi^{(n)}_1 ({\bf p})$ (\ref{1-50-33e}), and the parameter (\ref{1-50-33bise}) we find the inelastic IW function $\sigma_{\Lambda}(w)$ of Fig.~5.
%
%
%
%
%
The zero recoil value, the slope and the curvature of the inelastic IW function are
\beq
\label{1-50-34e} 
\sigma_\Lambda(1) = 1.11\ , \qquad \qquad \sigma'_\Lambda(1) = -1.89\ , \qquad \qquad \sigma''_\Lambda(1) = 4.16
\eeq

\noi This corresponds to the lowest excitation $n = 0$ of the inelastic $L= 0 \to L = 1$ IW function, and the Bjorken sum rule is almost saturated by it. Indeed, the r.h.s. of (\ref{1-50-17e}) has a large contribution from the $n = 0$ state,
\beq
\label{1-50-36e} 
\rho_\Lambda^2 = 1.27 \geq \left[\sigma_\Lambda(1)\right]^2 = 1.11^2 = 1.23
\eeq

\section{Spectrum and IW functions with harmonic oscillator wave functions}

As explained in the preceding Sections 3 and 4, we have not obtained a satisfactory description of the baryon IW function, neither using the internal wave function for three quarks deduced by Pervin {\itshape et al.} from a standard linear + Coulomb interquark potential, nor using the two-body wave functions we have deduced from the pointlike diquark model of Bing Chen {\itshape et al.}

Compared to the lattice QCD result, $\rho_\Lambda^2$ has been found either much too large with a three-quark wave function of Pervin {\itshape et al.}, or much too low for the diquark model of Bing Chen {\itshape et al.} This situation is quite different from the meson case where the various standard spectroscopic models with relativistic kinetic energy combined with the BT scheme give consistently $\rho^2 \simeq 1$, in good agreement with data. 

Why one fails in the case of the Bing Chen {\itshape et al.} is clear from the discussion : the 
pointlike diquark assumption directly contradicts the dynamical calculations of the three quark system, in particular those of lattice QCD, which show definitely that the diquark system has a large extension. In fact the model is close to a heavy meson, with a similar potential, and the BT formalism yields consistently $\rho_\Lambda^2$ not much above $\simeq 1.$

For the Pervin {\itshape et al.} wave function, one has no reason to suspect the linear + Coulomb spectroscopic Hamiltonian, and the reason is less obvious : the calculation of the wave function clearly requires larger bases, since the authors have observed a very large discrepancy between the HO and PC bases - all the more since in a three-quark system, as we have shown in Section 3, the $\rho_\Lambda^2$ deduced from BT is very sensitive to the details of the wave functions, in contrast with a non-relativistic treatment.

Then, leaving the correct solution of the three-quark case for a further investigation, we turn presently, for the phenomenological purpose of computing the observables, to a very simple model that is able to fit the observed $\rho_\Lambda^2$. It is a point-like diquark model, but quite different from the one above with a QCD-inspired potential, with now a harmonic oscillator potential of arbitrary strength, which can be fitted both to the low-lying spectrum and to the desired $\rho_\Lambda^2$. Such a model is analogous in spirit with HO models used in the beginning of the quark model, except for the further simplification of using a pointlike diquark picture.
The reason to expect sensible results from such a rough model is the fact that, in a first approximation, $\rho_\Lambda^2$ seems the main parameter controlling the heavy limit of the form factors, because dipole fits describe well the overall shape of $\xi_\Lambda(w)$ both in the model and on the lattice.

Let us assume harmonic-oscillator wave functions for the ground state $\Lambda_Q\ (Q = b, c)$ and for the tower of the radially excited $L = 1$ states ($n \geq 0$) according to the Hamiltonian
\beq
\label{1-50-23bise}
\left( {{\bf p}^2 \over 2 \mu} + {1 \over 2} K r^2 + C \right)\psi = E \psi
\eeq

\noi where the reduced mass $\mu$ is given by (\ref{1-50-30-2e}) and $m_D$ is the light diquark mass. The spring tension $K$ in eqn. (\ref{1-50-23bise}) is flavor-independent, the usual hypothesis for the harmonic oscillator Hamiltonian, well satisfied for meson and baryon spectra.\par 
From (\ref{1-50-23bise}), in terms of the reduced mass $\mu$ and the $n = 0$ level spacing 
\beq
\label{1-50-23tere}
\omega = m_{\Lambda_c} {\scriptstyle \left({1 \over 2}^-\right)} - m_{\Lambda_c} {\scriptstyle \left({1 \over 2}^+\right)}
\eeq

\noi the spring tension is given by
\beq
\label{1-50-23-4e}
K = \mu \omega^2
\eeq

For very large $m_Q$, $\mu \simeq m_D$, where $m_D$ is the diquark mass $m_D \simeq 2 m$, and $m$ the light quark mass. For finite $m_Q$ one has $\mu < m_D$, and $\mu \simeq 0.4$ GeV in the case of charmed quarks.\par
According to (\ref{1-50-23bise}) the wave functions read
\beq
\label{1-50-23e}
\varphi_0({\bf p}) = (4\pi)^{3/4} R^{3/2} \exp \left(- { {\bf p}^2 R^2 \over 2} \right)
\eeq
\beq
\label{1-50-24e}
\varphi^{(n)}_1 ({\bf p}) = (-1)^n (4\pi)^{3/4} 2^{n+1} \sqrt{{n!(n+1)! \over (2n+3)!}}\ R^{5/2} \mid {\bf p}\mid L^{3/2}_n ({\bf p}^2 R^2) \exp \left(- { {\bf p}^2 R^2 \over 2} \right)
\eeq

\noi that are normalized according to 
$$\int {d {\bf p} \over (2 \pi)^3} \mid \varphi({\bf p}) \mid^2\ = 1$$

\noi We will consider also the wave functions for $n > 0$ and $L = 1$ in order to verify that Bjorken SR holds.\par 

Let us consider harmonic oscillator parameters that describe qualitatively the spectrum data, namely
\beq
\label{1-50-24bise}
m_{\Lambda_c} {\scriptstyle \left({1 \over 2}^+\right)_{n= 0}} = 2.286\ {\rm GeV} \ , \qquad \qquad  m_{\Lambda_c} {\scriptstyle \left({1 \over 2}^- \right)_{n= 0 }} = 2.595\  {\rm GeV}
\eeq

\noi that gives the level spacing and reduced mass
\beq
\label{1-50-24-3e}
\omega = m_{\Lambda_c} {\scriptstyle \left({1 \over 2}^-\right)_{n= 0}} - m_{\Lambda_c} {\scriptstyle \left({1 \over 2}^+\right)_{n= 0}}\simeq 0.309\ {\rm GeV} \ , \ \  \mu = {m_D m_c \over m_D+m_c} \simeq 0.40\  {\rm GeV} 
\eeq

\noi for a light quark mass $m \simeq {m_D \over 2} \simeq 0.30\  {\rm GeV}$ and a charm quark mass like in Section 2, $m_c \simeq 1.25$ {\rm GeV}.
Therefore, the spring tension (\ref{1-50-23tere}) will be
\beq
\label{1-50-24-3-1e}
K = 0.038\ {\rm GeV}^3
\eeq

Although the quantum numbers are still not confirmed, we consider now the natural candidate for the radial excitation, as assumed in \cite{BING-CHEN}. 
\beq
\label{1-50-24-2e}
m_{\Lambda_c} {\scriptstyle {\left({1 \over 2}^+ \right)_{n= 1}}} = 2.767\ {\rm GeV}
\eeq

\noi This gives the level spacing $m_{\Lambda_c} {\scriptstyle {\left({1 \over 2}^+ \right)_{n= 1}}}-m_{\Lambda_c} {\scriptstyle {\left({1 \over 2}^+ \right)_{n= 0}}} = 0.481\ {\rm GeV}$, while our simple model predicts 
\beq
\label{1-50-24-3bise}
m_{\Lambda_c} {\scriptstyle {\left({1 \over 2}^+ \right)_{n= 1}}}-m_{\Lambda_c} {\scriptstyle {\left({1 \over 2}^+ \right)_{n= 0}}} = 2 \omega = 0.618\ {\rm GeV}\ ,
\eeq

\noi some $20 \%$ higher.\par
However, since the IW function is defined in the heavy quark limit, we should take the reduced mass for $m_Q \to \infty$. To summarize, the spring tension $K$ is kept fixed and the reduced mass becomes in the heavy quark limit $\mu \to m_D$ where $m_D$ is the diquark mass. One has then, in the heavy quark limit, the radius squared of the wave function,
\beq
\label{1-50-24-2terte}
R^2 = {1 \over (m_D K)^{1/2}}
\eeq

\noi  One finds for some illustrative cases, for $m_c=1.25$ GeV, the radius squared, the elastic slope $\rho_\Lambda^2$ and, using formula (\ref{1-50-13e}), the square of the $n = 0$ inelastic IW function $L = 0 \to L = 1$ at zero recoil $\mid\sigma^{(0)}_\Lambda(1)\mid^2$,
\beq
\label{1-50-24-4bisae}
m_D = 0.6\ {\rm GeV}\ , \ \ \ R^2 = 6.76\ {\rm GeV^{-2}}\ , \ \ \ \rho_\Lambda^2 = 1.99\ , \ \ \ \mid\sigma^{(0)}_\Lambda(1)\mid^2\ = 1.93 
\eeq
\beq
\label{1-50-24-2-4bisbe}
m_D = 0.8\ {\rm GeV}\ , \ \ \  R^2 = 5.34\ {\rm GeV^{-2}}\ , \ \ \  \rho_\Lambda^2 = 2.48\ , \ \ \ \mid\sigma^{(0)}_\Lambda(1)\mid^2\ = 2.42 
\eeq

\noi Notice that within HQET one has, for baryons, the parameter $\overline{\Lambda}$,
\beq
\label{1-50-24-2-5e}
\overline{\Lambda} \simeq m_D
\eeq

\noi and the value $\overline{\Lambda} = m_D \simeq 0.8\ {\rm GeV}$ is precisely the one adopted in the HQET study of Leibovich and Stewart \cite{LEIBOVICH-STEWART}.\par

Here, to compute the interesting observables, we would like to adjust $m_D$ in order to obtain the central value for the slope obtained from the lattice data. We get roughly,
\beq
\label{1-50-24-2-4bisce}
m_D = 0.67\ {\rm GeV}\ , \ \ \  R^2 = 6.20\ {\rm GeV^{-2}}\ , \ \ \  \rho_\Lambda^2 = 2.15\ , \ \ \ \mid\sigma^{(0)}_\Lambda(1)\mid^2\ = 2.10 
\eeq

We plot in Fig.~6 the elastic IW function for the set of parameters (\ref{1-50-24-2-4bisce}).\par
\begin{figure}[htb]
\centerline{\includegraphics[height=\grapheheight]{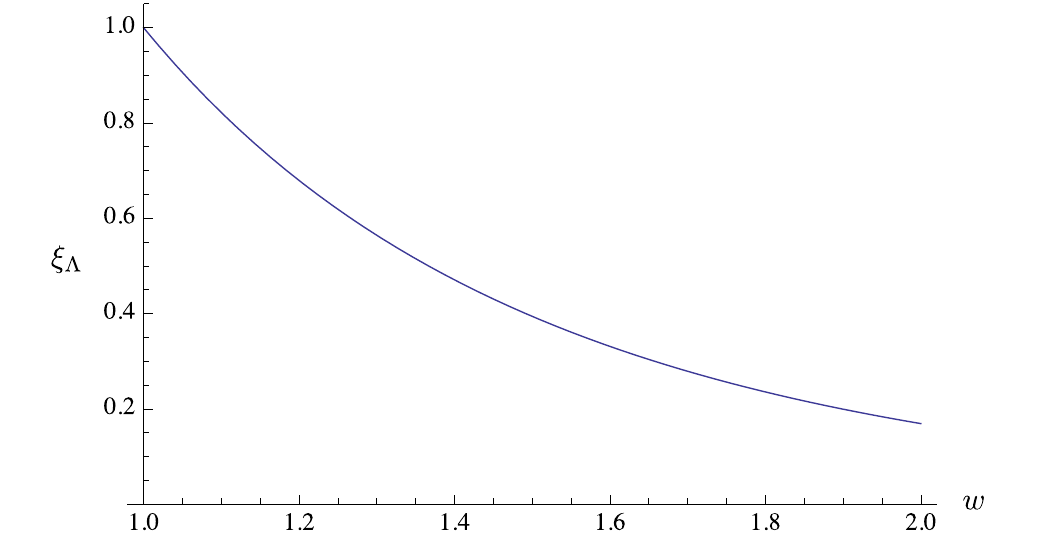}}
\caption{Elastic Isgur-Wise function $\xi_{\Lambda}(w)$ obtained in the Q-diquark model with the harmonic oscillator wave function (\ref{1-50-23e}) and the parameters (\ref{1-50-24-2-4bisce}).}
\end{figure}
%
%
%
%
%
A very good fit to the IW function of Fig. 6 is given by the ``dipole'' form with $\rho^2_\Lambda = 2.15$. Comparing the values for $\rho_\Lambda^2$ and $\mid \sigma^{(0)}_\Lambda(1)\mid^2$, we observe that the lowest radial excitation $n = 0$ largely dominates the r.h.s. of Bjorken SR (\ref{1-50-17e}).\par 
Therefore, we conclude that {\it the lowest inelastic IW function $(L = 0, n = 0) \to (L = 1, n = 0)$ is large, and thus there is a good prospect for this transition to be well observed at LHCb}.

\vskip 6 truemm

We plot in Fig.~7 the inelastic IW function $\sigma_\Lambda(w)$ with the set of parameters (\ref{1-50-24-2-4bisce}).\par
%
%
%
%
%
%
A reasonable ``dipole'' fit to Fig. 7 is given by
\beq
\label{C-13-1e} 
\sigma_\Lambda(w) = \sigma_\Lambda(1) \left({2 \over w+1}\right)^{2\sigma'_\Lambda (1)}
\eeq

\noi with
\beq
\label{C-13-2e} 
\sigma_\Lambda(1) = 1.44 \ ,  \qquad \qquad \sigma'_\Lambda(1) = 2.57
\eeq
\begin{figure}[htb]
\centerline{\includegraphics[height=\grapheheight]{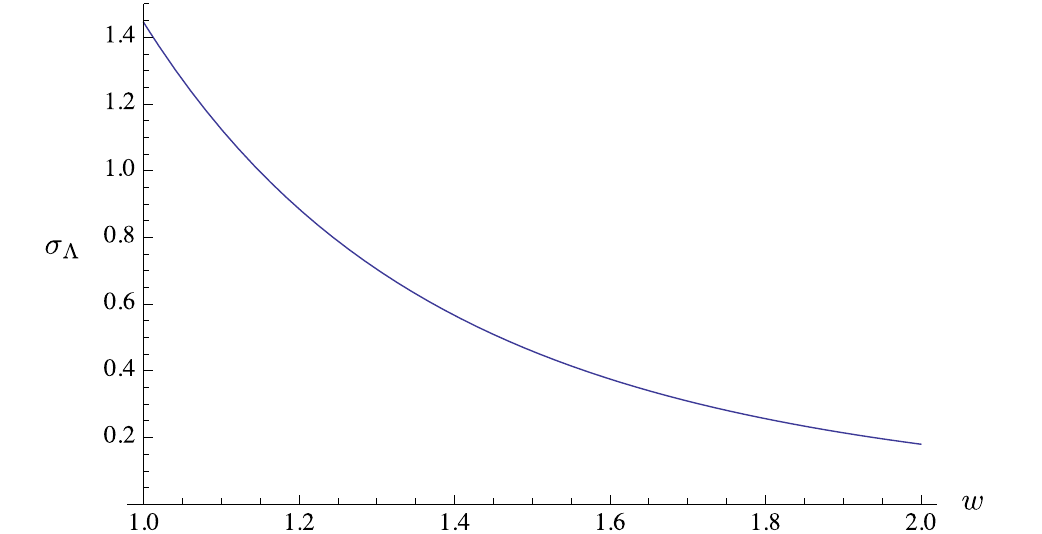}}
\caption{Inelastic Isgur-Wise function $(L = 0, n = 0) \to (L = 1, n = 0)$ obtained in the Q-diquark model with the harmonic oscillator wave functions (\ref{1-50-23e},\ref{1-50-24e}) and the parameters  (\ref{1-50-24-2-4bisce}).}
\end{figure}
%
%
\section{Observables in $\boldsymbol{\Lambda_b \to \Lambda_c\left({1 \over 2}^\pm \right) \ell \overline{\nu}}$ transitions}

The Mainz group has extensively formulated a number of different observables that could allow to test Lepton Flavor Universality Violation \cite{MAINZ OBSERVABLES}.
The expressions for the observables in terms of helicity amplitudes are given below in Appendix E.

\subsection{Observables for $\boldsymbol{\Lambda_b \to \Lambda_c \left({1 \over 2}^+ \right) \ell \nu\ (\ell = e, \tau)}$ transitions}

\vskip 2 truemm

For the numerical calculations of the form factors and helicity amplitudes we adopt the ``dipole'' shape expression for the IW function, with the slope (\ref{2-26tere}) determined from the lattice data in Section 2, 
\beq
\label{B-8e}
\xi_\Lambda(w)=\left( {2 \over w+1} \right)^{2 \rho_\Lambda^2} , \qquad \qquad \rho_\Lambda^2 = 2.15 \pm 0.10
\eeq

\noi The ansatz (\ref{B-8e}) is close to the numerical calculation in the BT model within the Q-diquark scheme with HO NR internal wave function (\ref{1-50-24e}) with parameters (\ref{1-50-24-2-4bisce}).

For the function $A(w)$ we adopt
\beq
\label{B-9e}
A(w)=A'(1)(w-1)f(w) \ , \qquad \qquad A'(1) = -0.35 \pm 0.15
\eeq

\noi where the function $f(w)$ satisfies $f(1)=1$ and is introduced to soften the behaviour of $A(w)$ for large $w$, near $w_{max}$, because lattice data give only the slope (\ref{2-26bise}). As an example, we could use simply $f(w) = \xi_{\Lambda}(w)$. So, we take
\beq
\label{B-9-1e}
A(w)=A'(1)(w-1)\xi_{\Lambda}(w) \ , \qquad \qquad A'(1) = -0.35 \pm 0.15
\eeq
 
We will below comment further the role of the function $f(w)$ in (\ref{B-9e}), when discussing the comparison of the spectrum with experiment in section 6.1.1.
 Moreover, for $m_c, \ m_b$ we use the central values (\ref{2-26bise}), and for $\overline{\Lambda}$ we adopt the value of our model (\ref{1-50-24-2-4bisce}), that agrees within errors with the lattice determination (\ref{2-26bise}), 
\beq
\label{B-9bise}
m_c = 1.25\ {\rm GeV}\ , \qquad m_b = 4.25\ {\rm GeV}\ , \qquad \overline{\Lambda} = m_D = 0.67\ {\rm GeV} 
\eeq

The observables are given in Appendix E in terms of the helicity amplitudes $H^{V/A}_{\lambda_2,\lambda_W}$, that are given in terms of the form factors by the expressions,
$$H^{V/A}_{+{1\over2}t} = {\sqrt{Q_\pm} \over \sqrt{q^2}} \left( M_\mp f_1^{V/A} \pm q^2 f_3^{V/A} \right)$$
\beq
\label{C-13e} 
H^{V/A}_{+{1\over2}0} = {\sqrt{Q_\mp} \over \sqrt{q^2}} \left( M_\pm f_1^{V/A} \pm q^2 f_2^{V/A} \right)
\eeq
$$H^{V/A}_{+{1\over2}+1} = \sqrt{2Q_\mp}  \left (f_1^{V/A} \pm M_\pm f_2^{V/A} \right)$$

In the physical processes the $V-A$ chiral combination (\ref{C-1e}) appears, and one has the parity relations between the $V/A$ helicity amplitudes
\beq
\label{C-14e} 
H^V_{-\lambda_2,-\lambda_W} = H^V_{\lambda_2,\lambda_W}, \qquad \qquad H^A_{-\lambda_2,-\lambda_W} =- H^A_{\lambda_2,\lambda_W}
\eeq

\noi The form factors in (\ref{C-13e}) are described in Appendix A.1.\par

\subsubsection{The normalized theoretical rate compared to LHCb data}

\vskip 3 truemm

Among the observables, only the shape of the LHCb data on the differential rate is known, but not the absolute magnitude \cite{IW-LHCb}. We compare the LHCb rate normalized to one with the predictions of our model.\par

As a first remark, let us notice that our model of form factors up to $O(1/m_Q)$ included, with essentially a single main parameter $\rho_\Lambda^2$ can well reproduce the LHCb normalized rate, as shown in Fig.~8.
%
%
%
%
%
We have used the ``dipolar'' shape (\ref{B-8e}) with a slope slightly lower than the domain obtained from the lattice, $\rho_\Lambda^2 = 2$, the mass parameters (\ref{B-9bise}), and the assumption $A(w) = 0$.
\begin{figure}[htb]
\centerline{\includegraphics[height=\grapheheight]{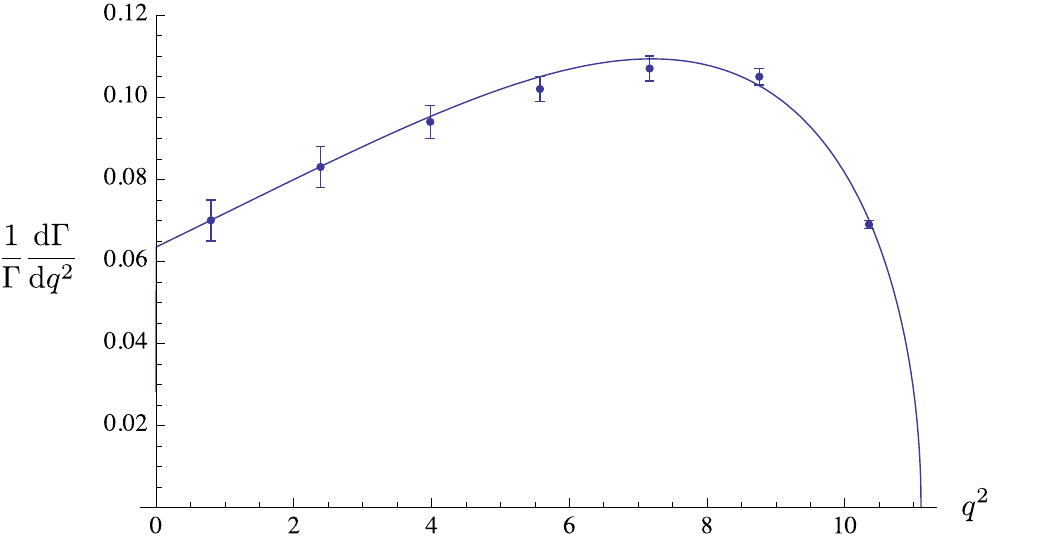}}
\caption{Normalized rate ${d\Gamma \over dq^2}/\Gamma$ compared to the LHCb data. We use the ``dipolar'' shape with $\rho_\Lambda^2 = 2$, the mass parameters (\ref{B-9bise}), and we assume $A(w) = 0$.}
\end{figure}

We consider next the comparison with the parameters obtained from the lattice. Since the value of the slope $\rho_\Lambda^2 = 2$ is at the lower edge of the domain (\ref{B-8e}) and the assumption $A(w) = 0$ is at odds with the values (\ref{B-9e}), we need to check the effect of the range of the lattice values.\par

We now compare the lattice parameters (\ref{B-8e},\ref{B-9e},\ref{B-9bise}) with the LHCb data. With the aim of clarifying the discussion, we choose three sets of parametrizations, all of them with the mass parameters (\ref{B-9bise}).\par
\noi (1) The lattice (\ref{B-8e},\ref{B-9e}) central values $\rho_\Lambda^2 = 2.15$, A'(1) = -0.35, and the linear approximation $A(w) = -0.35(w-1)$.\par
\noi (2) The lattice (\ref{B-8e},\ref{B-9e}) central values $\rho_\Lambda^2 = 2.15$, A'(1) = -0.35, and softened $A(w)$ as $w$ increases, $A(w) = -0.35(w-1)\xi_\Lambda(w)$.\par
\noi (3) The lattice (\ref{B-8e},\ref{B-9e}) smallest values $\rho_\Lambda^2 = 2.05$, A'(1) = -0.20, and softened $A(w)$ as $w$ increases, $A(w) = -0.20(w-1)\xi_\Lambda(w)$.\par

\vskip 3 truemm

We compare these different choices to the data in Fig.~9.
%
%
%
%
We observe that the set of parameters (1) describes the data very poorly, in particular due to the linear behaviour of $A(w)$ and also due to a slightly too large slope. The set (2) is somewhat better, due to the softening of $A(w)$ at large $w$. Finally, the parameters (3) describe the data rather well, although not as well as the naive choice of Fig. 8.\par
\begin{figure}[htb]
\centerline{\includegraphics[height=\grapheheight]{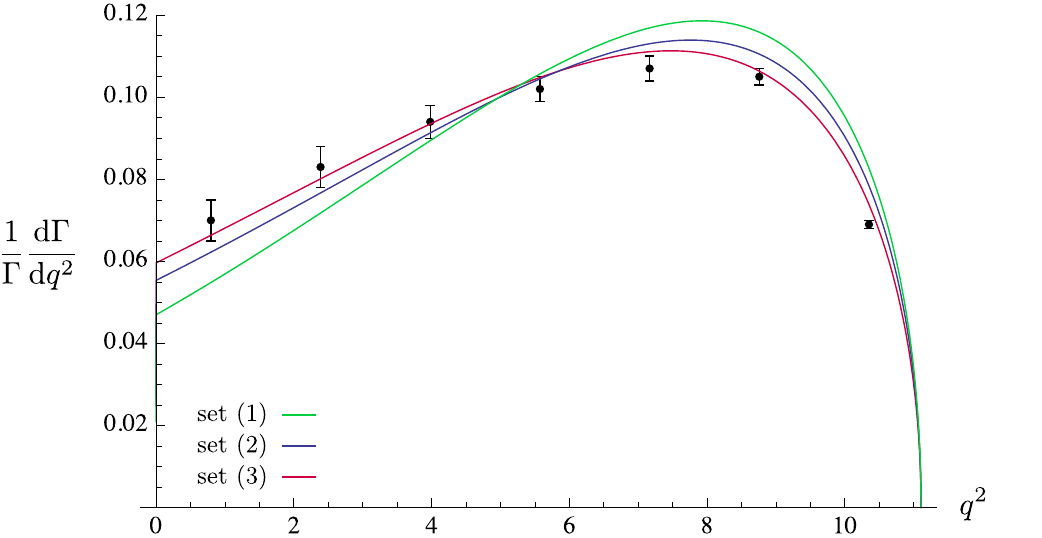}}
\caption{Normalized rate ${d\Gamma \over dq^2}/\Gamma$ compared to the LHCb data for the three sets of parameters (1), (2) and (3), respectively lower, middle and upper curves at $w = 1$  (or $q^2=0$).}
\end{figure}
The main conclusion of this discussion is that the LHCb normalized rate agrees within errors with the fit to the lattice data of form factors performed in Section 2, that are summarized in formulas (\ref{B-8e},\ref{B-9e},\ref{B-9bise}).

\subsubsection{Other observables}

We have seen that there are no sizeable differences between the set of parameters $(\rho_\Lambda^2, A'(1)) = (2.05, -0.20)$ (Fig. 9) and the naive ansatz $(\rho_\Lambda^2, A'(1)) = (2, 0)$ (Fig. 8). For the calculation of the rest of the observables we will use for simplicity the latter set. Moreover, all observables that are given by ratios of squared of helicity amplitudes are not sensitive to the small differences between the parameters used in Fig. 8 and the set (3) in Fig. 9.

\begin{figure}[htbp]
\centerline{\includegraphics[height=\grapheheight]{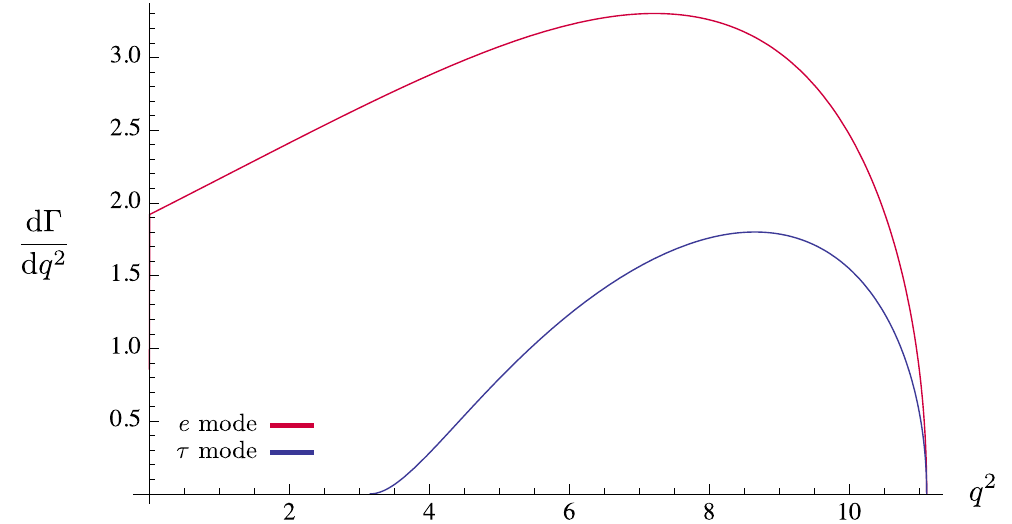}}
\caption{${d\Gamma \over dq^2}$ for the electron and tau modes. In the electron case, one has ${d\Gamma \over dq^2} \to 0$ for $q^2 \to 0$.}
\end{figure}
%
%
%
\begin{figure}[!tbp]
\centerline{\includegraphics[height=\grapheheight]{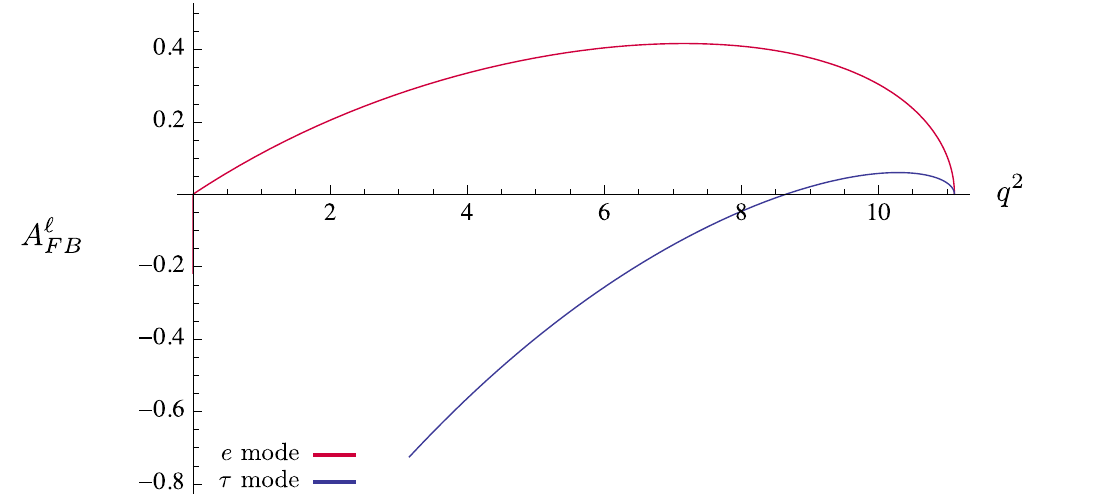}}
\caption{Forward-backward asymmetry $A^\ell_{FB}(q^2)$ for the electron and tau modes.}
\end{figure}
%
%
%
%
\begin{figure}[!tbp]
\centerline{\includegraphics[height=\grapheheight]{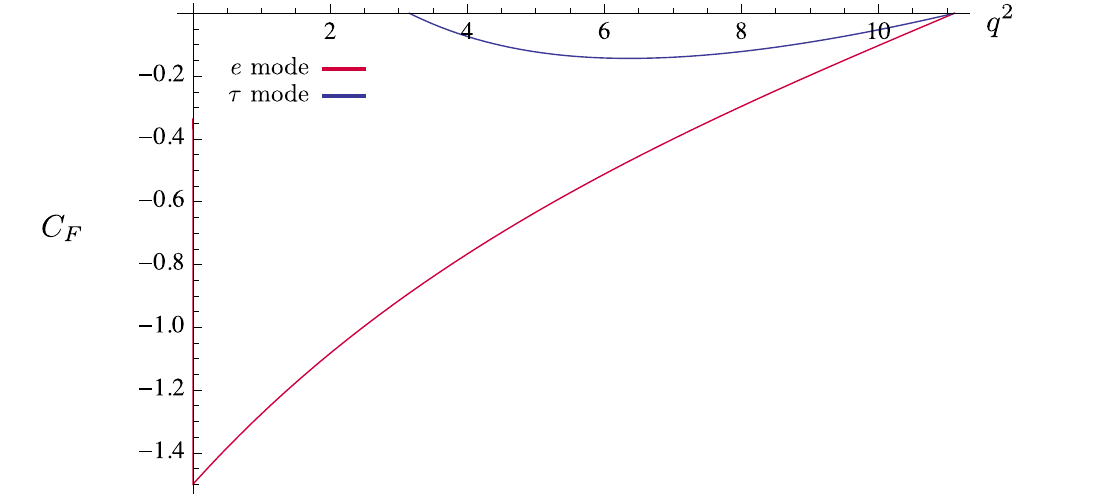}}
\caption{Convexity parameter $C_F(q^2)$ for the electron and tau modes.}
\end{figure}
%
%
%
%
\begin{figure}[!tbp]
\centerline{\includegraphics[height=\grapheheight]{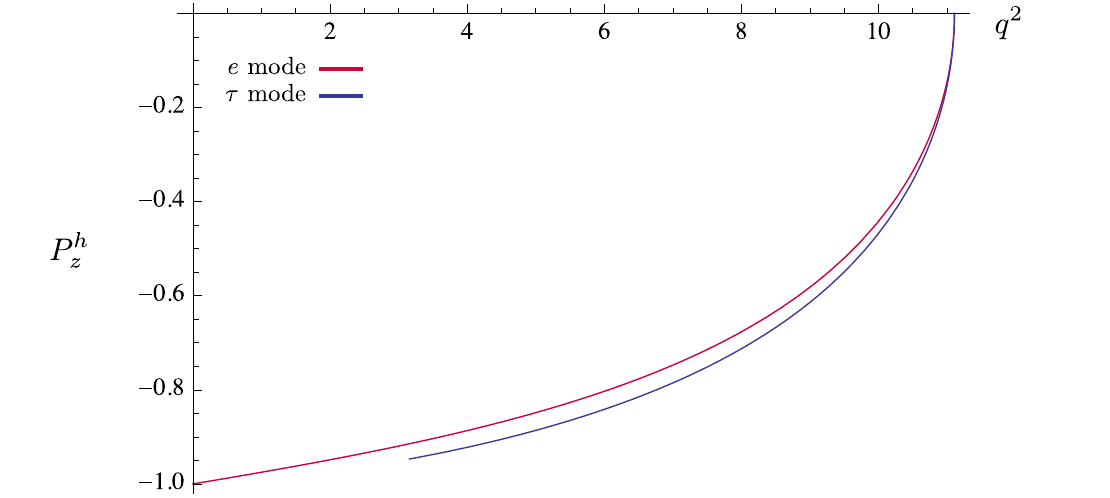}}
\caption{Longitudinal hadron polarization $P_z^h(q^2)$ for the electron and tau modes.}
\end{figure}
%
%
%
%
\begin{figure}[!tbp]
\centerline{\includegraphics[height=\grapheheight]{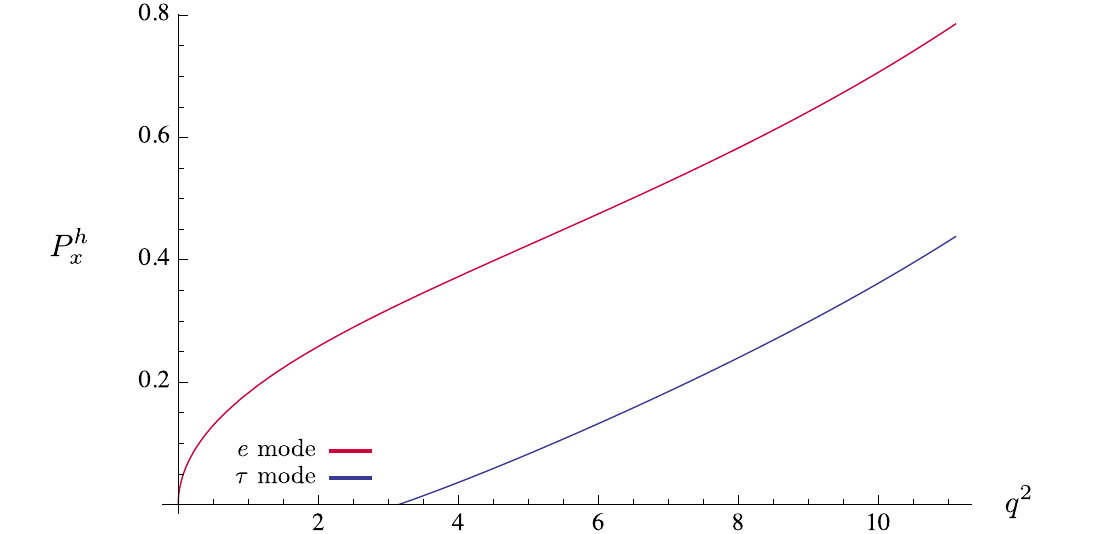}}
\caption{Transverse hadron polarization $P_x^h(q^2)$ for the electron and tau modes.}
\end{figure}
%
%
%
%
\begin{figure}[!tbp]
\centerline{\includegraphics[height=\grapheheight]{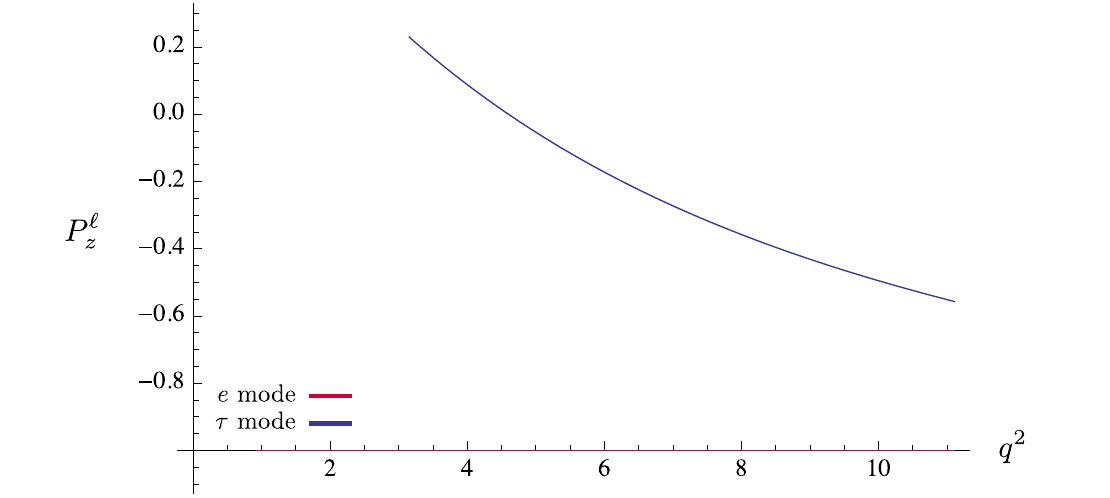}}
\caption{Longitudinal lepton polarization $P_z^\ell(q^2)$ for the electron and tau modes. $P_z^e(q^2)$ is very close to $-1$, unlike $P_z^\tau(q^2)$.}
\end{figure}
%
%
%
%
\begin{figure}[!tbp]
\centerline{\includegraphics[height=\grapheheight]{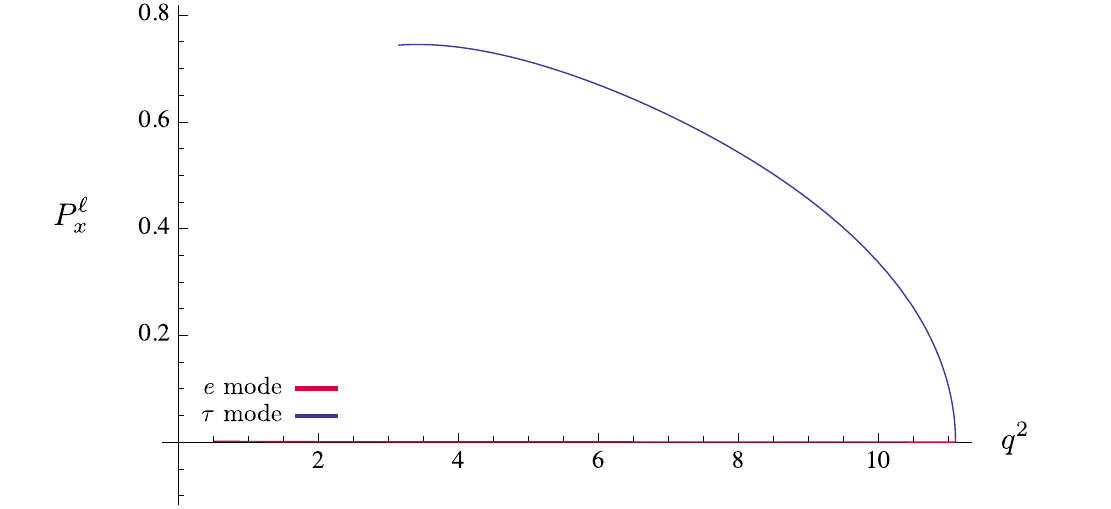}}
\caption{Transverse lepton polarization $P_x^\ell(q^2)$ for the electron and tau modes. $P_x^e(q^2)$ is very small, unlike $P_x^\tau(q^2)$.}
\end{figure}
%
%
%
%
%
\subsubsection{Comments on other observables $\boldsymbol{\Lambda_b \to \Lambda_c \left({1 \over 2}^+ \right) \ell \nu\ (\ell = e, \tau)}$}

It is interesting to observe the shape of the forward-backward asymmetry in Fig. 11 for the electron mass and for the $\tau$ mass, where we observe a zero in the distribution.\par
From (\ref{C-7e}), {\it for $m_\ell \simeq 0$, the electron case}, the FB asymmetry is given by

\beq
\label{C-15e}
A^e_{FB}(q^2) = - {3 \over 2} {{\cal H}_P\over {\cal H}_{tot}} = - {3 \over 2} {\mid H_{+{1\over2}+1} \mid^2 - \mid H_{-{1\over2}-1} \mid^2 \over {\cal H}_{tot}}
\eeq

\vskip 3 truemm

From the $V-A$ structure of the theory, the left-handed final baryon dominates, and therefore we expect to have the inequality 
\beq
\label{C-16e}
\mid H_{+{1\over2}+1} \mid^2\ <\ \mid H_{-{1\over2}-1} \mid^2
\eeq

\noi and similar inequalities for other helicity amplitudes. From (\ref{C-16e}) and  (\ref{C-15e}) we expect $A^e_{FB}(q^2)$ to be positive for all values of $q^2$, as we observe in Fig. 11.\par

Let us see how the inequality (\ref{C-16e}) holds in our model of the form factors described in Appendix A.1.  Just for illustration, keeping only the heavy quark limit terms, one finds indeed
\beq
\label{C-18e}
{\cal H}_P =\ \mid H_{+{1\over2}+1} \mid^2 - \mid H_{-{1\over2}-1} \mid^2\ = -8 \sqrt{\lambda(m_{\Lambda_b}^2,m_{\Lambda_b}^2,q^2)} \left[\xi_\Lambda(w)\right]^2 < 0
\eeq

\noi with\ $\lambda(a,b,c) = a^4+b^4+c^4-2a^2 b^2-2b^2c^2-2c^2a^2$, so that
\beq
\label{C-21e}
q^2 = m_{\Lambda_b}^2+m_{\Lambda_c}^2-2m_{\Lambda_b} m_{\Lambda_c} w, \qquad \lambda(m_{\Lambda_b}^2,m_{\Lambda_c}^2,q^2) = 2 m_{\Lambda_b} m_{\Lambda_c} \sqrt{w^2-1}
\eeq

In the presence of {\it a non-vanishing lepton mass $m_\ell$}, the FB asymmetry (\ref{C-7e}) presents a zero.\par
In particular, for the $\tau$ case, one has a zero in the $FB$ asymmetry as shown in Fig. 11. It is interesting to have a theoretical idea of the position of this zero, which, keeping only the heavy quark limit terms, is
\beq
\label{C-22e}
{q^2_0}(A^\tau_{FB}) = m_\tau \sqrt{m_b^2-m_c^2}
\eeq

\noi which qualitatively agrees with the one of Fig. 11, computed taking into account $1/m_Q$ subleading terms.\par
We observe in Fig. 15 that in the $\tau$ case, the longitudinal lepton polarization $P_z^\tau (q^2)$ has a zero in the neighborhood of $q^2 \simeq 4.\ \rm{GeV} ^2$. Indeed, performing an expansion in powers of ${m_\tau \over mb}$ and ${m_c \over mb}$, one finds the position of this zero
\beq
\label{C-22bise}
q^2_0(P_z^\tau) \simeq 2 m_\tau^2 \left( 1 + 2\ {m_c^2 \over m_b^2} - 3\ {m_\tau^2 \over m_b^2} \right) 
\eeq

\noi that is numerically reasonable.

\vskip 10 truemm

\subsection{Observables for $\boldsymbol{\Lambda_b \to \Lambda_c \left({1 \over 2}^- \right) \ell \nu\ (\ell = e, \tau)}$ transitions}

To compute the form factors we refer to the expressions and discussion of Appendix A.2, taken from Leibovich and Stewart \cite{LEIBOVICH-STEWART}. We neglect the subleading Lagrangian perturbations (\ref{A.2-4e}), that amounts to take (cf. (\ref{A.2-6e})) $\tilde \sigma (w) \equiv \sigma_\Lambda(w)$ and the central value of (\ref{A.2-10e}), ${\hat \sigma}_1 = 0$. We are left with the leading and subleading contributions proportional to the inelastic IW function.\par 
For the inelastic IW function $\sigma(w) \equiv \sigma_\Lambda(w)$ we use the calculation (\ref{C-13-1e} ,\ref{C-13-2e}) done with the same parameters used in the elastic case (\ref{1-50-24-4bisae}),
\beq
\label{C-24bise} 
\sigma_\Lambda(w) = 1.44 \left({2 \over w+1}\right)^{5.14}
\eeq

\noi Moreover, for the rest of the parameters we also use the central values (\ref{B-9bise}), and 
\beq
\label{B-9-2e}
\overline{\Lambda'} = 0.95\ {\rm GeV} 
\eeq

The helicity amplitudes $H^{V/A}_{\lambda_2,\lambda_W}$ are in this case,
$$H^{V/A}_{+{1\over2}t} = {\sqrt{Q_\mp} \over \sqrt{q^2}} \left( M_\pm g_1^{V/A} \mp q^2 g_3^{V/A} \right)$$
\beq
\label{C-23e}
 H^{V/A}_{+{1\over2}0} = {\sqrt{Q_\pm} \over \sqrt{q^2}} \left( M_\mp g_1^{V/A} \mp q^2 g_2^{V/A} \right)
\eeq
$$H^{V/A}_{+{1\over2}+1} = \sqrt{2Q_\pm}  \left (- g_1^{V/A} \pm M_\mp g_2^{V/A} \right)$$

\noi In the physical processes the $V-A$ chiral combination (\ref{C-1e}) appears, and the parity relations between helicity amplitudes are now,
\beq
\label{C-24e}
H^V_{-\lambda_2,-\lambda_W} = -H^V_{\lambda_2,\lambda_W}, \qquad \qquad H^A_{-\lambda_2,-\lambda_W} = H^A_{\lambda_2,\lambda_W}
\eeq

The interesting observables of Appendix E are given in Figs. 17-23.\par
For the electron case, the FB asymmetry is given by (\ref{C-15e}) and again, although the final state parity has changed, from the $V-A$ structure of the theory, the left-handed final baryon dominates, and we expect to have the inequality (\ref{C-16e}) and $A^e_{FB}(q^2)$ to be positive for all values of $q^2$, as we indeed observe in Fig. 18.\par

It is interesting to see how the inequality (\ref{C-16e}) holds in our model of the form factors described in Appendix A.2.  Similarly to what we have done above for the ground state, keeping only the heavy quark limit terms, one finds indeed
\beq
\label{C-25e}
{\cal H}_P =\ \mid H_{+{1\over2}+1} \mid^2 - \mid H_{-{1\over2}-1} \mid^2 = - {16 \over 3} m_b m_c (w^2-1)^{3/2} \left[\sigma_\Lambda(w)\right]^2 < 0
\eeq

In the presence of a non-vanishing lepton mass $m_\ell$, the inequality (\ref{C-25e}) does not follow for all values of $q^2$.\par
\begin{figure}[htbp]
\centerline{\includegraphics[height=\grapheheight]{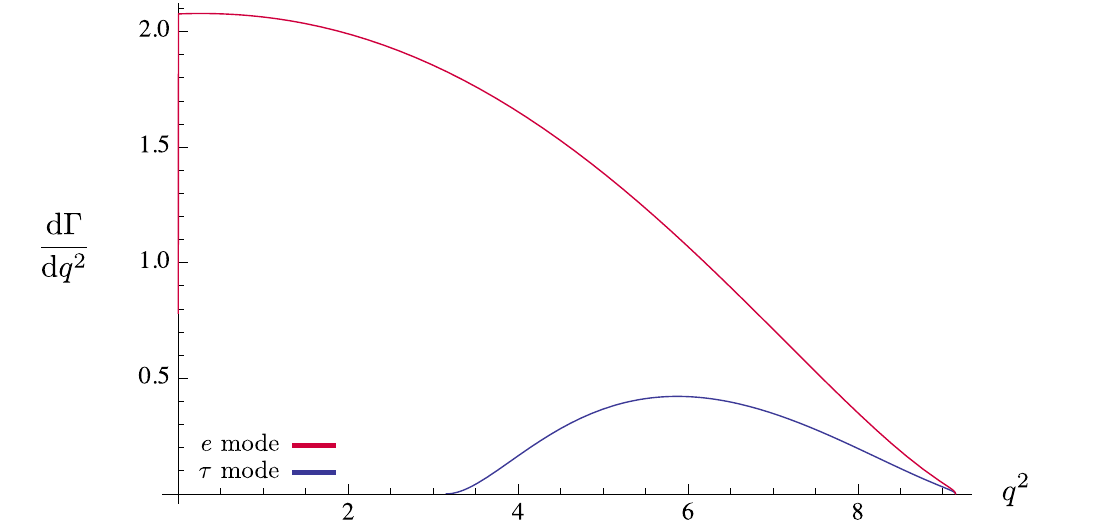}}
\caption{${d\Gamma \over dq^2}$ for the electron and tau modes. In the electron case, one has ${d\Gamma \over dq^2} \to 0$ for $q^2 \to 0$.}
\end{figure}
%
%
%
%
%
\begin{figure}[htbp]
\centerline{\includegraphics[height=\grapheheight]{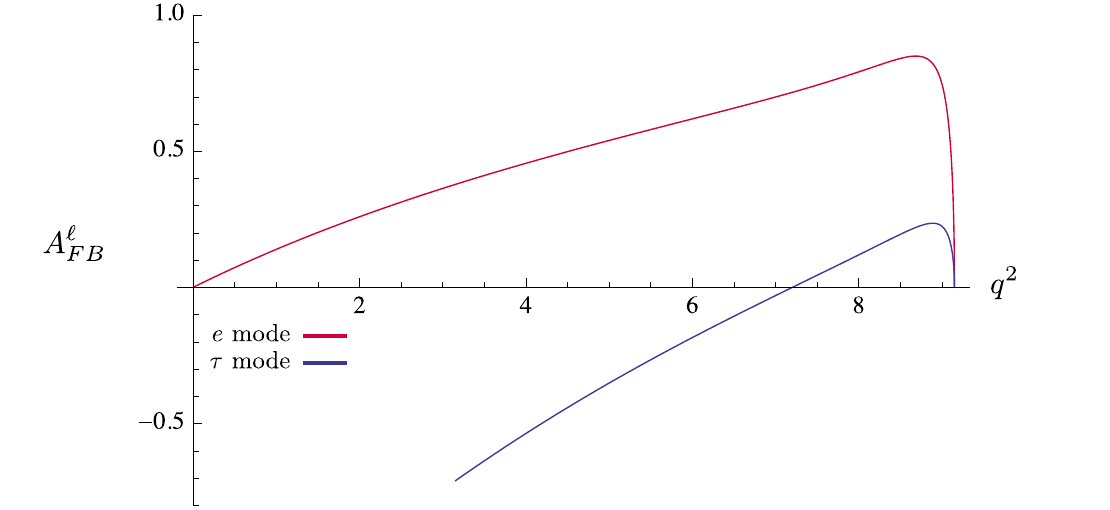}}
\caption{Forward-backward asymmetry $A^\ell_{FB}(q^2)$ for the electron and tau modes.}
\end{figure}
%
%
%
%
\begin{figure}[htbp]
\centerline{\includegraphics[height=\grapheheight]{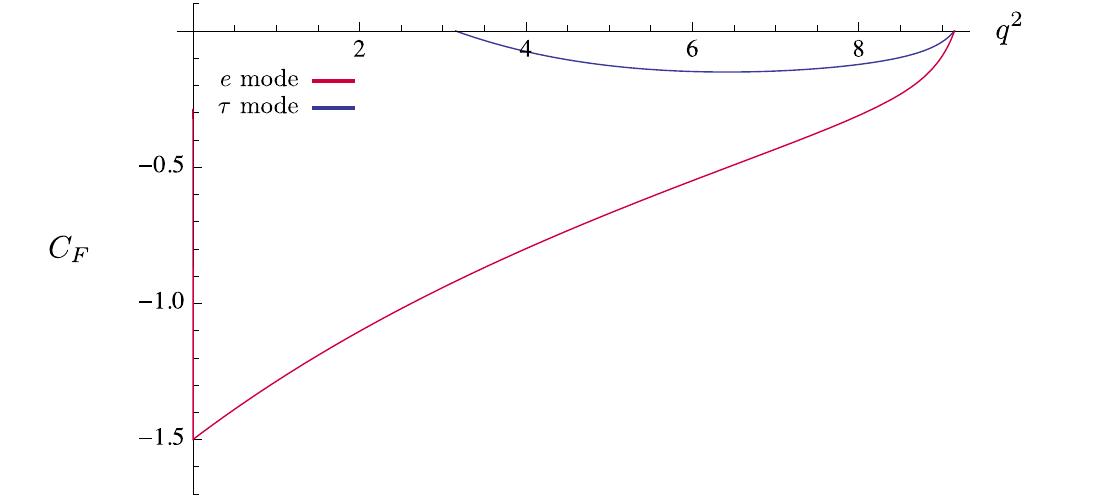}}
\caption{Convexity parameter $C_F(q^2)$ for the electron and tau modes.}
\end{figure}
%
%
%
\begin{figure}[htbp]
\centerline{\includegraphics[height=\grapheheight]{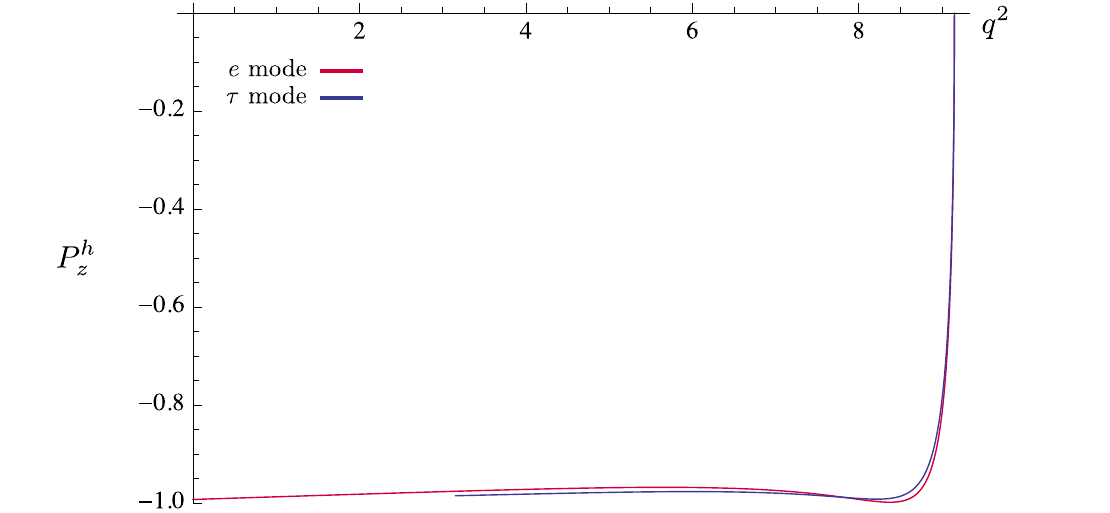}}
\caption{Longitudinal hadron polarization $P_z^h(q^2)$ for the electron and tau modes.}
\end{figure}
%
%
%
%
\begin{figure}[htbp]
\centerline{\includegraphics[height=\grapheheight]{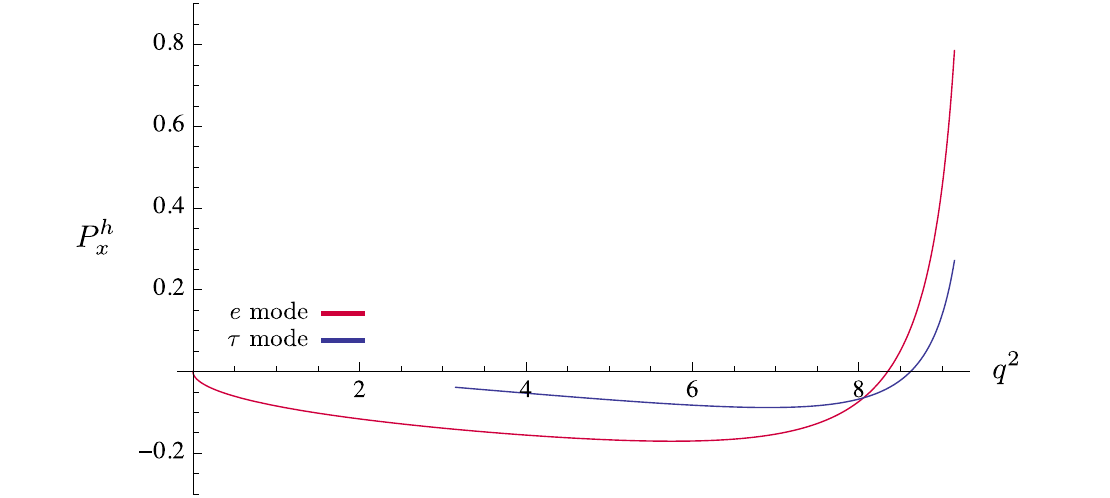}}
\caption{Transverse hadron polarization $P_x^h(q^2)$ for the electron and tau modes.}
\end{figure}
%
%
%
%
\begin{figure}[htbp]
\centerline{\includegraphics[height=\grapheheight]{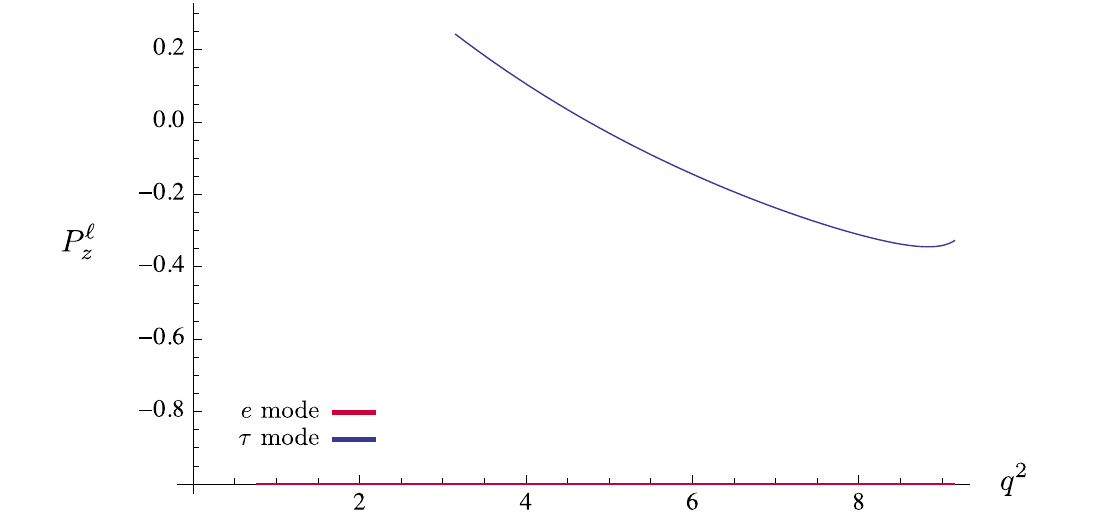}}
\caption{Longitudinal lepton polarization $P_z^\ell(q^2)$ for the electron and tau modes. $P_z^e(q^2)$ is very close to $-1$, unlike $P_z^\tau(q^2)$.}
\end{figure}
%
%
%
%
\begin{figure}[htbp]
\centerline{\includegraphics[height=\grapheheight]{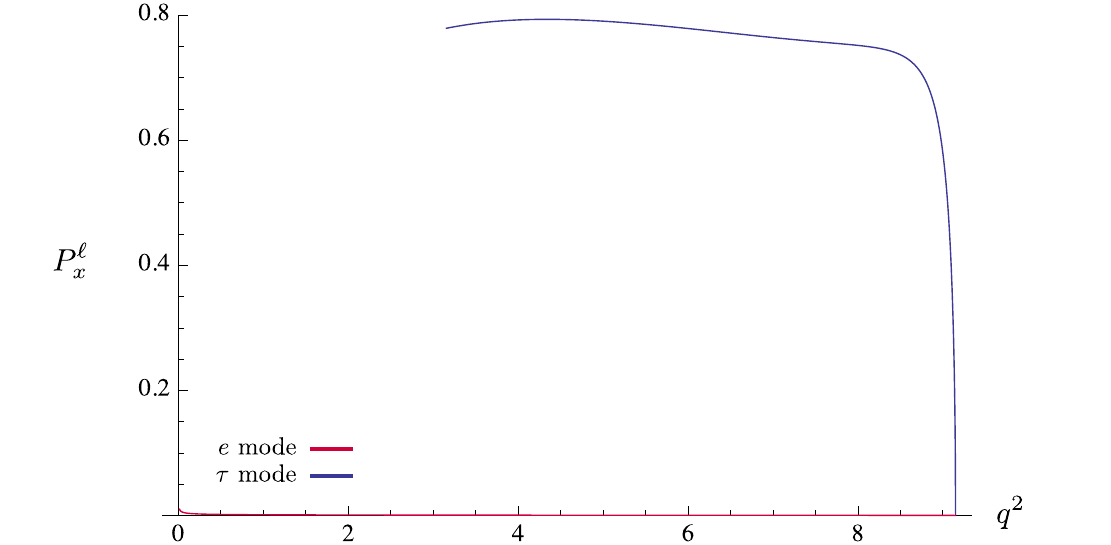}}
\caption{Transverse lepton polarization $P_x^\ell(q^2)$ for the electron and tau modes. $P_x^e(q^2)$ is very small, unlike $P_x^\tau(q^2)$.}
\end{figure}
%
%
%
%
%

For the $\tau$ case, one has a zero in the $FB$ asymmetry. Keeping only the heavy quark limit terms one finds the same value for the position of this zero as in the elastic case (\ref{C-22e}),
\beq
\label{C-28e}
q^2_0(A_{FB}^\tau) = m_\tau \sqrt{m_b^2-m_c^2}
\eeq

Also, we observe in Fig.~22~that, in the $\tau$ case, the longitudinal lepton polarization $P_z^\tau (q^2)$ has a zero in the neighborhood of $q^2 \simeq 4.\ \rm{GeV} ^2$. However, in this case, unlike the ground state, we have not found a simple analytic expression for the position of this zero because the expansion in powers of ${m_\tau^2 \over m_b^2}$ and ${m_c^2 \over m_b^2}$ converges slowly.

\subsection{$\boldsymbol{\tau/\ell}$ observables sensitive to LFUV}

We now compute the relevant ratio of rates to test LFU (Lepton Flavor Universality), 
\beq
R_{\tau/\ell} \left({1 \over 2}^\pm \right) = {\Gamma (\Lambda_b \to \Lambda_c \left({1 \over 2}^\pm \right) \tau \nu) \over \Gamma (\Lambda_b \to \Lambda_c \left({1 \over 2}^\pm \right) e \nu)}
\eeq
  
\noi and we find, for the ground state,
\beq
R_{\tau/\ell} \left({1 \over 2}^+ \right) \simeq 0.317
\eeq

\noi and for the transition to the excited state, 
\beq
R_{\tau/\ell} \left({1 \over 2}^-\right) \simeq 0.141
\eeq

An interesting observable is the forward-backward asymmetry $A^\ell_{FB}$, that has a very different behaviour for the light leptons and for the $\tau$. In this latter case $A^\tau_{FB}$, unlike $A^e_{FB}$, presents a zero at $q^2_0(A^\tau_{FB}) \simeq m_\tau \sqrt{m_b^2-m_c^2}$, for both ${1 \over 2}^+$ and ${1 \over 2}^-$ quantum numbers (Figs. 11, 18). It would be very interesting to have a measurement of the position of the zero, that could be a test of the SM.\par
For both cases ${1 \over 2}^+$ and ${1 \over 2}^-$ (Figs. 13, 20), in the region of common phase space, the longitudinal hadron polarization is very similar for light leptons and for the $\tau$. This could be also an interesting test of the SM.\par
The transverse hadron polarization is very different for ${1 \over 2}^+$ and ${1 \over 2}^-$ (Figs. 14, 21). This shows clearly that this observable strongly depends on the internal wave function, as it is quite different for ${1 \over 2}^+$ and ${1 \over 2}^-$ states, that have very different wave functions.\par
For both cases ${1 \over 2}^\pm$ (Figs. 15, 22), the longitudinal electron polarization is very close to -1, while it has a very different behaviour for the $\tau$, that presents a zero at $q^2 \simeq 4\ \rm{GeV}^2$. These features could also provide interesting tests of the SM.\par
Also for both ${1 \over 2}^\pm$ (Figs. 16, 23) the transverse electron polarization is very small while for the $\tau$ it is positive and sizeable for most of the phase space.

\subsection{Comparison with the work of Gutsche \textbf{\textit{et al.}} }

Our calculation of the observables relies on the helicity formalism of the Mainz group paper by Gutsche {\itshape et al.} \cite{MAINZ-LAMBDAb-LAMBDAc-TAU, MAINZ OBSERVABLES}, where the calculations for the ground state transitions $\Lambda_b \to \Lambda_c\left({1 \over 2}^\pm \right) \ell \overline{\nu}$ for $\ell = e, \tau$ were done in their Covariant Confined Quark Model (CCQM). This work was recently extended to the $\Lambda_b \to \Lambda_c\left({1 \over 2}^-, {3 \over 2}^- \right) \ell \overline{\nu}$ transitions for $\ell = e, \tau$ \cite{MAINZ LAMBDAb LAMBDAcSTAR}.\par
We would like to compare some aspects of their approach with our own.\par
First, in the CCQM, baryons are composites of $Qqq$, with the quantum numbers of the different states given by local interpolating fields with the correct quantum numbers, a compositeness condition, and a simple universal gaussian form for the vertex functions. Moreover, the calculation is done at finite mass \cite{MAINZ LAMBDAb LAMBDASTAR}.\par
Our approach is a naive quark model in a Q-diquark scheme with a harmonic-oscillator potential, reproducing qualitatively the spectrum and giving within the BT scheme a reasonable slope for the IW function, consistent with the lattice data. The quantum numbers for the $L = 0$ and the $L = 1$ states are then related by the Schr$\ddot {\rm o}$dinger equation giving the wave functions for both states, the main parameters being the HO radius and the masses.\par
It is encouraging that for the ground state transitions $\Lambda_b \to \Lambda_c\left({1 \over 2}^+ \right) \ell \overline{\nu}\ (\ell = e, \tau)$ we find plots that are close to the ones of ref. \cite{MAINZ-LAMBDAb-LAMBDAc-TAU}, in particular the position of the zero in the forward-backward asymmetry for the $\tau$ case. It would be very interesting to find the position of this zero in a theoretical scheme as model-independent as possible, to put it on solid grounds as a test of the Standard Model.\par
On the other hand, our numbers obtained for the ratios $R_{\tau/\ell} \left({1 \over 2}^+ \right) \simeq 0.317$ and $R_{\tau/\ell} \left({1 \over 2}^-\right) \simeq 0.141$ are very close to the predictions of ref. \cite{MAINZ LAMBDAb LAMBDAcSTAR}, so that they seem to be on a firm ground. 

\section{Conclusions}

Our objective has been the calculation of the observables in the decays $\Lambda_b \to \Lambda_c^\pm \ell \overline{\nu}$, that could provide tests of Lepton Flavor Universality Violation.\par 
We have done this in a quark model that, unlike present Lattice QCD calculations, allows not only the computation of transitions within the ground state, but also those to the $L = 1$ excitations.\par
The BT method is very suited for such a calculation, the wave functions are three-dimensional but the result is covariant in the heavy quark limit.\par
The BT approach can explain the value of the slope of the baryon IW function $\rho_\Lambda^2$, very different of the non-relativistic value, as it happens for mesons.\par
The slope $\rho_\Lambda^2$ is a very important parameter to describe the form factors, and it is at the same time a discriminant of the different models. The adopted dipolar fit for the IW function satisfies a number of theorems that the different derivatives of the IW function must satisfy.\par
To test the different models, we have analysed the lattice data on the form factors, and we have shown that the slope of the IW function is of the order $\rho_\Lambda^2 \simeq 2$. On the other hand, we have demonstrated that the slope of the subleading form factor $A'(1)$ can be determined independently of $\rho_\Lambda^2$.\par
We have found a number of difficulties of standard QCD-inspired Hamiltonian approaches, both in the three quark $Qqq$ model and also in quark pointlike-diquark models. We have thoroughly discussed these problems.\par
In view of these difficulties, we have adopted a preliminary quark pointlike-diquark model that allows a qualitative description of the spectrum, and of the IW slope $\rho_\Lambda^2 \simeq 2$, in agreement with lattice data.\par
We have computed the different observables proposed by the Mainz group for both transitions $\Lambda_b \to \Lambda_c^\pm \ell \overline{\nu}$, emphasizing the differences between $e$ and $\tau$ transitions.\par
Using Bjorken sum rule, we find in our model that {\it the lowest inelastic IW function $\sigma_\Lambda(w)$, that describe the  $(L = 0) \to (L = 1)$ transitions, is large and thus there is a good prospect for the decays $\Lambda_b \to \Lambda_c\left({1 \over 2}^-, {3 \over 2}^- \right) \ell \overline{\nu}$ to be studied in detail at LHCb}. Both decays depend on $\sigma_\Lambda(w)$ because the states ${1 \over 2}^-, {3 \over 2}^-$ belong to the same doublet in the heavy quark limit.\par
We have seen that some observables, the forward-backward asymmetries and the longitudinal lepton polarization, present a zero at some characteristic value of $q^2$ for the $\tau$ case. The positions of these zeros could provide tests of physics beyond the Standard Model. In particular, the forward-backward asymmetry for both $\Lambda_b \to \Lambda_c \left({1 \over 2}^\pm \right) \tau \nu$ cases presents a zero for $q^2 \simeq m_\tau \sqrt {m_b^2-m_c^2}$. \par
We do not study for the moment the case of the inelastic transitions $\Lambda_b \to \Lambda_c\left({3 \over 2}^{-}\right) \ell \overline{\nu} \ (\ell = e, \tau)$. In the quark model one has to consider a $S = 1$ diquark coupled to $L = 1$. We postpone this study since we would like to analyze and compute the different observables that could be interesting for our purpose, besides the one computed in ref. \cite{MAINZ LAMBDAb LAMBDAcSTAR}.\par
As a word of caution, we have to say that our results for the observables are preliminary, as we will need in the future to treat systematically the three quark system $Qqq$ to study the spectrum and the IW function, and then turn to phenomenological applications.

\vskip 10 truemm
\appendix
\renewcommand{\thesubsubsection}{\Alph{section}.\arabic{subsection}.\alph{subsubsection}}
\section*{Appendix A}
\setcounter{section}{1}
%
%
%
%
%
\subsection{Baryon form factors $\boldsymbol{\Lambda_b \to \Lambda_c \left({1 \over 2}^+\right)}$ up to order $\boldsymbol{1 / m_Q}$}

From Falk and Neubert \cite{FALK-NEUBERT}, the 6 conventional form factors $f_i, g_i\ (i = 1, 2, 3)$ write
%
$$< \Lambda_c(p',s') \mid \overline{c} \gamma^\mu b \mid \Lambda_b(p,s) >\ = \overline{u}_{\Lambda_c}(p',s') \left[f_1 \gamma^\mu - i f_2 \sigma^{\mu \nu} q - \nu + f_3 q^\mu \right] u_{\Lambda_b}(p,s)$$
\beq
\label{B-1e}
< \Lambda_c(p',s') \mid \overline{c} \gamma^\mu \gamma_5 b \mid \Lambda_b(p,s) >\ = \overline{u}_{\Lambda_c}(p',s') \left[g_1 \gamma^\mu - i g_2 \sigma^{\mu \nu} q - \nu + g_3 q^\mu \right] \gamma_5 u_{\Lambda_b}(p,s)
\eeq
%
%
\noi The form factors $f_3, g_3$ will contribute to $\Lambda_b \to \Lambda_c \tau \nu$.\par
The alternative notation, convenient for HQET, is given in terms of the four-velocities
$$< \Lambda_c (v',s') \mid \overline{c} \gamma^\mu b \mid \Lambda_b (v,s) >\ = \overline{u}_{\Lambda_c}(p',s') \left[ F_1 \gamma^\mu + F_2 v^\mu + F_3 v^{' \mu} \right] u_{\Lambda_b}(v,s)$$
\beq
\label{B-2e}
< \Lambda_c (v',s') \mid \overline{c} \gamma^\mu b \gamma_5 \mid \Lambda_b (v,s) >\ = \overline{u}_{\Lambda_c}(v',s') \left[ G_1 \gamma^\mu + G_2 v^\mu + G_3 v^{' \mu} \right]\gamma_5 u_{\Lambda_b}(v,s)
\eeq
%
%
The form factors $f_i, g_i$ write, in terms of the $F_i, G_i$ :
$$f_1 = F_1 + (m_{\Lambda_b}+m_{\Lambda_c}) \left({F_2 \over {2m_{\Lambda_b}}} + {F_3 \over {2m_{\Lambda_c}}} \right)$$
$$f_2 = - {F_2 \over {2m_{\Lambda_b}}} - {F_3 \over {2m_{\Lambda_c}}}\ , \qquad \qquad f_3 = {F_2 \over {2m_{\Lambda_b}}} - {F_3 \over {2m_{\Lambda_c}}}$$
$$g_1 = G_1 - (m_{\Lambda_b}-m_{\Lambda_c}) \left({G_2 \over {2m_{\Lambda_b}}} + {G_3 \over {2m_{\Lambda_c}}} \right)$$
\beq
\label{B-3e}
g_2 = - {G_2 \over {2m_{\Lambda_b}}} - {G_3 \over {2m_{\Lambda_c}}}\ , \qquad \qquad g_3 = {G_2 \over {2m_{\Lambda_b}}} - {G_3 \over {2m_{\Lambda_c}}} 
\eeq

\vskip 3 truemm

In terms of HQET form factors up to order ${1 \over m_Q}$ one has
$$F_1(w) = \xi_\Lambda(w) + \left( {1 \over {2m_b}} +  {1 \over {2m_c}} \right) \left[B_1(w) - B_2(w)\right]$$
$$G_1(w) = \xi_\Lambda(w) + \left( {1 \over {2m_b}} +  {1 \over {2m_c}} \right) B_1(w)$$
$$F_2(w) = G_2(w) = {1 \over {2m_c}}\ B_2(w)$$
\beq
\label{B-4e}
F_3(w) = - G_3(w) = {1 \over {2m_b}}\ B_2(w)
\eeq

\noi where the ${1 \over m_Q}$ corrections read
\beq
\label{B-5e}
B_1(w) = \overline{\Lambda}\ {w-1 \over w+1}\ \xi_\Lambda(w) + A(w)\ , \qquad B_2(w) = - \overline{\Lambda}\ {2 \over w+1}\ \xi_\Lambda(w) 
\eeq

The terms proportional to $\overline{\Lambda}\xi_\Lambda(w)$ correspond to the first order Current perturbation in HQET, while the form factor $A(w)$ corresponds to the Lagrangian insertion perturbation
\beq
\label{B-6e}
< \Lambda_c(p',s') \mid i \int dx T\{J(0), L_1(x) \} \mid \Lambda_b(p,s) >\ = A(w)\ \overline{u}_{\Lambda_c} \Gamma u_{\Lambda_b}
\eeq

Luke's theorem \cite{LUKE-THEOREM} implies at zero recoil
\beq
\label{B-7e}
B_1(1) = A(1) = 0
\eeq

An interesting feature of formulas (\ref{B-3e},\ref{B-4e}) is that the form factors $f_2,f_3$ and $g_2,g_3$ are of order $1 / m^2_Q$. This fact can have consequences for the comparison between $\Lambda_b \to \Lambda_c \tau \nu$ and $\Lambda_b \to \Lambda_c \ell \nu$.

\vskip 4 truemm

Let us finally give the notation for the form factors used in the Lattice calculations \cite{DETMOLD ET AL.}.
In terms of the form factors (\ref{B-1e},\ref{B-2e}) the defintion used in \cite{DETMOLD ET AL.} is the following. \par
\noi For the vector form factors : 
$$f_+ = f_1 + {q^2 \over m_{\Lambda_b}+m_{\Lambda_c}} f_2$$
\beq
\label{B-10e}
f_\perp = f_1 + (m_{\Lambda_b}+ m_{\Lambda_c}) f_2
\eeq
$$f_0 = f_1 + {q^2 \over m_{\Lambda_b}-m_{\Lambda_c}} f_3$$

\noi and for the axial form factors :

$$g_+ = g_1 - {q^2 \over m_{\Lambda_b}-m_{\Lambda_c}} g_2$$
\beq
\label{B-11e}
g_\perp = g_1 - (m_{\Lambda_b}-m_{\Lambda_c}) g_2
\eeq
$$g_0 = g_1 - {q^2 \over {m_{\Lambda_b}+ m_{\Lambda_c}}} g_3$$
%
%
\subsection{Baryon form factors $\boldsymbol{\Lambda_b \to \Lambda_c \left({1 \over 2}^-\right)}$ up to order $\boldsymbol{1 / m_Q}$}

The matrix elements read, 
$$< \Lambda_c \mid \overline{c} \gamma_\mu b \mid \Lambda_b >\ = \overline{u}(p_2,s_2)\left[ \gamma_\mu g_1^V(q^2) - i\sigma_{\mu \nu} q^\nu g_2^V(q^2) + q_\mu g_3^V(q^2) \right]\gamma_5 u(p_1,s_1)$$
\beq
\label{A.2-1bise}
< \Lambda_c \mid \overline{c} \gamma_\mu \gamma_5 b \mid \Lambda_b >\ = \overline{u}(p_2,s_2)\left[ \gamma_\mu g_1^A(q^2) - i\sigma_{\mu \nu} q^\nu g_2^A(q^2) + q_\mu g_3^A(q^2) \right] u(p_1,s_1)
\eeq

\vskip 3 truemm

\noi Notice the presence (absence) of $\gamma_5$ in the $V (A)$ matrix elements for $\Lambda_b \to \Lambda_c \left( {1 \over 2}^- \right)$ due to the intrinsic negative parity of the final state.

\vskip 3 truemm

The alternative notation in terms of the four-velocities is given by \cite{LEIBOVICH-STEWART}
$$< \Lambda_c^{1/2} (v',s') \mid \overline{c} \gamma^\mu b \mid \Lambda_b (v,s) >\ = \overline{u}_{\Lambda_c}(v',s') \left[d_{V_1} \gamma^\mu + d_{V_2} v^\mu + d_{V_3} {v'}^\mu \right] \gamma_5 u_{\Lambda_b}(v,s)$$
\beq
\label{A.2-1e}
< \Lambda_c (v',s') \mid \overline{c} \gamma^\mu b \gamma_5 \mid \Lambda_b (v,s) >\ = \overline{u}_{\Lambda_c}(v',s') \left[ d_{A_1} \gamma^\mu + d_{A_2} v^\mu + d_{A_3} {v'}^\mu \right] u_{\Lambda_b}(v,s)
\eeq

\vskip 3 truemm

\noi and the relation between both notations is

$$g^V_1 = d_1^V - (m_{\Lambda_b}-m_{\Lambda_c})\ \left( {d_2^V  \over {2 m_{\Lambda_b}}} + {d_3^V  \over {2 m_{\Lambda_c}}} \right)$$ 
$$g^V_2 = - {d_2^V  \over {2 m_{\Lambda_b}}} - {d_3^V  \over {2 m_{\Lambda_c}}}\ , \qquad \qquad g^V_3 = {d_2^V  \over {2 m_{\Lambda_b}}} - {d_3^V  \over {2 m_{\Lambda_c}}}$$
\beq
\label{A.2-2e}
g^A_1 = d_1^A + (m_{\Lambda_b}+m_{\Lambda_c})\ \left( {d_2^A  \over {2 m_{\Lambda_b}}} + {d_3^A  \over {2 m_{\Lambda_c}}} \right)
\eeq 
$$g^A_2 = - {d_2^A  \over {2 m_{\Lambda_b}}} - {d_3^A  \over {2 m_{\Lambda_c}}}\ , \qquad \qquad 
g^A_3 = {d_2^A  \over {2 m_{\Lambda_b}}} - {d_3^A  \over {2 m_{\Lambda_c}}}$$

\vskip 6 truemm

Neglecting for the moment the subleading terms dependent on Lagrangian insertions and keeping only the subleading $1/m_Q$ terms that are proportional to the inelastic IW function $\sigma(w)$, the form factors (\ref{A.2-1e}) are given by the expressions of Leibovich and Stewart \cite{LEIBOVICH-STEWART},

$$d_1^V = {1 \over \sqrt{3}} \left[ (w-1)\sigma + \epsilon_c 3(w \overline{\Lambda '}-\overline{\Lambda}) \sigma - \epsilon_b (\overline{\Lambda '}-w\overline{\Lambda}) \sigma \right] $$
$$d_2^V = {1 \over \sqrt{3}} \left[ -2 \sigma - \epsilon_b (\overline{\Lambda '}+\overline{\Lambda}) \sigma \right]\ , \qquad \qquad 
d_3^V = {1 \over \sqrt{3}} \left[\epsilon_b (\overline{\Lambda '}+\overline{\Lambda}) \sigma \right]$$
\beq
\label{A.2-3e}
d_1^A = {1 \over \sqrt{3}} \left[ (w+1)\sigma + \epsilon_c 3(w \overline{\Lambda '}-\overline{\Lambda}) \sigma - \epsilon_b (\overline{\Lambda '}-w\overline{\Lambda}) \sigma \right]
\eeq
$$d_2^A = {1 \over \sqrt{3}} \left[ -2 \sigma + 2 \epsilon_b (\overline{\Lambda '}-\overline{\Lambda}) \sigma \right]\ , \qquad \qquad d_3^A = {1 \over \sqrt{3}} \left[ 2 \epsilon_b (\overline{\Lambda '}-\overline{\Lambda}) \sigma \right] $$

\vskip 6 truemm

The subleading Lagrangian perturbations give the following extra contributions to the preceding form factors \cite{LEIBOVICH-STEWART},
$$\Delta d_1^V = {1 \over \sqrt{3}} \left\{ \epsilon_c \left[-2(w^2-1) \sigma_1 + (w-1)(\phi_{kin}^{(c)}-2\phi_{mag}^{(c)}) \right] - \epsilon_b \left[- (w-1) \phi_{kin}^{(b)}\right] \right\} $$
$$\Delta d_2^V = {1 \over \sqrt{3}} \left\{ 2 \epsilon_c \left[-(\phi_{kin}^{(c)}-2\phi_{mag}^{(c)}) \right] - 2 \epsilon_b \left[- (w+1) \sigma_1 + \phi_{kin}^{(b)} + \phi_{mag}^{(b)}\right] \right\} $$
\beq
\label{A.2-4e}
\Delta d_3^V = {1 \over \sqrt{3}}\ 2\epsilon_b \left[- (w+1) \sigma_1 - \phi_{mag}^{(b)}\right]
\eeq
$$\Delta d_1^A = {1 \over \sqrt{3}} \left\{ \epsilon_c \left[-2(w^2-1) \sigma_1 + (w+1)(\phi_{kin}^{(c)}-2\phi_{mag}^{(c)}) \right] - \epsilon_b \left[- (w+1) \phi_{kin}^{(b)}\right] \right\} $$
$$\Delta d_2^A = {1 \over \sqrt{3}} \left\{ - 2\epsilon_c \left[\phi_{kin}^{(c)}-2\phi_{mag}^{(c)} \right] + 2 \epsilon_b \left[ -\phi_{kin}^{(b)}+\phi_{mag}^{(b)}\right] \right\} $$
$$\Delta d_3^A = {1 \over \sqrt{3}}\ 2 \epsilon_b \left[-(w-1) \sigma_1 - \phi_{mag}^{(b)} \right] $$

\vskip 3 truemm

According to Leibovich and Stewart, the chromomagnetic functions $\phi_{mag}^{(Q)}$ are expected to be small because the $j^P = 1^-$ doublet mass splittings are small, and they are taken 
\beq
\label{A.2-5e}
\phi_{mag}^{(Q)} = 0\ \qquad \qquad (Q = c, b)
\eeq
The functions $\phi_{kin}^{(Q)}$ can be absorbed by the Isgur-Wise function by replacing $\sigma$ with
\beq
\label{A.2-6e}
{\tilde \sigma}(w) = \sigma(w) + \epsilon_c \phi_{kin}^{(c)}(w) + \epsilon_b \phi_{kin}^{(b)}(w)
\eeq

\noi Moreover \cite{LEIBOVICH-STEWART} assume
\beq
\label{A.2-7e}
\phi_{kin}^{(c)} (1) = 0
\eeq

\noi as predicted by QCD in the large $N_c$ limit, and therefore
\beq
\label{A.2-8e}
{\tilde \sigma}(1) \simeq \sigma(1)
\eeq

\noi One is left then with two IW functions, $\tilde \sigma (w)$ and $\sigma_1(w)$ and defining the ratio 
\beq
\label{A.2-9e}
{\hat \sigma}_1(w) = {\sigma_1(w) \over {\tilde \sigma} (w)}
\eeq

\noi Leibovich and Stewart assume a constant ratio for ${\hat \sigma}_1(w) = {\rm constant} = {\hat \sigma}_1$ within the range
\beq
\label{A.2-10e}
-1\ \rm{GeV} <  {\hat \sigma}_1 < 1\ \rm{GeV}
\eeq    
\vskip 6 truemm
%
\section*{Appendix B}
%
%
%
\subsection*{The $\boldsymbol{Qqq}$ elastic IW function $\boldsymbol{\xi_{\Lambda}(w)}$ in the BT scheme}

Let us begin with the general formula for a transition matrix element in the Bakamjian-Thomas relativistic quark model in terms of $2 \times 2$ matrices \cite{COVARIANT-QM} :

\beq
\label{1-1e}
< {\bf P}' \mid O \mid {\bf P} >\ = \int \prod_{i=2}^{n} {d {\bf p}_i \over (2 \pi)^3}\ \sqrt{{\sum_j p_j^{'0} \sum_k p_k^{0} \over M'_0 M_0}}\ \prod_{i=1}^{n} \sqrt{k_i^{'0}k_i^{0} \over p_i^{'0}p_i^{0}}\ \sum_{s'_1 ... s'_n} \sum_{s_1 ... s_n}
\eeq
$$ \varphi'_{s'_1 ... s'_n}({\bf k}'_2 ... {\bf k}'_n)^*\ \left[D'_1({\bf R}^{'-1}_1) O({\bf p}'_1,{\bf p}_1) D_1({\bf R}_1\right]_{s'_1,s_1}\ \prod_{i=2}^n D_i({\bf R}_i^{'-1}{\bf R}_i)_{s'_i ... s_i}\ \varphi_{s_1 ... s_n}({\bf k}_2 ... {\bf k}_n)$$

\noi where 1 labels the active quark, the matrix element of the currrent operator $O$ is :
\beq
\label{1-2e}
O({\bf p}',{\bf p})_{s',s} =\ < {\bf p}',s' \mid O \mid {\bf p},s >
\eeq

\noi and the vectors ${\bf k}_i$, the 0-components $k_i^0$ and $p_i^0$, $M_0$ and the Wigner rotations ${\bf R}_i$ are functions of the ${\bf p}_i$ defined as follows :
\beq
\label{1-3e}
p_i^0 = \sqrt{{\bf p}_i^2+m_i^2}\ , \qquad M_0 = \sqrt{(\Sigma p_j)^2}\ , \qquad k_i = {\bf B}^{-1}_{\Sigma p_j} p_i\ , \qquad {\bf R}_i = {\bf B}^{-1}_{p_i}  {\bf B}_{\Sigma p_j} {\bf B}_{k_i}
\eeq

\noi where ${\bf B}_p$ is the boost $(\sqrt{p^2},{\bf 0}) \to p$ and $D_i({\bf R})$ is the matrix of rotation ${\bf R}$ for the spin $S_i$.

The internal wave function of the baryon $\Lambda_Q$ with heavy quarks $Q = b$ or $c$ and polarization $\mu$ will write
\beq
\label{1-4e}
\varphi^{(\mu)}_{s_1,s_2,s_3}({\bf k}_2, {\bf k}_3) = \chi^{(\mu)}_{s_1}\ {i \over \sqrt{2}}\ (\sigma_2)_{s_2,s_3}\ \varphi({\bf k}_2, {\bf k}_3)
\eeq

\noi because the spectator quarks 2, 3 are in a state of spin and isospin 0, and the notation $\chi^{(\mu)}_{s_1}$ for the active quark means $\chi^{(+1/2)}_{+1/2} = \chi^{(-1/2)}_{-1/2} = 1$ and $\chi^{(+1/2)}_{-1/2} = \chi^{(-1/2)}_{+1/2} = 0$.

Considering the polarized states $\Lambda_Q$, the matrix element (\ref{1-1e}) writes then
\beq
\label{1-5e}
< {\bf P}', \mu' \mid O \mid {\bf P}, \mu >\ = \int {d {\bf p}_2 \over (2 \pi)^3}\ {d {\bf p}_3 \over (2 \pi)^3}\ \sqrt{{\sum_j p_j^{'0} \sum_k p_k^{0} \over M'_0 M_0}}\ \prod_{i=1}^{3} \sqrt{k_i^{'0}k_i^{0} \over p_i^{'0}p_i^{0}}\ \sum_{s'_1, s'_2, s'_3} \sum_{s_1, s_2, s_3}
\eeq
$$\varphi^{(\mu')}_{s'_1,s'_2,s'_3}({\bf k}'_2, {\bf k}'_3)^*\ \left[D'_1({\bf R}^{'-1}_1) O({\bf p}'_1,{\bf p}_1) D_1({\bf R}_1)\right]_{s'_1,s_1}\ \prod_{i=2}^3 D_i({\bf R}_i^{'-1}{\bf R}_i)_{s'_i ... s_i}\ \varphi^{(\mu)}_{s_1, s_1, s_3}({\bf k}_2, {\bf k}_3)$$

\noi From the wave function (\ref{1-4e}) one gets
$$< {\bf P}', \mu' \mid O \mid {\bf P}, \mu >\ =  {1 \over 2} \int {d {\bf p}_2 \over (2 \pi)^3}\ {d {\bf p}_3 \over (2 \pi)^3}\ \sqrt{{\sum_j p_j^{'0} \sum_k p_k^{0} \over M'_0 M_0}}\ \prod_{i=1}^{3} \sqrt{k_i^{'0}k_i^{0} \over p_i^{'0}p_i^{0}}\ \varphi({\bf k}'_2, {\bf k}'_3)^* \varphi({\bf k}_2, {\bf k}_3)$$
\beq
\label{1-6e}
\times\ \left(\chi^{(\mu')\dagger}D'_1({\bf R}^{'-1}_1) O({\bf p}'_1,{\bf p}_1) D_1({\bf R}_1)\chi^{(\mu)}\right)\ Tr\left[D_2({\bf R}_2^{'-1}{\bf R}_2)^t\sigma_2 D_3({\bf R}_3^{'-1}{\bf R}_3) \sigma_2 \right] 
\eeq

\noi and using the relation 
\beq
\label{1-6bise}
\sigma_2 D({\bf R}) \sigma_2 = D({\bf R}^{-1})^t
\eeq

\noi one obtains
$$< {\bf P}', \mu' \mid O \mid {\bf P}, \mu >\ =  {1 \over 2} \int {d {\bf p}_2 \over (2 \pi)^3}\ {d {\bf p}_3 \over (2 \pi)^3}\ \sqrt{{\sum_j p_j^{'0} \sum_k p_k^{0} \over M'_0 M_0}}\ \prod_{i=1}^{3} \sqrt{k_i^{'0}k_i^{0} \over p_i^{'0}p_i^{0}}\ \varphi({\bf k}'_2, {\bf k}'_3)^*\ \varphi({\bf k}_2, {\bf k}_3)$$
\beq
\label{1-7e}
\times\ \left(\chi^{(\mu')\dagger}D({\bf R}^{'-1}_1) O({\bf p}'_1,{\bf p}_1) D({\bf R}_1)\chi^{(\mu)}\right)\ Tr\left[D({\bf R}_3^{-1}{\bf R}'_3{\bf R}_2^{'-1}{\bf R}_2) \right] 
\eeq

\noi where we have omitted the index indicating on which quark the Wigner rotation acts, and keep it only on the rotation, because all these matrices act on the spin ${1 \over 2}$. Equation (\ref{1-6e}) is the final formula in the $2 \times 2$ matrix formalism and at finite mass.\par
We now pass to a $4 \times 4$ matrix formulation :
\beq
O({\bf p}'_1,{\bf p}_1) \to \sqrt{{m'_1m_1 \over p^{'0}_1 p^{0}_1}}\ {{1+\gamma^0} \over 2}\ {\bf B}^{-1}_{p'_1} O {\bf B}^{-1}_{p_1}\ {{1+\gamma^0} \over 2}\ , \qquad \ \ \ \ \chi^{(\mu)} \to {{1+\gamma^0} \over 2} \chi^{(\mu)}
\eeq
\noi and will have, for the spinor matrix element :
$$\left(\chi^{(\mu')\dagger}D'_1({\bf R}^{'-1}_1) O({\bf p}'_1,{\bf p}_1) D_1({\bf R}_1)\chi^{(\mu)}\right)$$
\beq
\label{1-8e}
= {1 \over 4}\ \sqrt{{m'_1m_1 \over p^{'0}_1 p^{0}_1}}\ \left(\chi^{(\mu')\dagger}(1+\gamma^0){\bf B}_{k'_1}^{-1}{\bf B}_{u'}^{-1}O{\bf B}_u{\bf B}_{k_1}(1+\gamma^0)\chi^{(\mu)}\right)
\eeq

\noi since ${{1+\gamma^0} \over 2}$ commutes with the Wigner rotations, and we have made explicit the rotations in terms of boost matrices according to (\ref{1-3e}). In the last equation $O$ denotes simply the Dirac matrix in the current $\overline{c}Ob$.\par
In terms of the boosted spinors 
\beq
\label{1-9e}
\chi^{(\mu)}_u = {\bf B}_u \chi^{(\mu)}\ ,  \qquad \qquad \overline{\chi}_u^{(\mu')} = \overline{\chi}^{(\mu')}{\bf B}_{u'}^{-1} 
\eeq

\noi the spinor matrix element in (\ref{1-8e}) writes 
$$\left(\chi^{(\mu')\dagger}(1+\gamma^0){\bf B}_{k'_1}^{-1}{\bf B}_{u'}^{-1}O{\bf B}_u{\bf B}_{k_1}(1+\gamma^0)\chi^{(\mu)}\right)$$
\beq
\label{1-10e}
= \left(\overline{\chi}_{u'}^{(\mu')}{\bf B}_{u'}(1+\gamma^0){\bf B}_{k'_1}^{-1}{\bf B}_{u'}^{-1}O{\bf B}_u{\bf B}_{k_1}(1+\gamma^0){\bf B}_u^{-1}\chi^{(\mu)}_u\right)
\eeq

\noi and using the identities 
\beq
\label{1-11e}
{\bf B}_u{\bf B}_{k_1}(1+\gamma^0){\bf B}_u^{-1} = {(m_1+ {/\hskip - 2 truemm p}_1)(1+ {/\hskip - 2 truemm u}) \over \sqrt{2m_1(k_1^0+m_1)}}\ , \ {\bf B}_{u'}(1+\gamma^0){\bf B}_{k'_1}^{-1}{\bf B}_{u'}^{-1} = {(1+ {/\hskip - 2 truemm u}')(m'_1+ {/\hskip - 2 truemm p}'_1) \over \sqrt{2m'_1(k_1^{'0}+m'_1)}}
\eeq 

\noi one gets the formula in the $4 \times 4$ form 
\beq
\label{1-12e}
\left( \chi^{(\mu')\dagger}D'_1({\bf R}^{'-1}_1) O({\bf p}'_1,{\bf p}_1) D_1({\bf R}_1)\chi^{(\mu)} \right) = {1 \over 4}\ \sqrt{{m'_1m_1 \over p^{'0}_1 p^{0}_1}} 
\eeq 
$$\times {1 \over \sqrt{2m_1(k_1^0+m_1)}}\ {1 \over \sqrt{2m'_1(k_1^{'0}+m'_1)}} \left(\overline{\chi}_{u'}^{(\mu')} (1+ {/\hskip - 2 truemm u}')(m'_1+ {/\hskip - 2 truemm p}'_1) O (m_1+ {/\hskip - 2 truemm p}_1)(1+ {/\hskip - 2 truemm u}) \chi^{(\mu)}_u\right)$$

We have now to compute the trace in formula (\ref{1-7e}) that reads, in the $4 \times 4$ Dirac matrix formalism and in terms of the boost matrices :
\beq
\label{1-13e}
Tr\left[D({\bf R}_3^{-1}{\bf R}'_3{\bf R}_2^{'-1}{\bf R}_2) \right] \to {1 \over 2} \ Tr \left[(1+\gamma^0){\bf R}_3^{-1}{\bf R}'_3{\bf R}_2^{'-1}{\bf R}_2 \right] 
\eeq 
$$ = {1 \over 8}\ Tr\left[(1+\gamma^0){\bf B}_{k_3}^{-1}{\bf B}_{u}^{-1}{\bf B}_{p_3}{\bf B}_{p'_3}^{-1}{\bf B}_{u'}{\bf B}_{k'_3}(1+\gamma^0){\bf B}_{k'_2}^{-1}{\bf B}_{u'}^{-1}{\bf B}_{p'_2}{\bf B}_{p_2}^{-1}{\bf B}_{u}{\bf B}_{k_2}(1+\gamma^0) \right]$$
$$ = {1 \over 16}\ Tr\left[{\bf B}_{u}(1+\gamma^0){\bf B}_{k_3}^{-1}{\bf B}_{u}^{-1}{\bf B}_{u'}{\bf B}_{k'_3}(1+\gamma^0){\bf B}_{u'}^{-1}{\bf B}_{u'}(1+\gamma^0){\bf B}_{k'_2}^{-1}{\bf B}_{u'}^{-1}{\bf B}_{u}{\bf B}_{k_2}(1+\gamma^0){\bf B}_{u}^{-1} \right]$$

\noi because $1+\gamma^0$ commutes with the Wigner rotations, the quarks 2, 3 are spectators and then one has $p_2 = p'_2$ and $p_3 = p'_3$ and we have inserted the products ${\bf B}_{u}^{-1}{\bf B}_{u} = {\bf B}_{u'}^{-1}{\bf B}_{u'} = 1$ within the trace.\par

We now use relations of the type (\ref{1-11e}) and $(1+ {/\hskip - 2 truemm u})(1+ {/\hskip - 2 truemm u}) = 2(1+ {/\hskip - 2 truemm u})$, $(m+ {/\hskip - 2 truemm p}_2)(m+ {/\hskip - 2 truemm p}_2) = 2m(m+ {/\hskip - 2 truemm p}_2)$..., and one finally gets
$$Tr\left[D({\bf R}_3^{-1}{\bf R}'_3{\bf R}_2^{'-1}{\bf R}_2) \right] = {1 \over 4} {1 \over \sqrt{(k_2^0+m)(k_2^{'0}+m)(k_3^0+m)(k_3^{'0}+m)}}$$
\beq
\label{1-14e}
\times\ Tr\left[ (1+ {/\hskip - 2 truemm u}){(m+/\hskip - 2 truemm p}_2)(1+ {/\hskip - 2 truemm u'})(m+{/\hskip - 2 truemm p}_3) \right]
\eeq

\noi The computation of the trace finally gives
\beq
\label{1-15e}
Tr\left[D({\bf R}_3^{-1}{\bf R}'_3{\bf R}_2^{'-1}{\bf R}_2) \right] = {1 \over \sqrt{(k_2^0+m)(k_2^{'0}+m)(k_3^0+m)(k_3^{'0}+m)}}
\eeq
$$\times\ \left[m^2(1+u.u')+m(u+u').(p_2+p_3)+p_2.p_3+(u.p_2)(u'.p_3)+(u.p_3)(u'.p_2)-(u.u')(p_2.p_3) \right]$$

So one gets finally the matrix element
$$< {\bf P}', \mu' \mid O \mid {\bf P}, \mu >\ =  {1 \over 8} \int {d {\bf p}_2 \over (2 \pi)^3}\ {d {\bf p}_3 \over (2 \pi)^3}\ \sqrt{{\sum_j p_j^{'0} \sum_k p_k^{0} \over M'_0 M_0}}\ \prod_{i=1}^{3} \sqrt{k_i^{'0}k_i^{0} \over p_i^{'0}p_i^{0}}\ \varphi({\bf k}'_2, {\bf k}'_3)^* \varphi({\bf k}_2, {\bf k}_3)$$
$$\times\ {1 \over \sqrt{p^{'0}_1 p^{0}_1}}\ {1 \over \sqrt{(k_1^0+m_1)(k_1^{'0}+m'_1)}}\ {1 \over \sqrt{(k_2^0+m)(k_2^{'0}+m)(k_3^0+m)(k_3^{'0}+m)}}$$  
$$\left[m^2(1+u.u')+m(u+u').(p_2+p_3)+p_2.p_3+(u.p_2)(u'.p_3)+(u.p_3)(u'.p_2)-(u.u')(p_2.p_3) \right]$$
\beq
\label{1-16e}
\times \left(\overline{\chi}_{u'}^{(\mu')} (m'_1+ {/\hskip - 2 truemm p}'_1) O (m_1+ {/\hskip - 2 truemm p}_1) \chi^{(\mu)}_u\right) 
\eeq 

\noi because $(1+ {/\hskip - 2 truemm u}) \chi^{(\mu)}_u = 2\chi^{(\mu)}_u\ , \overline{\chi}_{u'}^{(\mu')} (1+ {/\hskip - 2 truemm u}') = 2\overline{\chi}_{u'}^{(\mu')}$.

In the heavy quark limit \cite{COVARIANT-QM} one has
$$(u, u') \to (v, v')\ ,\qquad \ \ \  \left({p_1 \over m_1}\ ,\ {p'_1 \over m'_1} \right) \to (v, v')\ , \qquad \ \ \  {\sum_j p_j^{'0} \sum_k p_k^{0} \over M'_0 M_0} \to v^0v'^0$$
$${k_1^0 \over m_1}\ ,\ {k_1^{'0} \over m'_1} \to 1\ ,\qquad \ \ \ (k_2^0, k_2^{'0}) \to (p_2.v, p_2.v')\ , \qquad \ \ \ (k_3^0, k_3^{'0}) \to (p_3.v, p_3.v')$$

\noi and since $p_2 = p'_2,\ p_3 = p'_3$ for the spectator quarks, one gets the heavy quark limit matrix element :
$$< {\bf P}', \mu' \mid O \mid {\bf P}, \mu >\ =  {1 \over \sqrt{v^0v'^0}} \int {d {\bf p}_2 \over (2 \pi)^3}\ {1 \over p_2^0} {d {\bf p}_3 \over (2 \pi)^3}\ {1 \over p_3^0}\ \varphi({\bf k}'_2, {\bf k}'_3)^* \varphi({\bf k}_2, {\bf k}_3)$$
$$\times\ {1 \over 2}\ {\sqrt{(p_2.v)(p_3.v)(p_2.v')(p_3.v')} \over \sqrt{(p_2.v+m)(p_3.v+m)(p_2.v'+m)(p_3.v'+m)}}$$  
$$\left[m^2(1+v.v')+m(v+v').(p_2+p_3)+(p_2.v)(p_3.v')+(p_3.v)(p_2.v')+(p_2.p_3)(1-v.v') \right]$$
\beq
\label{1-17e}
\times \left(\overline{\chi}_{v'}^{(\mu')} O \chi^{(\mu)}_v\right) 
\eeq 

Finally, identifying with the definition of the Isgur-Wise function within the same normalization convention
\beq
\label{1-18e}
< \Lambda_b({\bf P}', \mu') \mid O \mid \Lambda_c({\bf P}, \mu) >\ =  {1 \over \sqrt{v^0v'^0}}\ \xi_\Lambda(v.v') \ \left(\overline{\chi}_{v'}^{(\mu')}O\chi^{(\mu)}_v\right)
\eeq 

\noi one gets
\beq
\label{1-19e}
\xi_\Lambda(v.v') =  \int {d {\bf p}_2 \over (2 \pi)^3}\ {1 \over p_2^0} {d {\bf p}_3 \over (2 \pi)^3}\ {1 \over p_3^0}\  \sqrt{(p_2.v)(p_3.v)(p_2.v')(p_3.v')}
\eeq
$$\times\ \varphi({\bf k}'_2, {\bf k}'_3)^* \varphi({\bf k}_2, {\bf k}_3)$$
$$\times\ {\left[m^2(1+v.v')+m(v+v').(p_2+p_3)+(p_2.v)(p_3.v')+(p_3.v)(p_2.v')+(p_2.p_3)(1-v.v') \right] \over 2 \sqrt{(p_2.v+m)(p_3.v+m)(p_2.v'+m)(p_3.v'+m)}}$$

\noi where the arguments of the internal wave function are the three-dimensional parts of the four-vectors
\beq
\label{1-20e}
k_i = {\bf B}_v^{-1} p_i\ , \qquad \qquad k'_i = {\bf B}_{v'}^{-1} p_i \qquad \qquad (i = 2, 3)
\eeq

The factor $\sqrt{(p_2.v)(p_3.v)(p_2.v')(p_3.v')}$ in the first line of (\ref{1-19e}) comes from the Jacobian, and the last line comes from the Wigner rotations.\par
One can observe that the expression of the Isgur-Wise function (\ref{1-19e}) is fully covariant, in particular due to the Lorentz invariant measures ${d {\bf p}_i \over p_i^0}\ (i = 2, 3)$.\par 
As we will see below, to get covariance of the IW function one needs in (\ref{1-19e}) rotational invariance of the internal wave functions.\par
For $v.v' = 1$ one finds that, due to the normalization of the internal wave function, the Isgur-Wise function is correctly normalized : 
\beq
\label{1-21e} 
\xi_\Lambda(1) = \int {d {\bf p}_2 \over (2 \pi)^3}\ {d {\bf p}_3 \over (2 \pi)^3} \mid \varphi({\bf p}_2, {\bf p}_3) \mid^2\ = 1
\eeq

The expressions of the IW functions in the baryon case (\ref{1-19e}) contain Lorentz invariant measures $ {d {\bf p}_i \over p_i^0}$ and a Lorentz invariant kernel. However, we have the product of wave functions $\varphi({\bf k}'_2, {\bf k}'_3)^* \varphi({\bf k}_2, {\bf k}_3)$ and to show that both IW functions are Lorentz invariant, we need to demonstrate that these products are also Lorentz invariant.\par

Consider now the product of wave functions
\beq
\label{1-31e}
\varphi({\bf k}'_2,{\bf k}'_3)^* \varphi({\bf k}_2,{\bf k}_3) = \varphi(\overrightarrow{{\bf B}^{-1}_{v'} p_2}, \overrightarrow{{\bf B}^{-1}_{v'} p_3})^* \varphi(\overrightarrow{{\bf B}^{-1}_v p_2}, \overrightarrow{{\bf B}^{-1}_v p_2})
\eeq

\noi The radial wave functions are rotational invariant, so that they can be redefined as follows :
\beq
\label{1-32e}
\varphi({\bf k}_2,{\bf k}_3) = \psi({\bf k}_2^2,{\bf k}_3^2,{\bf k}_2.{\bf k}_3)
\eeq

\noi and similarly for $\varphi({\bf k}'_2,{\bf k}'_3)$.\par

One has
\beq
{\bf k}_2^2 = (\overrightarrow{{\bf B}^{-1}_{v} p_2})^2 = (({\bf B}^{-1}_{v} p_2)^0)^2-m^2 = (p_2.v)^2-m^2
\label{1-28e}
\eeq

\noi where the last equalities follow from the invariance of the scalar product because, defining the four-vector $v_0 = (1,{\bf 0})$, one has 
\beq
\label{1-29e}
k_2^0 = ({\bf B}^{-1}_{v} p_2)^0 = ({\bf B}^{-1}_{v} p_2).v_0 = p_2.({\bf B}_{v}v_0) = p_2.v
\eeq

What is missing are the three-dimencional scalar products like
\beq
\label{1-33e}
{\bf k}_2.{\bf k}_3 = (\overrightarrow{{\bf B}^{-1}_{v} p_2}).(\overrightarrow{{\bf B}^{-1}_{v} p_3}) = ({\bf B}^{-1}_{v} p_2)^0 ({\bf B}^{-1}_{v} p_3)^0-({\bf B}^{-1}_{v} p_2).({\bf B}^{-1}_{v} p_3)
\eeq

\noi and using relations (\ref{1-29e}) and the invariance of the scalar product, we have
\beq
{\bf k}_2.{\bf k}_3 = (p_2.v)(p_3.v)-(p_2.p_3)
\eeq

Finally, we have the Lorentz scalar wave function
\beq
\varphi({\bf k}_2,{\bf k}_3) = \psi((p_2.v)^2-m^2,(p_3.v')^2-m^2,(p_2.v)(p_3.v)-(p_2.p_3))
\eeq

\noi and similarly for $\varphi({\bf k}'_2,{\bf k}'_3)$.

Finally the baryon Isgur-Wise function writes in the explicit Lorentz invariant form
$$\xi_\Lambda(v.v') = \int {d {\bf p}_2 \over (2 \pi)^3}\ {1 \over p_2^0} {d {\bf p}_3 \over (2 \pi)^3}\ {1 \over p_3^0}\ \sqrt{(p_2.v)(p_3.v)(p_2.v')(p_3.v')}$$
$$\times\ \psi((p_2.v')^2-m^2,(p_3.v')^2-m^2,(p_2.v')(p_3.v')-(p_2.p_3))^*$$
\beq
\label{1-35e}
\times\ \psi((p_2.v)^2-m^2,(p_3.v)^2-m^2,(p_2.v)(p_3.v)-(p_2.p_3))
\eeq
$$\times\ {\left[m^2(1+v.v')+m(v+v').(p_2+p_3)+(p_2.v)(p_3.v')+(p_3.v)(p_2.v')+(p_2.p_3)(1-v.v') \right] \over 2 \sqrt{(p_2.v+m)(p_3.v+m)(p_2.v'+m)(p_3.v'+m)}}$$

\vskip 8 truemm
%
%
\section*{Appendix C}
\setcounter{section}{3}
\setcounter{subsection}{0}
%
\subsection*{The Q pointlike-diquark IW functions in the BT scheme}
%
%
%
\subsection{The elastic IW function}
%
%
This case is much simpler than the three-quark one $Qqq$ because the diquark is in a $S = 0, L = 0$ state, and there are no Wigner rotations on the spectator diquark.\par
The matrix element reads
\beq
\label{D-1e}
< {\bf P}', \mu' \mid O \mid {\bf P}, \mu >\ = \int {d {\bf p}_2 \over (2 \pi)^3}\ \sqrt{{\sum_j p_j^{'0} \sum_k p_k^{0} \over M'_0 M_0}}\ \prod_{i=1}^{2} \sqrt{k_i^{'0}k_i^{0} \over p_i^{'0}p_i^{0}}\  \varphi'({\bf k}'_2)^* \varphi({\bf k}_2)
\eeq
$$\times\ Tr\left[\overline{\chi}^{(\mu')+}D({\bf R}^{'-1}_1) O({\bf p}'_1,{\bf p}_1) D({\bf R}_1) \chi^{(\mu)}\right]$$

\noi that gives, in the heavy quark limit
\beq
\label{D-2e}
< {\bf P}', \mu' \mid O \mid {\bf P}, \mu >\ = {1 \over \sqrt{v^0 v^{'0}}} \left(\overline{\chi}^{(\mu')+}_{v'} O \chi^{(\mu)}_v \right) \int {d {\bf p}_2 \over (2 \pi)^3}\ {1 \over p_2^0} \sqrt{(p_2.v)(p_2.v')}\ \varphi'({\bf k}'_2)^* \varphi({\bf k}_2)
\eeq

\noi corresponding to the simple expression of the IW function
\beq
\label{D-3e}
\xi_{\Lambda}(v.v') = \int {d {\bf p}_2 \over (2 \pi)^3}\ {1 \over p_2^0} \sqrt{(p_2.v)(p_2.v')}\ \varphi'({\bf k}'_2)^* \varphi({\bf k}_2)
\eeq

\noi that is covariant because ${\bf k}_2^2 = (p_2.v)^2-m^2, {\bf k}_2^{' 2} = (p_2.v')^2-m^2$, where $m$ is the diquark mass, and is correctly normalized, $\xi_{\Lambda}(1) = 1$.
%
%
\subsection{The inelastic $\boldsymbol{L = 0 \to L = 1}$ IW function}
\beq
\label{D-4e}
< {\bf P}', \mu' \mid O \mid {\bf P}, \mu >\ = \int {d {\bf p}_2 \over (2 \pi)^3}\ \sqrt{{\sum_j p_j^{'0} \sum_k p_k^{0} \over M'_0 M_0}}\ \prod_{i=1}^{2} \sqrt{k_i^{'0}k_i^{0} \over p_i^{'0}p_i^{0}}\ \varphi({\bf k}_2)
\eeq
$$\times\  \sum_{s'_1,s_1} \varphi'^{(\mu')*}_{s'_1}({\bf k}'_2)\left[\overline{\chi}^{(\mu')+}D({\bf R}^{'-1}_1) O({\bf p}'_1,{\bf p}_1) D({\bf R}_1) \chi^{(\mu)} \right]_{s'_1 s_1} \varphi^{(\mu)}_{s_1}({\bf k}_2)$$

\noi where
$$\varphi^{(\mu)}_{s_1}({\bf k}_2)  = \chi^{(\mu)}_{s_1} \varphi({\bf k}_2)$$
\beq
\label{D-5e}
\varphi'^{(\mu')*}_{s'_1}({\bf k}'_2) = \sum_{m'} < 1\ \mu'-m', {1 \over 2} m' \mid J\ \mu' > Y_1^{\mu'-m} \chi_{s'_1}^{(m')} 
\eeq

\noi where $J =  {1 \over 2}$ or $J =  {3 \over 2}$.\par
The sum over the Clebsch-Gordan coefficients can be written as
$$\sum_{m'} < 1\ \mu'-m', {1 \over 2} m' \mid J\ \mu' > Y_1^{\mu'-m} \chi^{(m')}$$
$$= {1 \over \sqrt{4 \pi}} {1 \over \mid{\bf k}'_2\mid} \sum_{m'} (-1)^{{1 \over 2} + m'} \left[{\bf \sigma}^{(\mu'-m')} i \sigma_2 \right]_{\mu',-m'} \chi^{(m')} \left({\bf k}'_2 \right)^{\mu'-m'}$$
$$= - {1 \over \sqrt{4 \pi}} {1 \over \mid{\bf k}'_2\mid} \left({\bf \sigma}.{\bf k}'_2 \right) \chi^{(\mu')}$$

\noi Passing now to the $4 \times 4$ matrix formulation and taking the heavy quark limit, one finds, after some algebra,
\beq
\label{D-6e}
< {\bf P}', \mu' \mid O \mid {\bf P}, \mu >\ = \ {1 \over \sqrt{v^0 v^{'0}}}\ {1 \over 4}\ \int {d {\bf p}_2 \over (2 \pi)^3}\ {1 \over p_2^0}\ \varphi'({\bf k}'_2)^* \varphi({\bf k}_2)
\eeq
$$\times\  {1 \over \sqrt{(p_2.v')^2-m^2}}\ \overline{\chi}^{(\mu')}_{v'}\left\{ \left[ {/ \hskip - 2 truemm p_2} - (p_2.v'){/ \hskip - 2 truemm v'} \right] \gamma_5 (1+{/ \hskip - 2 truemm v'}) O (1+{/ \hskip - 2 truemm v}) \right\} \chi^{(\mu)}_{v}$$

Particularizing to $J = {1 \over 2}$ and identifying to the HQET matrix element defining the Isgur-Wise function $\sigma(w)$ \cite{LEIBOVICH-STEWART}
\beq
\label{D-7e}
< {\bf P}', \mu' \mid O \mid {\bf P}, \mu >\ = {\sigma(w) \over \sqrt{3}}\ \left[ \overline{\chi}^{(\mu')}_{v'} \gamma_5 ({/ \hskip - 2 truemm v}+w) O \chi^{(\mu)}_{v} \right]
\eeq

\noi one finds
\beq
\label{D-8e}
\sigma(w) = {\sqrt{3} \over w^2-1}  \int {d {\bf p}_2 \over (2 \pi)^3}\ {1 \over p_2^0}\ \sqrt{(p_2.v)(p_2.v')} \varphi'({\bf k}'_2)^* \varphi({\bf k}_2)
\eeq
$$ \times  {p_2.(v-wv') \over \sqrt{(p_2.v')^2-m^2)^2 - m^2}}$$
\vskip 6 truemm
\section*{Appendix D}
%
\subsection*{Bing Chen {\bfseries{\itshape et al.}} $\boldsymbol{L = 0}$ and $\boldsymbol{L = 1}$ wave functions in the quark-diquark model}
%
%
In an expansion in terms of $L = 0$ and $L = 1$ on harmonic oscillator bases,
\beq
\label{1-50-30e}
\varphi^{(n)}_0 ({\bf p}) = (-1)^n (4\pi)^{3/4} 2^n \sqrt{{(n!)^2 \over (2n+1)!}}\ {1 \over \beta^{3/2}}\ L^{1/2}_n \left({{\bf p}^2 \over \beta^2}\right) \exp \left(- { {\bf p}^2 \over {2 \beta^2}} \right)
\eeq
\beq
\label{1-50-31e}
\varphi^{(n)}_1 ({\bf p}) = (-1)^n (4\pi)^{3/4} 2^{n+1} \sqrt{{n!(n+1)! \over (2n+3)!}}\ {1 \over \beta^{5/2}} \mid {\bf p}\mid L^{3/2}_n \left({{\bf p}^2 \over \beta^2}\right) \exp \left(- { {\bf p}^2 \over {2 \beta^2}} \right)
\eeq

\noi the ground state wave function reads, with the calculation of the wave function in the heavy quark limit,
$$\varphi_0 ({\bf p}) = -\ 0.9940325\ \varphi^{(0)}_0 ({\bf p}) - 8.5672485 \times 10^{-3}\ \varphi^{(1)}_0 ({\bf p})$$
$$ -\ 9.9527270 \times 10^{-2}\ \varphi^{(2)}_0 ({\bf p}) - 2.4497384 \times 10^{-2}\ \varphi^{(3)}_0 ({\bf p})$$  
\beq
\label{1-50-32e}
-\ 2.7361497 \times 10^{-2}\ \varphi^{(4)}_0 ({\bf p}) - 1.4908912 \times 10^{-2}\ \varphi^{(5)}_0 ({\bf p})
\eeq
$$-\ 
 1.2411494 \times 10^{-2}\ \varphi^{(6)}_0 ({\bf p}) - 9.4764605 \times 10^{-3}\ \varphi^{(7)}_0 ({\bf p})$$
$$ -\ 
 6.3898186 \times 10^{-3}\ \varphi^{(8)}_0 ({\bf p})- 8.0367858 \times 10^{-3}\ \varphi^{(9)}_0 ({\bf p})$$

\noi and the $L = 1$ wave function
$$\varphi_1 ({\bf p}) = 0.9482319\ \varphi^{(0)}_1 ({\bf p}) - 0.2740721\ \varphi^{(1)}_1 ({\bf p})$$
$$+\ 0.1497750\ \varphi^{(2)}_1 ({\bf p}) - 4.7684737 \times 10^{-2}\ \varphi^{(3)}_1 ({\bf p})$$
\beq
\label{1-50-33e} 
+\ 3.0210067 \times 10^{-2}\ \varphi^{(4)}_1 ({\bf p}) - 7.8150993 \times 10^{-3}\ \varphi^{(5)}_1 ({\bf p})
\eeq
$$+\ 7.4121789 \times 10^{-3}\ \varphi^{(6)}_1 ({\bf p}) - 5.9317378 \times 10^{-4}\ \varphi^{(7)}_1 ({\bf p})$$ 
$$+\ 2.1176776 \times 10^{-3}\ \varphi^{(8)}_1 ({\bf p}) + 
 9.3134667 \times 10^{-4}\ \varphi^{(9)}_1 ({\bf p})$$
\vskip 6 truemm
\section*{Appendix E}
%
\subsection*{Helicity amplitudes and observables}
%
%
The expressions for the helicity amplitudes and observables as formulated by Gutsche {\itshape et al.} \cite{MAINZ LAMBDAb LAMBDAcSTAR} are summarized here.\par
In terms of $V-A$ chiral helicity amplitudes
\beq
\label{C-1e}
H_{\lambda_2,\lambda_W} = H^V_{\lambda_2,\lambda_W}-H^A_{\lambda_2,\lambda_W}
\eeq

\noi ($\lambda_2$ : helicity of the final $\Lambda_c$, $\lambda_W$ : helicity of the final virtual $W$)

\vskip 6 truemm

Gutsche {\itshape et al.} define the following bilinears in terms of helicity amplitudes
$${\cal H}_U =\ \mid H_{+{1\over2}+1} \mid^2 + \mid H_{-{1\over2}-1} \mid^2, \qquad \qquad {\cal H}_P =\ \mid H_{+{1\over2}+1} \mid^2 - \mid H_{-{1\over2}-1} \mid^2$$
$${\cal H}_L =\ \mid H_{+{1\over2}0} \mid^2 + \mid H_{-{1\over2}0} \mid^2, \qquad \qquad \qquad \ \ {\cal H}_{L_P} =\ \mid H_{+{1\over2}0} \mid^2 - \mid H_{-{1\over2}0} \mid^2 \ \ \ \ \ \ \ \ \ $$
\beq
\label{C-2e}
{\cal H}_S =\ \mid H_{+{1\over2}t} \mid^2 + \mid H_{-{1\over2}t} \mid^2, \qquad \qquad \qquad \ \ \ {\cal H}_{S_P} =\ \mid H_{+{1\over2}t} \mid^2 - \mid H_{-{1\over2}t} \mid^2 \ \ 
\eeq
$${\cal H}_{LT} = Re\left( H_{+{1\over2}+1} H_{-{1\over2}0}^\dagger + H_{+{1\over2}0} H_{-{1\over2}-1}^\dagger \right), \qquad {\cal H}_{LT_P} = Re\left( H_{+{1\over2}+1} H_{-{1\over2}0}^\dagger - H_{+{1\over2}0} H_{-{1\over2}-1}^\dagger \right)$$
$${\cal H}_{ST} = Re\left( H_{+{1\over2}+1} H_{-{1\over2}t}^\dagger + H_{+{1\over2}t} H_{-{1\over2}-1}^\dagger \right), \qquad \ {\cal H}_{LT_P} = Re\left( H_{+{1\over2}+1} H_{-{1\over2}t}^\dagger - H_{+{1\over2}t} H_{-{1\over2}-1}^\dagger \right)$$
$${\cal H}_{SL} = Re\left( H_{+{1\over2}0} H_{+{1\over2}t}^\dagger + H_{-{1\over2}0} H_{-{1\over2}t}^\dagger \right), \qquad \ \ \ \ {\cal H}_{SL_P} = Re\left( H_{+{1\over2}0} H_{+{1\over2}t}^\dagger - H_{-{1\over2}0} H_{-{1\over2}t}^\dagger \right)$$

\vskip 4 truemm

\noi where the left (right) column corresponds to parity conserving (parity violating) quantities, and
\beq
\label{C-3e}
{\cal H}_{tot} = {\cal H}_U + {\cal H}_L + \delta_\ell ({\cal H}_U + {\cal H}_L + 3 {\cal H}_S)
\eeq

\noi with the dependence on the lepton mass given by
\beq
\label{C-4e}
\delta_\ell = {m_\ell^2 \over {2 q^2}}
\eeq

\vskip 3 truemm

In terms of these quantities, the interesting observables read as below.
%
%
\subsubsection*{Differential rate}
\beq
\label{C-5e}
{d\Gamma \over dq^2} = \Gamma_0 {(q^2-m_\ell^2)^2 \mid {\bf p}_2 \mid \over M_1^7 q^2} {\cal H}_{tot}
\eeq

\noi where
\beq
\label{C-6e}
\Gamma_0 = {G_F^2\mid V_{cb} \mid^2 M_1^5 \over 192 \pi^3}
\eeq
%
\subsubsection*{Forward-backward asymmetry}
%
\beq
\label{C-7e}
A^\ell_{FB}(q^2) = {{d\Gamma(F)-d\Gamma(B)} \over {d\Gamma(F)+d\Gamma(B)}} = - {3 \over 2} {{\cal H}_P+4 \delta_\ell {\cal H}_{SL} \over {\cal H}_{tot}}
\eeq
%
\subsubsection*{Convexity parameter {\normalfont (second derivative of the zenithal angular distribution})}
%
\beq
\label{C-8e}
C_F(q^2) = {1 \over {\cal H}_{tot}} {d^2W(\theta) \over d(\cos \theta)^2} = {3 \over 4} (1-2 \delta_\ell ) {{\cal H}_U-2 {\cal H}_L \over {\cal H}_{tot}}
\eeq
%
\subsubsection*{Longitudinal hadron polarization}
%
\beq
\label{C-9e}
P_z^h(q^2) = {{\cal H}_P + {\cal H}_{L_P} + \delta_\ell ({\cal H}_P + {\cal H}_{L_P} + 3 {\cal H}_{S_P}) \over {\cal H}_{tot}}
\eeq
%
\subsubsection*{Tranverse hadron polarization}
%
\beq
\label{C-10e}
P_x^h(q^2) = - {3 \pi \over {4 \sqrt{2}}} {{\cal H}_{LT} - 2 \delta_\ell {\cal H}_{ST_P} \over {\cal H}_{tot}}
\eeq
%
\subsubsection*{Longitudinal lepton polarization}
%
\beq
\label{C-11e} 
P_z^\ell(q^2) = - {{\cal H}_U + {\cal H}_L - \delta_\ell ({\cal H}_U + {\cal H}_L + 3 {\cal H}_S) \over {\cal H}_{tot}}
\eeq
%
\subsubsection*{Transverse lepton polarization} 
%
\beq
\label{C-12e} 
P_x^\ell(q^2) = - {3 \pi \over {4 \sqrt{2}}} \sqrt{\delta_\ell} {{\cal H}_P - 2 {\cal H}_{SL} \over {\cal H}_{tot}}
\eeq
\vskip 10 truemm
\section*{Acknowledgment}
%

We are indebted to Misha Ivanov,  Jurgen Koerner and Valery Lyubovitskij for enlighting correspondence on their work on $\Lambda_b$ decays.

\end{document}